%% file: alma_sn1987a_velocity_v1.tex
\documentclass[fleqn,usenatbib]{mnras}

\usepackage{newtxtext,newtxmath}

\usepackage[T1]{fontenc}
\usepackage{ae,aecompl}
\usepackage{soul}
\usepackage[percent]{overpic}

\usepackage{graphicx}	
\usepackage{amsmath}	
\usepackage{multirow}
\usepackage[table]{xcolor}
\usepackage{tikz}
\usepackage{enumitem}

\newcommand{\makepanel}[2]{
\begin{overpic}[width=.28\textwidth]{#1/#2_0.png}
   \put(80,85){\textbf{-90°}}  
    \end{overpic}
    \begin{overpic}[width=.28\textwidth]{#1/#2_-90.png}
   \put(80,85){\textbf{0°}}  
    \end{overpic}
    \begin{overpic}[width=.28\textwidth]{#1/#2_-180.png}
   \put(80,85){\textbf{90°}}  
    \end{overpic}
    \begin{overpic}[width=.38\textwidth]{#1/#2_-45.png}
   \put(80,60){\textbf{-45°}}  
    \end{overpic}
   \begin{overpic}[width=.38\textwidth]{#1/#2_-135.png}
   \put(80,60){\textbf{-45°}}  
    \end{overpic}
     \ifthenelse{\equal{#1}{3d_views}}%
    {\includegraphics[width=.032\textwidth]{#1/colorbar_ocean.png} 
    \includegraphics[width=.032\textwidth]{#1/colorbar_turbo.png}}%
    {\begin{overpic}[width=.030\textwidth]{#1/colorbar_magma.png}\put(0,100){{CO}}\end{overpic} 
    \hspace{1mm}
    \begin{overpic}[width=.030\textwidth]{#1/colorbar_viridis.png}\put(0,100){{SiO}}\end{overpic}}%
}

\newcommand{\makepanelcombi}[2]{
\begin{overpic}[width=.34\textwidth]{#1/#2_0.png}
   \put(80,85){\textbf{-90°}}  
    \end{overpic}
    \begin{overpic}[width=.31\textwidth]{#1/#2_-90.png}
   \put(80,85){\textbf{0°}}  
    \end{overpic}
    \begin{overpic}[width=.34\textwidth]{#1/#2_-180.png}
   \put(80,85){\textbf{90°}}  
    \end{overpic}
    \begin{overpic}[width=.41\textwidth]{#1/#2_-45.png}
   \put(80,60){\textbf{-45°}}  
    \end{overpic}
   \begin{overpic}[width=.41\textwidth]{#1/#2_-135.png}
   \put(80,60){\textbf{-45°}}  
    \end{overpic}
     \ifthenelse{\equal{#1}{3d_views}}%
    {\includegraphics[width=.036\textwidth]{#1/colorbar_ocean.png} 
    \includegraphics[width=.036\textwidth]{#1/colorbar_turbo.png}}%
    {\begin{overpic}[width=.033\textwidth]{#1/colorbar_magma.png}\put(0,100){{CO}}\end{overpic} 
    \hspace{1mm}
    \begin{overpic}[width=.033\textwidth]{#1/colorbar_viridis.png}\put(0,100){{SiO}}\end{overpic}}%
}

\newcommand{\Msun}{{$M_{\odot}$}}

\title[3D Comparison of SN\,1987A: ALMA vs. Simulations]{3D insights into SN\,1987A: ALMA observations compared to hydrodynamical explosion simulations}

\author[R. Wesson et al.]{R. Wesson,$^{1, 2, 3}$\thanks{rw@nebulousresearch.org}
M. Gabler,$^{4}$
M. Lyons,$^{1}$
J. Wildman,$^{1}$
Mikako Matsuura,$^{1}$
H.-T. Janka,$^{5}$
B. Giudici,$^{4}$
\newauthor
P. Cigan,$^{6}$
H.L. Gomez,$^{1}$
R. Indebetouw,$^{7,8}$, 
A.M.S. Richards,$^{9}$, 
and A. Wongwathanarat$^{5}$
\\
$^{1}$ Cardiff Hub for Astrophysics Research and Technology (CHART),
School of Physics and Astronomy, Cardiff University, 
The Parade, Cardiff CF24 3AA, UK\\
$^{2}$
Department of Physics and Astronomy, University College London, Gower Street, London WC1E 6BT, UK \\
$^3$
Department of Physics, Maynooth University, Maynooth, Co Kildare,
Ireland \\
$^{4}$ Departament d'Astronomia i Astrof\'{\i}sica, Universitat de Val\`encia, Edifici d'Investigaci\'{o} Jeroni Mu\~{n}oz, C/Dr. Moliner, 50, E-46100 Burjassot (Val\`encia), Spain\\
$^{5}$Max-Planck-Institute for Astrophysics, Karl-Schwarzschild-Str. 1, 85748 Garching, Germany
$^{6}$ U.S. Naval Observatory, 3450 Massachusetts Ave. NW, Washington, DC 20392-5420, USA\\
$^{7}$ National Radio Astronomy Observatory, 520 Edgemont Road, Charlottesville, VA 22903, USA\\
$^{8}$ Department of Astronomy, University of Virginia, P.O. Box 3818, Charlottesville, VA 22903-0818, USA\\
$^{9}$ Jodrell Bank Centre for Astrophysics, Department of Physics and Astronomy, University of Manchester, Manchester, M13 9PL, UK\\
}

\date{Accepted XXX. Received YYY; in original form ZZZ}

\pubyear{2023}

\begin{document}
\label{firstpage}
\pagerange{\pageref{firstpage}--\pageref{lastpage}}
\maketitle

\begin{abstract}{
We obtain three-dimensional distributions of CO and SiO molecules from high spatial resolution (0.03--0.06") ALMA observations of SN\,1987A at two different epochs. The evolution between these two epochs is consistent with homologous expansion. From these 3D maps, we reconstruct the 3D mass distributions of the ejecta in CO and SiO molecules, which we compare with those obtained by state-of-the-art, long-time hydrodynamical supernova explosion models computed with the \textsc{Prometheus-HotB} code for 10 different progenitors, including both red and blue supergiants.
The models which best match the mass distributions correspond to explosions of binary-merger blue supergiant progenitors; 
at least two such models approximately reproduce the observed CO morphology. In contrast, the SiO velocity distribution and morphology are not as well reproduced in these models, indicating insufficient mixing of Si into the outer layers already at the progenitor stage. The theoretical models suggest a strong correlation between the centre of mass of the densest carbon- and oxygen-rich ejecta and the direction of the neutron-star kick. If such a correlation also applies to the CO emission in the ejecta of SN\,1987A, the kick of the compact remnant is expected to point towards the observer, at an angle of approximately $45^\circ$ to the north.}
\end{abstract}



\section{Introduction}
Supernovae (SNe) play key roles in the chemical and dynamical evolution of the interstellar medium (ISM) of galaxies. Heavy elements are synthesised in the stellar cores of their progenitors, and SN explosions enrich the ISM by expelling these elements into it.
The enormous SN explosion energy ($\sim10^{51}$\,ergs) generates shocks which propagate into the surrounding circumstellar and interstellar material, and hence, SNe are key to the dynamical evolution of the ISM. 

Due to their profound astrophysical significance, SNe have long been and continue to be the focus of extensive theoretical investigation. In core-collapse supernova (CCSN), the core of a massive star collapses under its own gravity and launches a shock wave, once the matter bounces back at nuclear densities. This shock stalls due to energy losses caused by the dissociation of infalling nuclei and intense emission of neutrinos. For most CCSNe, the stalled shock is revived by neutrinos \citep{colgate+1966,arnett1966,chevalier1976, bethe+1985}. Within this delayed neutrino-driven SN paradigm, hydrodynamical instabilities like convection \citep{bethe1990,herant1992,burrows1995,janka1995} and the standing accretion shock instability (SASI) \citep{blondin2003,foglizzo2006,foglizzo2007} play a crucial role in enhancing the neutrino-matter interactions behind the shock. A major breakthrough came with the calculation of self-consistent successful explosions in full 3D \citep[e.g.,][]{kuroda2012, Takiwaki2014, Melson2015,melson+2015, lentz+2015, mueller+2017, muller2019, burrows2019, powell2019, powell2020, bollig2021, vartanyan2022, burrows+2024, janka2024,nakamura+2025}. Such simulations are physically extremely complex and numerically expensive. Therefore, they are usually limited to a few seconds of evolution. To confront theoretical models with observations, however, the long-time evolution of the SN needs to be followed at least until the shock leaves the stellar progenitor, which can take hours to days. Such simulations often rely on simplified neutrino transport \citep{wongwathanarat2010,Wongwathanarat2013,Wongwathanarat:2015jv, Utrobin2021,Gabler2021} or continue more detailed simulations of the early phase \citep{muller2018,stockinger2020,sandoval2021,vartanyan+2025}. 
In parallel, phenomenologically calibrated explosion models have been used to compare to particular SN remnants, such as Cas A \citep{orlando2016} and SN\,1987A \citep{orlando2020,ono2020,nakamura2022,ono2024,2025arXiv250419896O} 

SN\,1987A is the nearest supernova explosion detected in the last 400 years, at a distance of 51.4$\pm1.2$\,kpc \citep{Panagia.1999}.
 Since its detection, observations of SN\,1987A have yielded many discoveries, significantly improving our understanding of the evolution of supernovae and supernova remnants.
 SN\,1987A is young enough that the main body of the ejecta is still freely expanding,
 so its elemental density distribution is unchanged from its state a few months after the explosion \citep{McCray:1993p29839}.
Unlike older SN remnants which have interacted with the interstellar and circumstellar material,
 the expanding ejecta retains the imprint of the dynamics during the early days after the supernova explosion, so that it can still be used to constrain the physical processes which took place at that time \citep{McCray:1993p29839}.

In the last decade, advances in 3D spectral mapping with large telescopes have enabled reconstruction of the 3D structure of the ejecta. In particular, the free expansion allows radial velocities to be translated into the third spatial dimension with much greater confidence than in e.g. turbulent molecular clouds.
Atacama Large Millimeter Array (ALMA) observations of SN\,1987A are a powerful probe of its physical conditions and evolution. Dust and molecules including CO, SiO and HCO$^{+}$ have been detected in the ejecta at millimetre wavelengths, and synchrotron radiation due to shock interaction in the circumstellar ring has been measured
\citep{2013ApJ...773L..34K, Indebetouw:2014bt, Zanardo:2014gu, Matsuura:2017cf, Cigan:2019cl}.
High angular resolution ($<$100mas) CO and SiO images show that the distributions of these molecules are clumpy, in contrast to smoothly distributed H$\alpha$ emission, and that different products (C, O and Si) of nucleosynthesis are located in different places within the ejecta
\citep{Abellan:2017by, Cigan:2019cl}.
In the optical and infrared, the Very Large Telescopes, \textit{Hubble Space Telescope} and \textit{JWST} observations can also be used to reconstruct the spatial distributions of atomic lines at $\sim$100 mas scales in 3D \citep{Larsson:2016bj, Kangas.2021, 2023ApJ...949L..27L,larsson2025}.

These 3D observations can be compared with theoretically predicted distributions of ejecta gas from hydrodynamic simulations. \citet{Abellan:2017by} analyzed the size and spacing of clumps in the ejecta, finding that none of the investigated models at that time matched the observations well. \citet{2025arXiv250419896O} compared model predictions with JWST NIRSpec 3D maps \citep{2023ApJ...949L..27L} and NIRCam images \citep{2024MNRAS.532.3625M}, suggesting that the ejecta consist of two ellipsoidal structures.
With the availability of improved hydrodynamical simulations with more input parameter space, we can identify more suitable explosion models for SN\,1987A.

This paper presents 3D maps of molecular lines, reconstructed from ALMA observations taken between 10,000 and 11,800 days after the explosion. We use these data to determine the CO and SiO mass fractions as a function of velocity relative to the centre of the explosion, for comparison with theoretically-predicted velocity distributions. While a number of the models we analyse are clearly incompatible with the observations, models of binary-merger progenitors which reproduce the observed light curve of SN1987A perform the best in reproducing the observed distribution of CO. However, these models underpredict the amount of SiO at high velocities. 

In Section\,\ref{sec:kick}, we reveal a strong correlation between the kick of the nascent neutron star and the centre of mass of the densest carbon- and oxygen-rich ejecta. Associating the densest ejecta with the region of the strongest emission allows us to tentatively estimate the kick direction expected in SN\,1987A to be towards the observer and roughly $45^\circ$ northward. Before concluding, we discuss our findings and the potential caveats of our analysis.

\section{ALMA observations}

High angular resolution ALMA observations of CO and SiO in SN\,1987A were taken in 2012--2015, and the details of the observations and the data reduction were reported in \citet{Abellan:2017by} and \citet{Cigan:2019cl}. We refer to these data as ``epoch 1'' in our analysis.
More recently, SiO $J$=6--5 observations were taken in 2019, while a second epoch of SiO $J$=5--4 was taken in 2017. We designate these data as ``epoch 2".
The observing logs for both epochs are summarised in Table\,\ref{observing_log}.

The observations were carried out using standard ALMA observing procedures, and processed as described in \citet{Cigan:2019cl}, using CASA \citep{2007ASPC..376..127M} and the ALMA calibration pipeline \citep{hunter23}. 
For epoch 1, the data from the program code 2015.1.00631S are high angular resolution images \citep{Abellan:2017by}, while those from the program code 2013.1.00280.S are from lower angular resolution images \citep{Matsuura:2017cf}, and these data were combined during the data reduction process to improve the data quality. 
The beam sizes of the combined images are listed in Table\,\ref{observing_log}.
Epoch 2 consists of high resolution observations from the program 2017.1.00789.S and lower resolution data from the program 2018.1.00717.S (SiO-5).
Thanks to more accurate telescope positions for data
observed in recent cycles, the astrometric accuracy at S/N 5--10 is
about 15 mas at the lower resolution and 10 mas at the higher
resolution and the flux scale accuracy is $\sim$7\% .

After applying all respective observatory calibrations at each epoch, line-free continuum channels  are identified and used to subtract continuum emission. This is necessary because there is some dust continuum emission from the ejecta at the observed frequencies \citep{Cigan:2019cl}. 
From these cleaned data  we construct a continuum image   
and a spectral line cube. Finally, the velocities are shifted to the reference frame of SN\,1987A, which has a heliocentric velocity of $+$287 km\,s$^{-1}$ (\citealt{Groningsson:2008jb}).

The calibration sources for SiO $J$=6--5 observations in 2019 were
J0601$-$7036 (phase calibrator),
J0533$-$7216 (check source), and
J0519$-$4546 (bandpass calibrator).
The observing conditions were good with precipitable water vapour (PWV) of 0.5\,mm. The baseline lengths ranged from 0.04--5.8\,km, with a maximum recoverable scale of 1\farcs8, sufficient to cover the entire ejecta.
The data cube was constructed at 300 km\,s$^{-1}$ resolution in the line of sight.
The natal weighting for the {\sc CASA} clean process 
produces a synthesised
beam of 115$\times$108 mas$^2$ at position angle (PA) 61$^{\circ}$. The root mean square (rms) of $\sim0.055$ mJy per beam was measured in the area which is clear of SiO emission. These parameters optimised the S/N for sensitivity to
emission on the scales present in SN\,1987A.

The calibration sources for SiO $J$=5--4 observations in 2017 were
J0601$-$7036 (phase calibrator),
J0529$-$7245 (check source) and
J0635-7514 (bandpass calibrator).
The observing night had
PWV 1\,mm.
The baseline lengths were 0.09--12.6\,km, resulting in a maximum recoverable scale of 1\farcs3, again large enough to cover the entire ejecta.
As before, the data cube was constructed for 300\,km\,s$^{-1}$ resolution in line of sight velocity. The synthesised beam was 56$\times$34 mas with PA 47$^{\circ}$.
 These parameters optimised the spatial resolution.
The rms level, measured in a region with no SiO emission, was $\sim0.033$ mJy per beam.


\begin{table*}
\begin{tabular}{lllllrlrrlll}
\hline
Line & $\nu_{\rm o}$ & Epoch & Date & Day & $t_{\rm exp}$ & Beam FWHM & PA & Pixel scale & Velocity & rms \\
&&&&&&&&&resolution\\
& [Hz] & & & & [min] & [arcsec] & [$^\circ$] & [mas] & [km s$^{-1}$]& [mJy/beam] \\
\hline
\input{observing_log}
\end{tabular}
\caption{Journal of the ALMA observational data sets. $t_{\rm exp}$ is the exposure time, $\nu_0$ the central frequency of the band observed; PA is the position angle (measured east from north) of the major axis of the beam; and rms the root mean square of signal-free regions of the images.}
\label{observing_log}
\end{table*}


\subsection{Spectral grids}
\label{sec:spectralgrids}

\begin{figure}
	 \includegraphics[width=.475\textwidth]
     {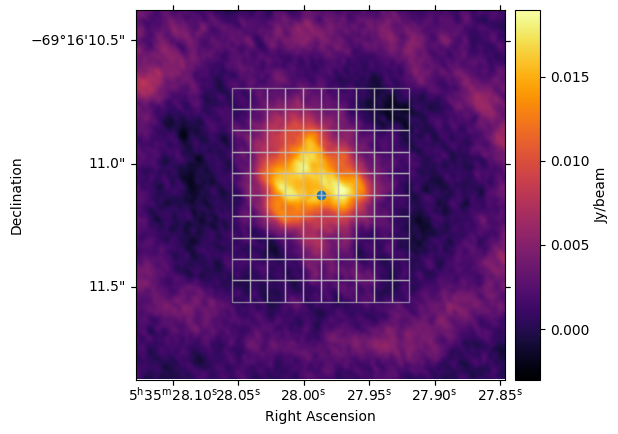}
        \caption{The analysed region, superimposed on the CO $J$=2--1 first epoch image, collapsed along the velocity axis. }
        \label{fig:ejecta_plot}
\end{figure}

The circumstellar medium (CSM) surrounding SN\,1987A's expanding ejecta is believed to have formed about 20,000 yr ago following the merger of a massive binary system which formed the immediate supernova progenitor (\citealt{Morris.2007}). This CSM has a triple-ringed structure, comprising an inner equatorial ring measuring $\sim$1.5$\times$1.7~arcsec, and two much fainter outer rings. The equatorial ring appears elliptical as it is inclined at $\sim$43$^\circ$ to the plane of the sky (\citealt{2011A&A...527A..35T}).

We analyse the velocity data in a region which covers the ejecta but excludes the equatorial ring (Fig.~\ref{fig:ejecta_plot}). The analysed region covers 0.6"$\times$0.6". We divide this region into a grid of 10$\times$10 cells, corresponding to 0.06"$\times$0.06" per cell, which is comparable to the beam size in all data sets.
At day 10,053, this angular size corresponds to a velocity of approximately 531\,km\,s$^{-1}$ at the distance of the supernova, so that the spatial resolution is the limitation in the measurements of the expanding velocities, rather than the velocity channel, which we bin to a resolution of 100 or 300\,km\,s$^{-1}$, depending on the dataset.

Before further analysis of each dataset, we mask out data below a noise threshold. We estimate the noise amplitude in four regions outside the equatorial ring, which are assumed to contain no signal from the circumstellar material. These regions were also divided into 10$\times$10 grid cells, and the mean absolute deviation (MAD) in each cell was calculated. The mean of all MAD values thus calculated in each dataset is taken as a base level of emission below which only noise contributes to the observed fluxes.

The determination of the noise threshold, and the effect of the adopted threshold on the subsequent analyses, are described in more detail in Appendix~\ref{ap:noise}.

\subsection{Integrated line profiles}
\begin{figure}
\includegraphics[width=0.475\textwidth]{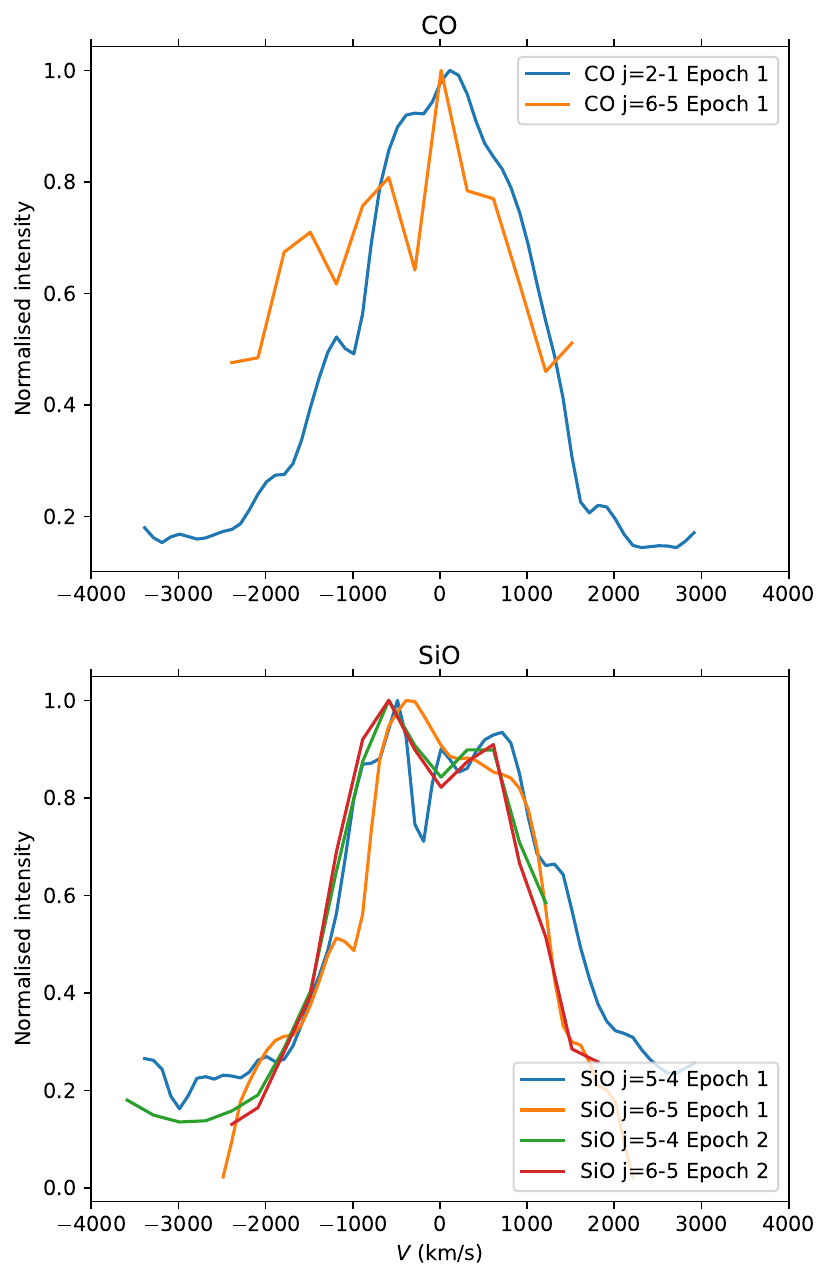}
\caption{
Spatially integrated emission line profiles for each of the six data cubes. Velocities are in the rest frame of the LMC (+287~km\,s$^{-1}$ relative to the Local Standard of Rest (LSR))}
\label{integratedlineprofiles}
\end{figure}
As a first step, we integrate the spectrum over all spatial cells shown in Fig.~\ref{fig:ejecta_plot}. During this integration, we remove data below the previously calculated noise limit. The integrated spectrum is shown in Fig.~\ref{integratedlineprofiles} for the different molecular lines and epochs. The lines are clearly asymmetric and show a local minimum close to zero velocity, indicating a relatively empty region in the very centre of the ejecta. We also note that CO (top panel) seems to be blueshifted, while SiO (bottom panel) has a larger intensity for negative velocities.
We further note that the absolute value of the shifts for CO are smaller than for SiO, indicating a potential different morphology, which is more extended in the plane of the sky for CO.

The lower signal-to-noise ratio of the CO 6--5 data (orange line, top panel) is apparent, and it can also be seen that the SiO 5--4 data from the second epoch of observations (green line, bottom panel) is truncated on the far side at a velocity of +2000~kms$^{-1}$, due to the bandwidth not being sufficient to cover the entire range of velocities. The effect of this on our derived mass-velocity distributions plots is discussed later.

\subsection{3D plots}

We constructed 3D maps of molecular emission by converting pixel coordinates into physical distances from the explosion site. To convert angular distances to physical distances in the $x$-$y$ plane, we adopt the distance to SN\,1987A of 51.4\,kpc found by \citet{Panagia.1999}, who used the timing of the peak of ultraviolet line emission from the equatorial ring to calculate its physical size. More recently, \citet{2019Natur.567..200P} measured a distance to the centre of the LMC of 49.59$\pm$0.09 (statistical)$\pm$0.54 (systematic)~kpc 
using Cepheids. The distance to the SN of 51.4$\pm$1.2\,kpc \citep{Panagia.1999} is consistent with this to within the uncertainties, so we adopt the distance estimated specifically to SN\,1987A.

In the $z$-direction, we first subtracted the systemic velocity of the LMC, $v_{\rm hel}$=+287$\pm$3\,km\,s$^{-1}$ \citep[$v_{\rm LSR}\sim+300$\,km\,s$^{-1}$;][]{Meaburn.1995, Groningsson:2008jb} from the observed velocity values. Then, we assume that the ejecta have been freely expanding since the explosion. The distance from a point to the $x$-$y$ plane is then simply the velocity multiplied by the time since explosion.

The position of the origin of the explosion of SN\,1987A is relatively poorly constrained. No compact object has been directly detected in the remnant of SN\,1987A, but its existence is confirmed by the presence of ionised gas detected via \textit{JWST} observations of spatially-unresolved narrow emission lines of argon and sulfur at the centre of the equatorial ring \citep{fransson2024}. The neutron star (NS) very likely experienced a kick during the explosion. Therefore, its current position may not correspond to the centre of the explosion site \citep{Wongwathanarat2013}.
We use the position determined by \citet{Alp.2018}, who calculated the geometric centre of the equatorial ring by fitting an ellipse to the locations of the centroids of its bright clumps. They calculated a best-fitting position of $\alpha$=05:35:27.9875(11), $\delta$=$-69$:16:11.107(4) (ICRF J2015.0). 3D plots were then generated for all datasets using this estimated centre of explosion.

\begin{figure*}
\makepanel{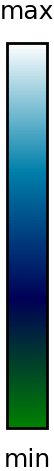}{CO_J=2-1_epoch_1}

\caption{3D intensity maps of CO ($J$=2--1) emission, epoch 1. In this and subsequent figures, the plane of the equatorial ring is indicated in blue, and different colour maps are used for material above and below that plane. An orange point indicates the position of the explosion, and a red arrow points from the SN towards the observer in each panel. The volume represented in each panel measures 40,000 AU on each side and is centred on the supernova explosion position.}
\label{3dmap_co21_c1}
\end{figure*}


\begin{figure*}
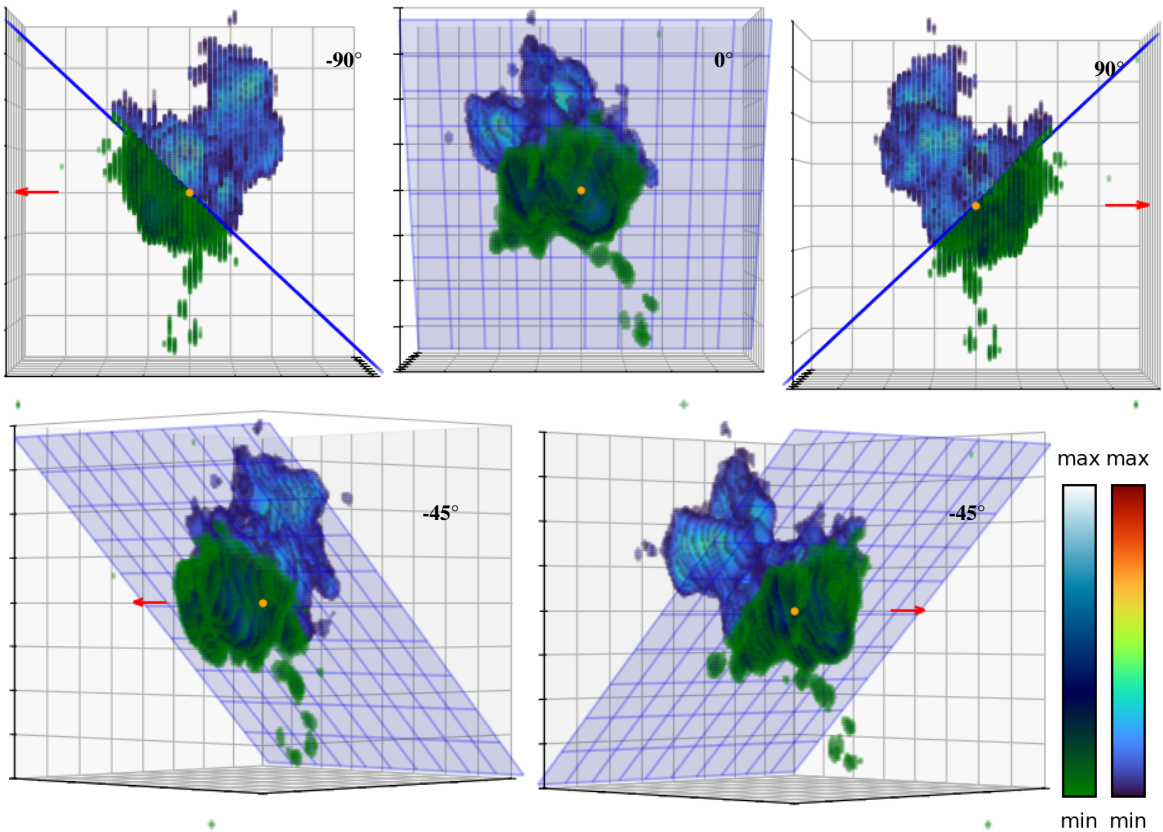

\makepanel{3d_views}{SiO_J=5-4_epoch_1}
\caption{3D intensity maps of SiO ($J$=5--4) emission, epoch 1. All panels show the equatorial ring plane (blue), explosion site (orange point), and observer direction (red arrow). Colour coding distinguishes material above and below the plane. Box size: 40,000 AU.}
\label{3dmap_sio54_c1}
\end{figure*}

\begin{figure*}
\makepanel{3d_views}{SiO_J=5-4_epoch_2}
\caption{3D maps of SiO ($J$=5--4), epoch 2.  All panels show the equatorial ring plane (blue), explosion site (orange point), and observer direction (red arrow). Colour coding distinguishes material above and below the plane. Box size: 40,000 AU. The lower sampling in velocity space compared to the epoch 1 data  causes the apparent gaps in the -90$^\circ$ and +90$^\circ$ panels.}
\label{3dmap_sio54_c2}
\end{figure*}

\begin{figure*}
\makepanel{3d_views}{SiO_J=6-5_epoch_2}
\caption{3D intensity maps of SiO ($J$=6--5) emission, epoch 2.  All panels show the equatorial ring plane (blue), explosion site (orange point), and observer direction (red arrow). Colour coding distinguishes material above and below the plane. Box size: 40,000 AU. The lower sampling in velocity space compared to the epoch 1 data causes the apparent gaps in the -90$^\circ$ and +90$^\circ$ panels.}
\label{3dmap_sio65_c2}
\end{figure*}

The 3D maps of CO and SiO that we obtain at the different epochs are shown in Figs.~\ref{3dmap_co21_c1}-\ref{3dmap_sio65_c2} from different viewing angles. In the plots, we also show the plane of the ring of SN1987A as a blue shaded grid, and the position of the observer is indicated by the red arrow. We omitted a plot for the CO 6--5 at epoch 1, which suffers from much higher noise than the other data cubes, and SiO 6--5 of epoch 1, which is very similar to the SiO 6--5 epoch 2 data but has a lower signal-to-noise ratio per volume element due to higher sampling in velocity space.
In general, the maps of both SiO and CO show qualitatively similar results in all transition lines and at both epochs. The bulk of the ejecta forms a double-lobed structure stretching from north-east (top, left) to south-west, while a high-velocity ``tail'' of material extends further to the south. This morphology makes the ejecta appear moderately asymmetric. Comparing CO ($J$=2--1) in Fig.\,\ref{3dmap_co21_c1} and SiO, exemplarily $J$=5--4 (epoch 1) in Fig.\,\ref{3dmap_sio54_c1}, we note that SiO is slightly less extended, meaning that it has a lower expansion velocity compared to CO. This is in contrast to the more extreme line-of-sight shifts of the spectral lines in Fig.\,\ref{integratedlineprofiles}, and hints at higher CO velocities in the plane of the sky.
The difference in expansion velocities will be discussed quantitatively in section\,\ref{sec:mass_distributions}. In all plots, the maximum emission is not centralized, and from the centre of the explosion there comes only very limited emission. This void region is particularly visible in the SiO data of epoch 2 (Figs.\,\ref{3dmap_sio54_c2} and \ref{3dmap_sio65_c2}).

To quantify the asymmetry of the different molecules at the different epochs, we divide the data cubes into north/south, east/west, and approaching/receding hemispheres with the position of the progenitor lying on the dividing plane. The fractions of the emission coming from the corresponding hemispheres for each data cube are listed in Table~\ref{hemispheres}. Except for CO $J$=6--5, which has the worst noise, the different hemispheric emission fractions agree approximately for the different molecular transition lines and epochs.

Disregarding CO $J$=6--5, epoch 1 data has consistent asymmetries with a ratio of approximately 1:2 for the emissions North:South, Near:Far and Below:Above. In some cases the asymmetry even exceeds 3:7 slightly. The East:West ratio is generally more symmetric at around $\sim$0.6:0.4. Similarly, the two SiO data sets of epoch 2 are more symmetric in the plane of the sky, while the Near:Far ratio is comparable to the other data sets.

\input{hemispheres_table}

\subsection{Mass distributions}\label{sec:mass_distributions}

From the 3D maps obtained in the previous section, we can construct the distribution of the fractional mass of each molecule as a function of the radial velocity. 
To derive these mass distributions, we assume that the ejecta has the following properties:

\begin{enumerate}[leftmargin=2em, itemsep=0.5ex]
  \item it is optically thin at the observed wavelengths;
  \item it is in local thermodynamic equilibrium (LTE); 
  \item it has a uniform temperature.
\end{enumerate}

Under these assumptions, the fractional mass at a given velocity is proportional to the fractional intensity. Accordingly, uncertainties on the fractional masses can be estimated by scaling the uncertainties on the fractional intensities. 

To translate the spatial 3D maps into radial velocities, we assume that the material has been expanding at a constant velocity since the explosion. 
Then, the radial velocity can be obtained simply by dividing the radial distance to each point by the time since explosion at the epoch of the observation (listed in Table~\ref{observing_log}). With these data, we can obtain the mass distributions as a function of the radial velocity.

\begin{figure*}
    \includegraphics[width=\textwidth]{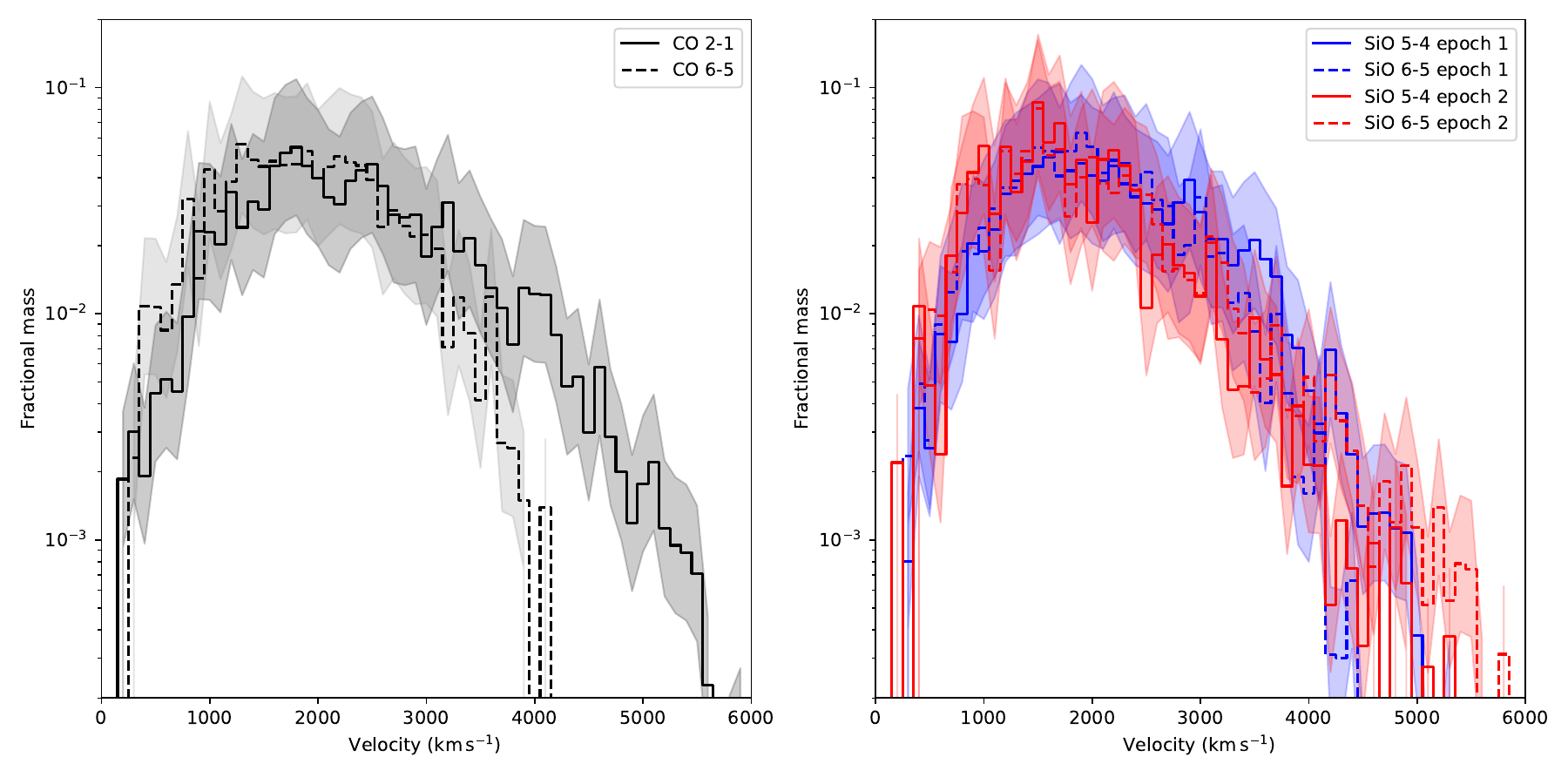}
    \caption{Observed mass-velocity distributions for CO (left) and SiO (right) for different line transitions. On the right panel, curves are plotted for epoch 1 (blue) and epoch 2 (red). 
    }
        \label{fig:observations}
\end{figure*}

Depending on the spectral resolution of the respective data set, the radial velocity was binned in increments of $100\,$km\,s$^{-1}$ or $300\,$km\,s$^{-1}$. The resulting histograms of the different fractional masses are given in Fig.\,\ref{fig:observations} for the different lines of CO and SiO.

To estimate the uncertainties on our mass-velocity distributions, we considered both statistical uncertainties and some systematic effects. These are described fully in Appendix~\ref{ap:noise}. The estimated uncertainty bands are shown in Fig.\,\ref{fig:observations} as shaded regions.

All obtained mass distributions are qualitatively similar with a broad peak between 1500 and 2000 km s$^{-1}$ and an extended tail reaching up to $\gtrsim 5500$ km s$^{-1}$ for the most extreme case (CO $J$=2--1, left panel in Fig.\,\ref{fig:observations}). The strong decrease of the mass distribution at the low-velocity side supports the existence of a void central region, as mentioned for the 3D maps in the previous section and by \citet{Abellan:2017by}. 
The low signal-to-noise ratio of the CO $J$=6--5 data is apparent, and causes the steeper decline of the mass distribution at higher velocities $\lesssim 4000$ km s$^{-1}$. Most of the higher velocity emission is too faint to exceed the noise threshold. Since it is more complete and less noisy than the CO 6--5, we use the CO 2--1 data as the reference of the CO distribution. Differences in the emission profiles of CO 2--1 and CO 6--5 were already pointed out by \cite{Cigan:2019cl}. 

For SiO, we have data from two different epochs, and can investigate a potential time evolution of the mass distribution. The data taken in epoch 2 in 2017/18 (red solid and dashed curves in Fig.\,\ref{fig:observations}) peak at slightly lower velocities ($v_\mathrm{peak}^\mathrm{SiO-2}\approx 1600$km\,s$^{-1}$) than the epoch 1 data ($v_\mathrm{peak}^\mathrm{SiO-1}\approx 1800$km\,s$^{-1}$). This results in a stronger skew towards lower velocities: at low velocities, the red curves of epoch 2 lie above the blue curves of epoch 1, whereas at velocities beyond the peak, the red curves fall below the blue ones. Only at very high velocities $\gtrsim 5000$ km s$^{-1}$ SiO $J$=6--5 of epoch 2 has a higher fraction of mass. This region of the graph is at very low amplitude and, hence, is very sensitive to the noise and the related cut-off of the data. The corresponding noise level of the data in the second epoch is slightly better than in epoch 1 (see Table\,\ref{observing_log}). Therefore, some of the emission from high-velocity ejecta is detected and the extreme velocities of SiO are higher than in the first epoch. We have analysed the influence of the noise in detail in Appendix\,\ref{ap:noise}.
In general, the differences between the velocity curves from the two epochs are of a similar order of magnitude to the estimated uncertainties. Therefore, we cannot draw any strong conclusions on the time evolution and the data are consistent with no significant time evolution. 

In summary, the agreement between the mass distributions and 3D structures at both epochs supports the assumption that the ejecta are freely expanding, and additionally provides evidence that they are optically thin. The difference in time between the epochs corresponds to expansion by a factor of about 1.12 for SiO 5--4 and 1.18 per cent for SiO 6--5. Consequently the volume of the ejecta increases, and the optical depth decreases, by factors of 1.40 and 1.64 respectively. The similarity between the derived mass-velocity curves at both epochs indicates that the ejecta were already optically thin before the first epoch.

To characterise the distributions quantitatively, we calculate three parameters: the velocity of the peak of the mass distribution $v_\mathrm{peak}$; the width of the velocity distribution defined as the difference between the 25th and 75th percentiles of the mass distribution; and a maximal velocity $v_\mathrm{max}$ defined as that within which 95\% of the molecular mass is contained. To estimate these parameters and their uncertainties for the observed mass distributions, we carry out a Monte Carlo analysis in which we generate 1000 variations of the mass distribution. For each data point, we draw values from a normal distribution centred on the observed value, with a standard deviation of 20\%, which corresponds to the uncertainty values estimated in Appendix~\ref{ap:noise} from both statistical and systematic effects. From these 1000 simulated distributions, we compute $v_\mathrm{peak}$, the width, and $v_\mathrm{max}$, taking the average and standard deviation of the resulting values as the final estimates and their uncertainties. To reduce the influence of small-scale noise, we apply a 3-point moving average before calculating $v_\mathrm{peak}$ and width. The values of $v_\mathrm{peak}$, width and $v_\mathrm{max}$ for each observational dataset are listed in Table~\ref{summarystatistics_obs} together with their uncertainties.

\input{distribution_parameters_observed}

Epoch 1 SiO data show peak velocities marginally lower than that of CO $v_\mathrm{peak}^\mathrm{SiO54-1}\lesssim v_\mathrm{peak}^\mathrm{SiO65-1}\lesssim1830$km\,s$^{-1}\approx v_\mathrm{peak}^\mathrm{CO21}$.
However, the maximum velocities differ significantly, $v_\mathrm{max}^\mathrm{SiO65-1}\approx 3220$km\,s$^{-1}$ and $v_\mathrm{max}^\mathrm{SiO54-1}\approx 3600$km\,s$^{-1}$, both below $v_\mathrm{max}^\mathrm{CO21}\approx 3980$km\,s$^{-1}$. 
Similarly to the dip in the CO 6--5 profile, this cut-off at lower $v_\mathrm{max}$ might be due to a lack of resolution in the observational data. The SiO epoch~1 datasets have higher noise levels than the CO 2--1. 

\section{Hydrodynamical models}
\label{sec:hydromodels}

\begin{table*}
\caption{Properties of the stellar models of the three groups used throughout this work. \label{tab:progenitors} }
\begin{tabular}{c c c c c c c c}
\hline
Group&Name& Type & ZAMS mass & E$_\mathrm{expl}$&$\beta$-decay &simulation time & Reference \\
&&&[\Msun]&$10^{51}$erg& & [d]&Progenitor\\
\hline\hline 
\multirow{5}{*}{1}& B15$_{\mathrm{no}\,\beta}$ & BSG & 15 & 1.39 & no $\beta-$decay & 361 & \cite{Woosley1988}\\
&B15 & BSG & 15 & 1.39 & without tracer & 361 & \cite{Woosley1988}\\
&B15$_\mathrm{X}$ & BSG & 15 & 1.39 & 100\% tracer X& 358 & \cite{Woosley1988}\\
&N20 & BSG & 20 & 1.65 & without tracer & 362 & \cite{Shigeyama1990}\\
&L15 & RSG & 15 & 1.75 & without tracer & 321 & \cite{Limongi2000}\\
&W15 & RSG & 15 & 1.47 &  without tracer & 373 & \cite{Woosley1995}\\
\hline
\multirow{5}{*}{2}&M15--7b--1&bBSG&15+7&1.40& 50\% tracer X & 368&\cite{menon2017}\\
&M15--7b--2&bBSG&15+7&1.43& 50\% tracer X & 364&\cite{menon2017}\\
&M15--7b--3&bBSG&15+7&1.43& 50\% tracer X & 368&\cite{menon2017}\\
&M15--7b--4&bBSG&15+7&1.78& 50\% tracer X&371&\cite{menon2017}\\
&M15--8b--1&bBSG&15+8&1.57& 50\% tracer X&362&\cite{menon2017}\\
\hline
\multirow{4}{*}{3}&WH12.5& RSG&12.5&0.70& 50\% tracer X & 373& \cite{Woosley2007} \\
&SW19.8& RSG&19.8&1.73& 50\% tracer X&363&\cite{Sukhbold2014}\\
&SW20.8& RSG&20.8&0.83& 50\% tracer X&363&\cite{Sukhbold2014}\\
&SW27.3& RSG&27.3&1.41& 50\% tracer X&363&\cite{Sukhbold2014}\\\hline
\end{tabular}
\\
ZAMS: zero-age main-sequence, (b)BSG: (binary merger) blue supergiant, RSG: red supergiant
\end{table*}

We compare our observations with the results of 15 simulations of the core-collapse supernova. The 10 underlying progenitor models that were exploded can be divided into three groups. The first group is formed by two classical single-star RSG \citep{Limongi2000,Woosley1995} and two classical single-star BSG \citep{Woosley1988,nomoto1988} progenitor models. The explosions of these stars have been described in detail in \citet{Wongwathanarat:2015jv} and \citet{Gabler2021}. We follow the notation of \citet{Gabler2021} as given in Table\,\ref{tab:progenitors}, where we provide some details of the models. The second group of models is based on more recent models optimised to reproduce the lightcurve of SN\,1987A \citep{Utrobin2021, Menon2019}. These progenitors are BSGs formed from a binary merger \citep{menon2017}. The mass of the primary is 15 M$_\odot$ at the zero-age main sequence for all five models of this group, while the mass of the secondary is 7 M$_\odot$ for the first four models and 8\,M$_\odot$ for the last model. The four explosions of the progenitor with the 7 M$_\odot$ companion were studied for four different set-ups described in \citep{Utrobin2021}. These binary-merger BSG (bBSG) models were evolved up to shock breakout in \citep{Utrobin2021}, and continued until about 1 year for this work. The third group is based on commonly used RSG progenitors from \citet{Woosley2007} and \citet{Sukhbold2014}. We present a selection of these models that span the widest ranges in the distributions of the elements Si, O and C \citep{Giudici2025a,Giudici2025b}.

For the hydrodynamic simulations of all the models, we use the \textsc{Prometheus-HOTB} code \citep{Gabler2021}. In this code, neutrinos responsible for reviving the stalled shock are evolved using an approximate ray-by-ray neutrino-transport scheme \citep{Scheck2006,Wongwathanarat2013,Wongwathanarat:2015jv}. This method potentially underestimates the production of radioactive $^{56}$Ni, due to uncertainties of the electron fraction $Y_e$ of
neutrino-processed ejecta \citep{Wongwathanarat2013}. A comparison with more sophisticated methods indicates that a significant fraction of the tracer nucleus $^{56}$X, which traces the production of neutron-rich elements for conditions when the electron fraction is below 0.49, might actually be $^{56}$Ni. To estimate the influence of this uncertainty, we also analyse a model B15$_\mathrm{X}$, in which we assume that all of the tracer is actually $^{56}$Ni. As a more realistic approximation, about 50 per cent of the tracer should be added to the yields of $^{56}$Ni. Hence, it is advisable to view the outcomes for models B15, N20, W15, and L15 as conservative estimates of the influences of the $\beta$-decay. The simulations of model groups 2 and 3 have been performed assuming that $50\%$ of the tracer material is $^{56}$Ni and contributes to the $\beta$ decay (see sixth column in Table\,\ref{tab:progenitors}). 

The early phases of the models, during which the ejecta asymmetries are seeded by hydrodynamic instabilities during the shock revival, are discussed in detail in \citet{Wongwathanarat2013} (group 1), and \citet{Giudici2025a} (group 3), respectively. The subsequent growth of Rayleigh-Taylor (RT) instabilities, which are responsible for the major part of the mixing of the different elements and their additional acceleration, are investigated in \citet{Wongwathanarat:2015jv} (group 1), \citet{Utrobin2021} (group 2) and \citet{Giudici2025a} and \citet{Giudici2025b} (group 3), respectively. They follow the evolution until shock breakout. During this phase, the initial large-scale asymmetries are fragmented and so-called RT fingers of heavy elements can grow into former progenitor shells of lighter elements. The last phase, until about 1~year of evolution time, is dominated by the energy input due to radioactive heating by $^{56}$Ni \citep[][group 1 and group 3, respectively]{Gabler2021,Giudici2025a}. The fine-structured fingers and the central bubble of high mass elements containing large amounts of $^{56}$Ni expand and get accelerated. At about 1~year, homology is reached. We then extrapolate the models to the current time of SN\,1987A and assume that the ejecta velocities do not change significantly.

Not all of the models considered here are specifically optimised to reproduce the observations of SN\,1987A. In particular, models in groups 1 and 3 do not match most of the expected progenitor and explosion properties. Group 1 models are classical simulations that have been extensively studied in the literature, while group 3 RSG models are not tailored to SN\,1987A but rather illustrate how differences in progenitor structure can produce distinct element morphologies. For several reasons discussed in \citet{Utrobin2021}, the bBSG models in group 2 provide the closest match to SN\,1987A’s light curve. Although they do not capture every detail in the light curves or gamma-ray lines, these bBSG progenitors remain the most promising candidates. In addition to these model-set limitations, our neutrino transport scheme is approximate, allowing the explosion energy to be chosen as a free parameter rather than computed self-consistently. This flexibility enables us to compensate for uncertainties in the explosion mechanism, which is not possible in fully self-consistent models. Consequently, our aim is not to match the molecular distributions of SN\,1987A in exact detail, but rather to explore general trends and assess how well generic or widely available models can reproduce the observed quantities. The group 2 models, in particular, give reason for optimism that their C+O and Si+O distributions may also agree reasonably well with the observations.

To compare the model results with the observations, we adopt the following procedure: In each numerical cell, we first estimate the maximal mass available for CO molecules depending on the relation of the mass fractions of carbon $X_\text{C}$ and oxygen $X_\text{O}$
\begin{equation}\label{eq:CO}
    \rho_\text{C+O} =
    \begin{cases}
        \rho X_\text{C} \frac{12+16}{12} & \text{if } 16 X_\text{C}<12 X_\text{O}\\
        \rho X_\text{O} \frac{12+16}{16} & \text{else } \\
    \end{cases}
\end{equation}
Then, the remaining oxygen fraction 
\begin{equation}
X_\text{O}^*=\max\left(0.0,X_\text{O}-\frac{16}{12} X_\text{C}\right)
\end{equation}
 can still react with Si. Analogously to Equation\,\ref{eq:CO} we obtain:
\begin{equation}\label{eq:SiO}
    \rho_\text{Si+O} =
    \begin{cases}
        \rho X_\text{Si} \frac{28+16}{28} & \text{if } 16 X_\text{Si}<28 X_\text{O}^*\\
        \rho X_\text{O}^* \frac{28+16}{16} & \text{else } \\
    \end{cases}
\end{equation}
The masses of C+O and Si+O within the numerical cells are obtained by integrating the densities over the respective volumes. Finally, the mass distribution is obtained by binning the obtained masses as function of velocities. Due to the very rough approach to estimating the molecule masses based on equations\,\eqref{eq:CO} and \eqref{eq:SiO}, we denote the corresponding combination as C+O and Si+O to differentiate them from the molecules CO and SiO. 

The total masses of the elements obtained in the simulations and the corresponding maximal masses of C+O and Si+O are tabulated in Table\,\ref{tab:masses}. 
Assuming that all available carbon and silicon atoms react with oxygen, we can estimate the maximum possible masses of CO and SiO that could form. To do this, we multiply the mass of C by 28/16 and the mass of Si by 44/28, which are the ratios of the molecular masses of CO and SiO to the atomic masses of C and Si, respectively. If these values are close to our estimates of equations\,\eqref{eq:CO} and \eqref{eq:SiO} then there is sufficient oxygen in the volume where C and Si are present. This is true for all RSG models of group 3, and models L15, N20 and W15 for Si+O and approximately also for C+O. 
In contrast, other groups of models have lower molecular masses than the possible maxima.

The estimated total molecule masses of the bBSG models are lower by $10\%$ for C+O and up to $22\%$ for Si+O, with respect to the maximum available C and Si, respectively. The largest differences are obtained for the variants of model B15, which have about $18\%$ lower Si+O and up to $45\%$ lower C+O masses compared to the estimated maximum masses of these molecules (see rows 5-8 in Table\,\ref{tab:masses}). This indicates that in this particular model, there is insufficient oxygen in regions rich in C to completely bind it into C+O. 

\begin{table}
\setlength{\tabcolsep}{5pt}
\begin{tabular}{l c c c c c c c}
\hline
Model & \multicolumn{3}{c}{Element} &\multicolumn{2}{c}{Combination} \\
\hline
&	C	&O	&Si	&Si+O &	C+O & $\frac{28}{12}$C &	$\frac{44}{28}$Si\\
&[M$_\odot$]&[M$_\odot$]&[M$_\odot$]&[M$_\odot$]&[M$_\odot$]&[M$_\odot$]&[M$_\odot$]\\
\hline		
B15	&0.118&	0.162&	0.071&	0.097&	0.161&	0.275&	0.112\\
B15$_\text{X}$&0.118&0.163&0.072&	0.093&	0.171&	0.275&	0.113\\
B15$_{\mathrm{no}\,\beta}$&0.118&	0.163&	0.072&	0.102&	0.152&	0.275&	0.113\\
N20	&0.101	&1.417	&0.085	&0.134	&0.215	&	0.236&	0.134\\
L15	&0.180	&0.583	&0.039	&0.059	&0.398	 & 0.420	&0.061\\
W15	&0.230	&0.730	&0.043	&0.061	&0.490	&	0.537&	0.068\\
\hline
M15--7b--1&0.127	&1.276&	0.027&	0.037&	0.274&		0.296	&0.042\\

M15--7b--2&0.099&	0.891&	0.023&	0.030&	0.212&		0.231&	0.036\\
M15--7b--3&	0.098	&0.846&	0.018&	0.022&	0.208		&0.229&	0.028\\
M15--7b--4&	0.103&	0.983&	0.031&	0.043&	0.220&		0.240	&0.049\\
M15--8b--1&	0.101&	0.978&	0.114&	0.173&	0.210&		0.236&	0.179\\
\hline
WH12.5	&0.125	&0.376	&0.019	&0.029	&0.267	&	0.292	&0.030\\
SW19.8	&0.233	&2.252	&0.409	&0.642	&0.542	&	0.544	&0.643\\
SW20.8	&0.247	&1.339	&0.035	&0.052	&0.559	&	0.576	&0.055\\
SW27.3	&0.438	&2.838	&0.081	&0.128	&1.022	&	1.022	&0.127\\
\hline
\end{tabular}
\caption{Total masses of the separate elements, C+O and Si+O as obtained by equations\,\eqref{eq:CO} and \eqref{eq:SiO} for the different models. Minor differences of the amounts of Si in the variants of model B15 are due to slightly different numerical set ups of the simulations for models with and without $\beta$ decay. The factors $\frac{28}{12}$ and $\frac{44}{28}$ in front of C and Si are defined as the ratio of the molecular mass (CO or SiO) to the atomic mass of the corresponding element (C or Si).
}
\label{tab:masses}
\end{table}

As stated earlier, we can investigate the effect of the $\beta$ decay on the morphology using the three variants of model B15, which have respectively no decay, without tracer and with $100\%$ tracer (Table~\ref{tab:progenitors}). Although all versions of this model produce the same amounts of the elements C, O and Si, the resultant C+O and Si+O masses differ depending on the treatment of beta decays. The higher the energy input, i.e. the more of the ejecta radioactively decays, the stronger the mixing. And the stronger the mixing, the greater the mass of C+O and the less the mass of Si+O obtained. The additional expansion due to the $\beta$ decay carries more oxygen from the inner core towards the C+O shell. Once more oxygen is mixed outward, the amount of C+O increases. During this process, some of the oxygen previously consumed in Si+O is no longer available for this reaction, and a larger fraction of Si cannot find O atoms. Consequently, the amount of Si+O decreases with increasing amount of $\beta$-decaying ejecta. 

In Table\,\ref{tab:masses}, we note how the amount of the different elements differs for different progenitor models (columns 2-4). While the total mass of C in the BSG models always is around $\sim0.1$~M$_\odot$,  it can reach up to $\gtrsim0.4$M$_\odot$ for RSG. Similarly, the yields of Si do not differ dramatically from model to model and reach from $0.017 - 0.113$ for most models. Only SW19.8 sticks out significantly with a total mass M$_\mathrm{Si}=0.409\,$M$_\odot$. This model explodes relatively late \citep{Giudici2025a}, and, hence, has a lot of time to produce Si during the shock revival phase. In addition to being a RSG, and, thus, not a viable progenitor for SN\,1987A, we shall see that this model cannot reproduce, in particular, the distribution of SiO (Sect.\,\ref{sec:compare_distr}).

The element that deviates the most in the total abundances between different models is oxygen. The lowest mass, M$_\mathrm{0}=0.162\,$M$_\odot$ is obtained for model B15, while the RSG reach up to M$_\mathrm{O}=2.838\,$M$_\odot$ for model SW27.3. The bBSG have M$_\mathrm{O}\sim0.82\dots0.96\,$M$_\odot$.

\section{Comparison of observations with models}
In this section we compare the observations with the different hydrodynamical explosion models. We first investigate the different mass distributions and then show how similar mass distributions may be caused by very different geometrical distributions of C+O or Si+O.

\subsection{Velocity vs mass distributions}\label{sec:compare_distr}
\input{distribution_parameters_both}

\begin{figure*}
	 \includegraphics[width=0.9\textwidth]{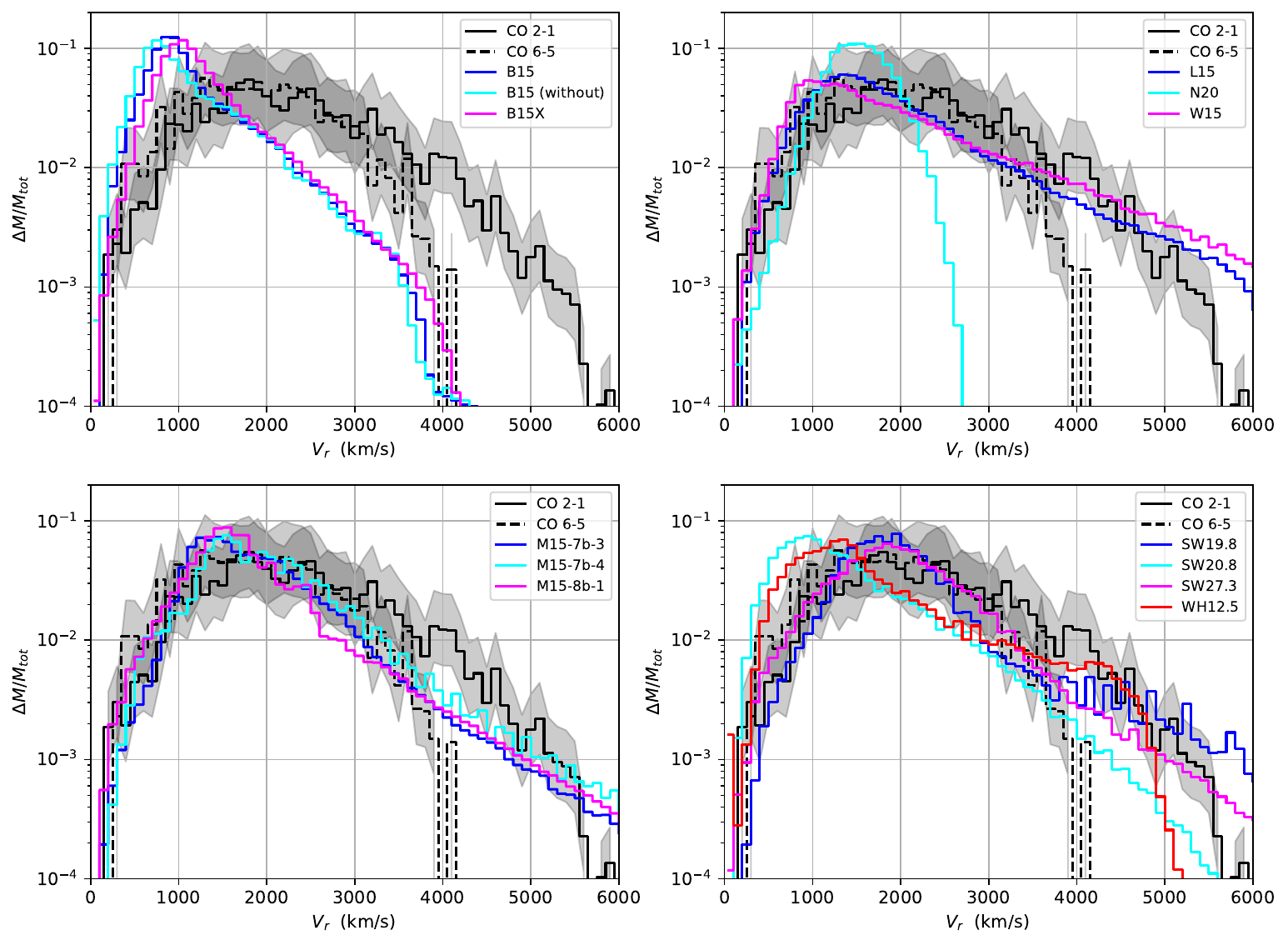}
        \caption{
       The mass fraction vs velocity 
        of CO $J$=2--1 (solid line) and $J$=6--5 (dashed), compared to distributions of C+O obtained in the hydrodynamic simulations. The names of the simulated models are indicated in the legend}.\label{fig:velocity_mass_co21}
\end{figure*}

\begin{figure*}
	 \includegraphics[width=.9\textwidth]{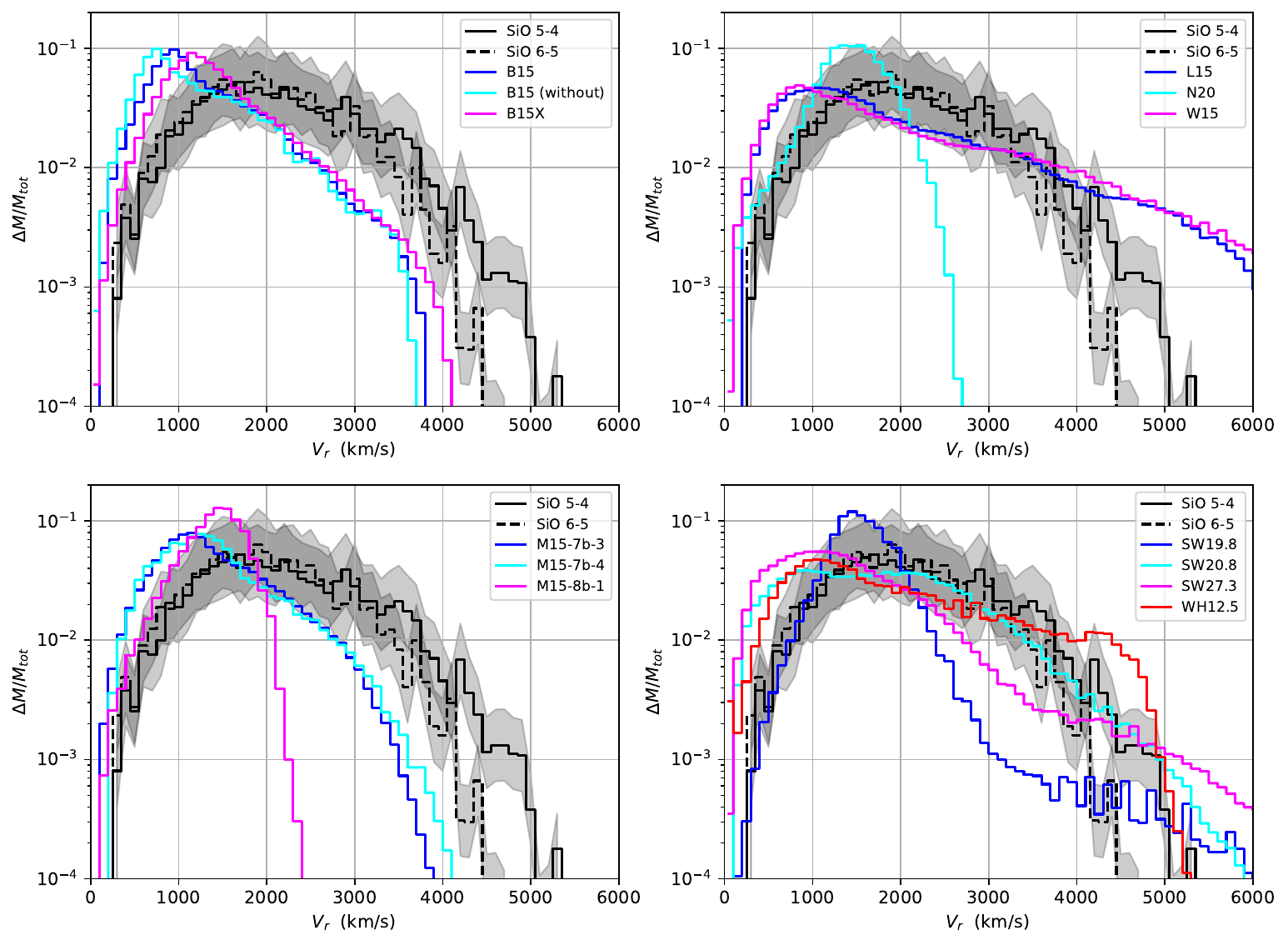}
        \caption{Velocity vs mass fraction of SiO $J$=6--5 (solid line) and 5--4 (dashed line) in epoch 1, compared to distributions of Si+O obtained in the simulations of the models indicated in each legend.
}
        \label{fig:velocity_mass_sio_c1}
\end{figure*}

\begin{figure*}
         \includegraphics[width=.9\textwidth]{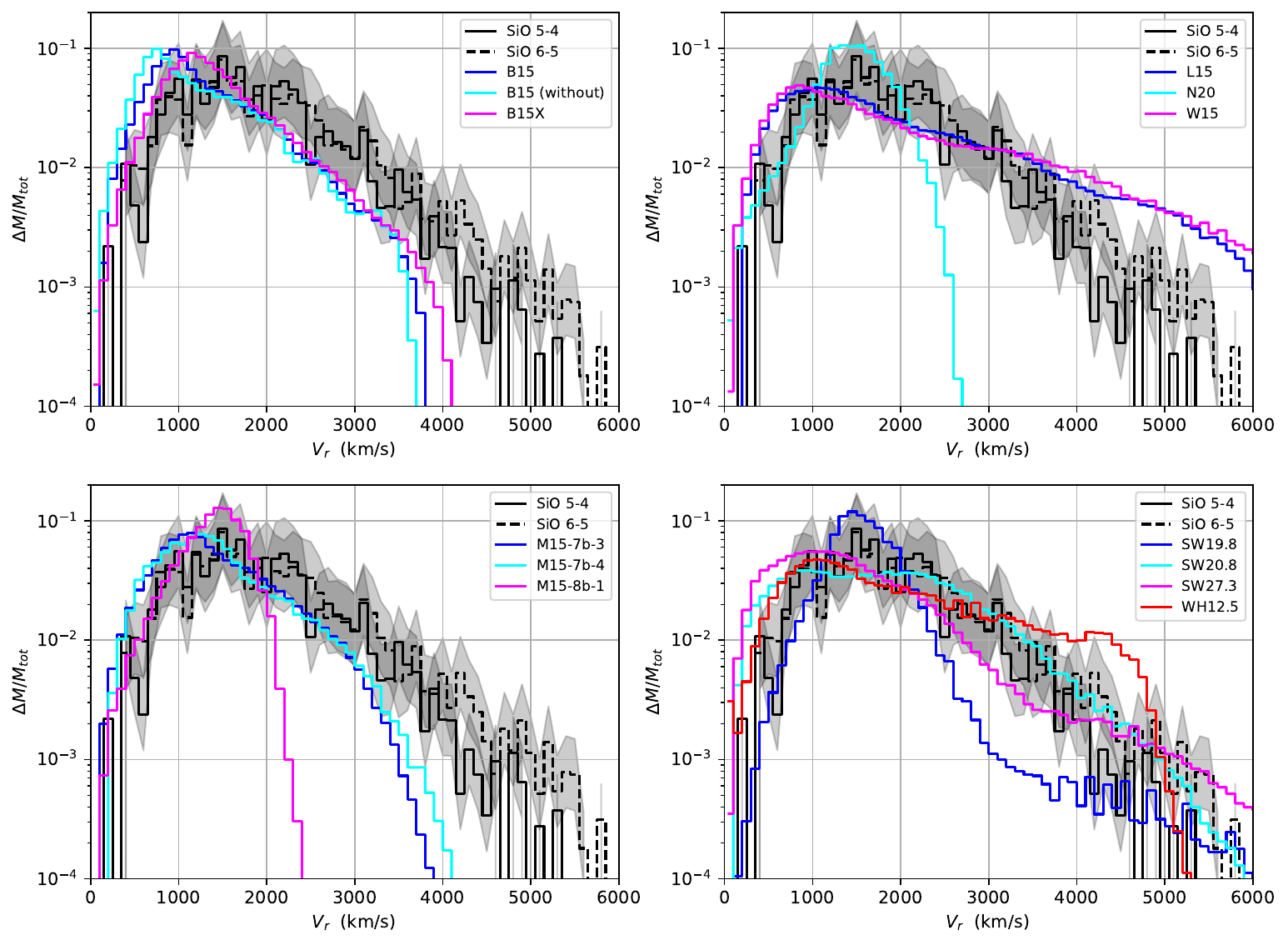}
        \caption{Velocity vs mass fraction of SiO $J$=6--5 (solid line) and 5--4 (dashed line) in epoch 2, compared to  distributions of Si+O obtained in the simulations of the models indicated in each legend.
}
        \label{fig:velocity_mass_sio_c2}
\end{figure*}

In order to make quantitive comparison,
the observed and simulated fractional mass distributions as a function of velocity are compared in Figs.~\ref{fig:velocity_mass_co21} (CO), \ref{fig:velocity_mass_sio_c1} (SiO epoch 1) and \ref{fig:velocity_mass_sio_c2} (SiO epoch 2). 
Each figure has four panels and we plot model B15 and the different $\beta$-decay treatments in the top left panel, the models of group 1 in the top right panel, the group 2 models in the bottom left and the RSG models in the bottom right panel. From these distributions, we measure the
 values of $v_\mathrm{peak}$, width and $v_\mathrm{max}$ as previously defined for observations. The corresponding values are compared to the observations in Table~\ref{summarystatistics_both}. 
The parameters of the theoretical distributions that lie within 3$\sigma$ of the observed values are highlighted in green.


We begin the comparison with the CO data plotted in Fig.~\ref{fig:velocity_mass_co21} and the model B15 from group 1. All three variants of this model have a narrow velocity distribution, and their peaks are located at lower velocities than the observed one (black curves).
The B15 model predicts $v_\mathrm{peak}^\mathrm{CO\text{-}B15}  \approx 850$\,km\,s$^{-1}$ and $v_\mathrm{peak}^\mathrm{SiO\text{-}B15} \approx 950$\,km\,s$^{-1}$ (see Table~\ref{summarystatistics_both}), but the observed peaks are $v_\mathrm{peak}^\mathrm{CO\text{-}ALMA} \approx 1800 \pm 210$\,km\,s$^{-1}$ and $v_\mathrm{peak}^\mathrm{SiO\text{-}ALMA} \approx 1500\text{--}1800 \pm 190$\,km\,s$^{-1}$ (Table\,\ref{summarystatistics_obs}). Moreover, the B15 model predicts a steep decline around $3800$\,km\,s$^{-1}$, beyond which there is neither C+O nor Si+O (top left panels of Figs.\,\ref{fig:velocity_mass_co21}-\ref{fig:velocity_mass_sio_c2}).  The cut-off is a direct consequence of the composition of the stellar progenitor that has no Si and very little O in the outer shells. 

Model B15 does not closely reproduce the ALMA measured velocity distributions of molecules in SN\,1987A, but is useful for evaluating the effect of $\beta$ decay by virtue of the explosions of this model employing three variations in the treatment of the $\beta$ decay (B15, B15$_{\textrm{no}\beta}$, and B15$_{\rm X}$).
Without $\beta$ decay (B15$_{\textrm{no}\beta}$), the inner ejecta which were subject to explosive nucleosynthesis reach up to $3800$\,km\,s$^{-1}$ (Figs.\ref{fig:velocity_mass_co21} and \ref{fig:velocity_mass_sio_c1}). This material is first accelerated during the explosion by the shock wave, and can then gets some boost during the action of Rayleigh-Taylor instabilities \citep{Wongwathanarat:2015jv}. After shock breakout, the additional energy deposition, caused by radioactive decay of $^{56}$Ni \citep{Gabler2021}, leads to a further acceleration of the bulk from $v_\mathrm{peak}^{\mathrm{CO-B15no}\,\beta}\approx\,750\,$km\,s$^{-1}$ to $v_\mathrm{peak}^\mathrm{CO-B15}\approx850$\,km\,s$^{-1}$). The fastest moving C+O is also accelerated slightly. In our enhanced $\beta$-decay model B15$_\text{X}$, the effect of $^{56}$Ni energy deposition is even stronger and the C+O reaches $v_\mathrm{peak}^\mathrm{CO-B15}\approx1050$\,km\,s$^{-1}$. The acceleration of Si+O is stronger than that of C+O, and the peak velocity of Si+O is pushed from $v_\mathrm{peak}^{\mathrm{SiO-B15no}\,\beta}\approx750$\,km\,s$^{-1}$ to  $v_\mathrm{peak}^\mathrm{SiO-B15X}\approx1150$\,km\,s$^{-1}$. 
This difference between Si+O and C+O is influenced by two factors: first, the Si is closer to the regions, which are rich in radioactive $^{56}$Ni, so that the effect of energy deposition and subsequent expansion should be stronger in these regions. Second, the mixing due to $\beta$ decay brings more carbon to the regions where Si reacts with O. In our approximation, we assume that C first attract sall available O, thus leaving less O for forming Si+O. This mainly affects the slow-moving Si, and hence is a sign of inward mixing of C rather than outward mixing of Si. We would again like to remind the reader that a more realistic treatment of the neutrino transport and the nuclear network should give results that lie between those of the two models B15 and B15$_\text{X}$. 

The distributions of the RSG models L15 and W15 (top right panels in Figs.\,\ref{fig:velocity_mass_co21}--\ref{fig:velocity_mass_sio_c2}) have similar shapes, extending to very high velocities $v_\mathrm{max}>4000$\,km\,s$^{-1}$for both C+O and Si+O. They match the observations between $3000$\,km\,s$^{-1}\lesssim v\lesssim5000$\,km\,s$^{-1}$ quite well.
The models overpredict the molecules at the highest velocities, but this may be caused by observational biases. As we have seen in Sect.\,\ref{sec:spectralgrids}, the ALMA data contain some noise, and noise cut-offs are applied. Therefore, it is possible that emission from fast but faint material is buried in the noise. Additionally, the fastest-moving ejecta may be excluded during processing to avoid contamination from the rings, as faster ejecta tends to approach the ring more closely. Hence, we do not consider this overabundance of C+O and Si+O in the models compared to the observations to be a significant issue. While the high velocity tail seems to agree well, the peaks of the C+O distribution $v_\mathrm{peak}^\mathrm{CO-L15}\approx 1350$\,km\,s$^{-1}$ and $v_\mathrm{peak}^\mathrm{CO-W15}\approx 950$\,km\,s$^{-1}$ are significantly offset from the observed $v_\mathrm{peak}^\mathrm{CO21}\approx 1830$\,km\,s$^{-1}$. The situation for Si+O $v_\mathrm{peak}^\mathrm{SiO-L15}\approx 1150$\,km\,s$^{-1}$ and $v_\mathrm{peak}^\mathrm{CO-W15}\approx 850$\,km\,s$^{-1}$ is not much better as $v_\mathrm{peak}^\mathrm{SiO}\approx 1600$\,km\,s$^{-1}$. Due to its higher explosion energy ($E_\mathrm{expl}=1.75$\,B) model L15 has higher peak velocities and thus a slightly better agreement with the observations. In addition to the problems with the too low peak velocities, the slope of the Si+O distributions of models L15 and W15 also does not match the observed slope of SiO at any epoch. The observed slopes are much steeper, indicating a clearly less extended Si+O ejecta. This is not surprising because the progenitor of SN\,1987A was a BSG, which is more compact. 

Model N20 shows a distinctly different C+O mass distribution compared to other models (cyan curve in the top right panels of Figs.\,\ref{fig:velocity_mass_co21}--\ref{fig:velocity_mass_sio_c2}) ). Its peak velocity $v_\mathrm{peak}^\mathrm{CO-N20}\approx 1450\,$km\,s$^{-1}$ is the highest among the considered models of group 1, and closest to the observations. However, its velocity distribution has a steep drop-off at 3,000\,km\,s$^{-1}$ and is much too narrow, with a width of only $300\,$km\,s$^{-1}$. In contrast, the observed CO distribution extends to at least 4,000 km s$^{-1}$.
The distributions for Si+O (cyan lines in the top right panels of Figs.~\ref{fig:velocity_mass_sio_c1} and \ref{fig:velocity_mass_sio_c2}) are very similar to that of C+O, and, thus, much narrower and with too low $v_\mathrm{max}$. In summary, this model’s velocity profile shows the poorest agreement with the observations.

The results of the RSG models in group 3 (bottom right panels in Figs.\,\ref{fig:velocity_mass_co21}-\ref{fig:velocity_mass_sio_c2}) span a wide range of mass distributions. Models SW19.8 and SW27.3 fit the peak of CO quite well and also produce a wide enough extension of the distribution at high velocities (blue and magenta lines in Fig.\,\ref{fig:velocity_mass_co21}). Model SW27.3 declines too fast, leading to a slightly too low $v_\mathrm{max}$. The C+O ejecta of models SW20.8 (cyan) and WH12.5 (red) are too slow in general. Note that model WH12.5 has a second bump at $v\approx 4200$\,km\,s$^{-1}$: this accumulation of mass at the bump is caused by strong Rayleigh-Taylor mixing which leads to a catching-up of the ejecta with the forward shock, even deforming the latter \citep{Giudici2025a,Giudici2025b}. Also the curves for Si+O of the explosions of the RSG progenitors show more diversity than the other models (Figs.\,\ref{fig:velocity_mass_sio_c1} and \ref{fig:velocity_mass_sio_c2}). First, the fractional mass distribution of the most spherical model SW19.8 (blue) has a sharp peak, with fast rising and declining flanks, and a very low mass tail up to high velocities. Model SW27.3 (magenta) has a very low velocity peak and a shallower decline with two bumps or shoulders in the declining tail. The very asymmetric model WH12.5 (red) also has too low a peak velocity $v^\mathrm{SiO-WH12.5}_\mathrm{peak}\approx1050$\,km\,s$^{-1}$ and similarly to the C+O curve, we observe a second bump at high velocity, caused by the high-velocity ejecta hitting the forward shock in this model. Finally, one of the most asymmetric models SW20.8 (cyan) shows two clear peaks in the distribution at low energies. This model best matches the observational SiO curve above the peak $v>2000$\,km\,s$^{-1}$. 

The last group of models presented here are BSG progenitors resulting from mergers in binary systems (middle group in Table\,\ref{tab:progenitors}). These are the most promising candidates for SN\,1987A, because they fulfil a number of constraints as discussed in \citep{Utrobin2021}. It is also the group of models that best matches the observed distribution of CO  (bottom left panel of Figs.\,\ref{fig:velocity_mass_co21} and middle group in Table\,\ref{summarystatistics_both}). 
Apart from slightly too low peak velocities $1350$\,km\,s$^{-1}$$<v^\mathrm{bBSG-CO}_\mathrm{peak}<1550$\,km\,s$^{-1}$ the tails of the distribution $v>2000$\,km\,s$^{-1}$ are matched well. However, the distributions of Si+O (Figs. \ref{fig:velocity_mass_sio_c1} and \ref{fig:velocity_mass_sio_c2}) are not close to those of the observations: the peak velocities are much too small, the tail declines too early, and substantially less mass is at high velocities. In this respect, model M15--8b--1 performs the worst, since the Si+O only reaches $v^\mathrm{SiO-M15-8b-1}_\mathrm{max} = 1750$\,km\,s$^{-1}$.

The results of models M15--7b--1 and M15--7b--2 are very similar to those of M15--7b--3 and M15--7b--4, and we therefore do not show them in Figs.\,\ref{fig:velocity_mass_co21}--\ref{fig:velocity_mass_sio_c2} and Table\,\ref{summarystatistics_both}

\begin{figure*}
   \begin{overpic}[width=.28\textwidth]{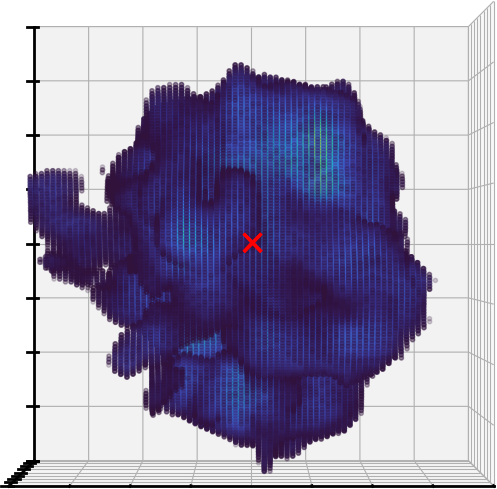}
   \put(80,85){\textbf{-90°}}  
    \end{overpic}
     \begin{overpic}[width=.28\textwidth]{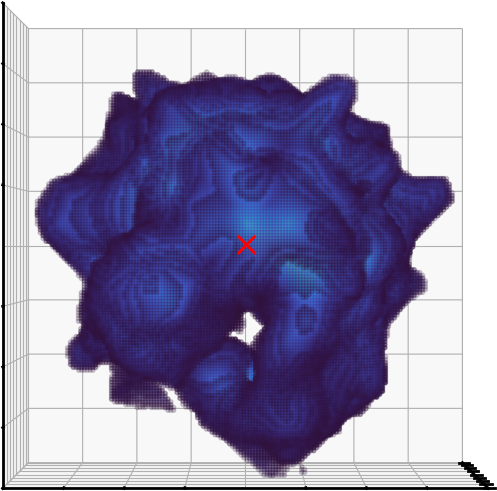}
     \put(80,85){\textbf{0°}}\end{overpic}
     \begin{overpic}[width=.28\textwidth]{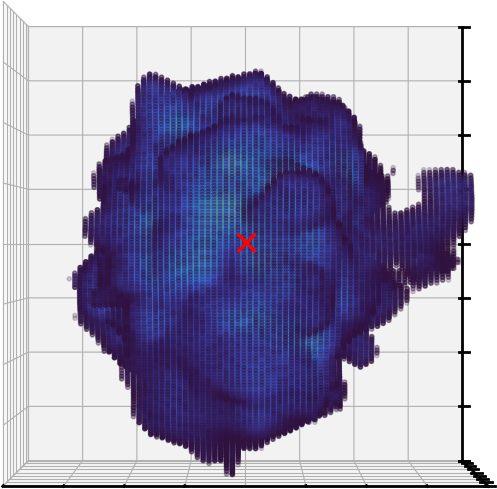}
     \put(80,85){\textbf{90°}}\end{overpic}
    \\
    \begin{overpic}[width=.38\textwidth]{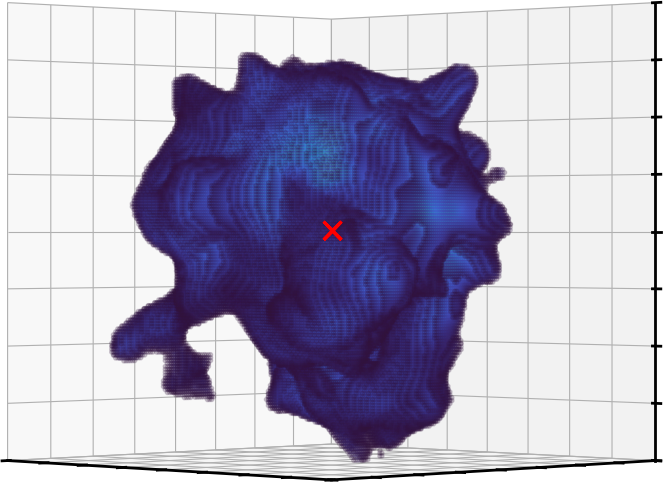} \put(80,60){\textbf{-45°}}\end{overpic}
    \begin{overpic}[width=.38\textwidth]{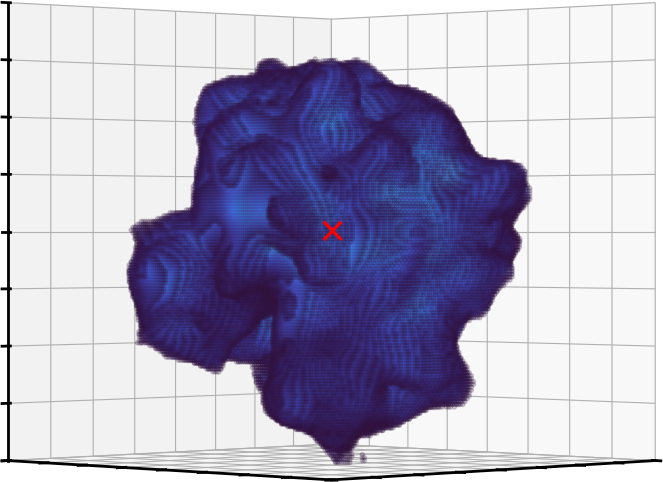}
    \put(80,60){\textbf{45°}}\end{overpic}
        \includegraphics[width=.033\textwidth]{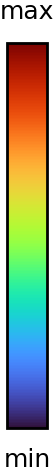}      
    \caption{3D density maps of C+O of model M15--7b--4 at $t\sim1$ yr. The minimal density (dark blue) plotted is 15\% of the maximum density (yellow-red) in the grid. The axis scale of the grid is $2\times10^{15}$ cm  and the maximum grid extension is $\pm8\times10^{15}$ cm. The red cross marks the center of the explosion.}
      \label{fig:model3D_M157b21CO}
\end{figure*}

\begin{figure*}
      \begin{overpic}[width=.28\textwidth]{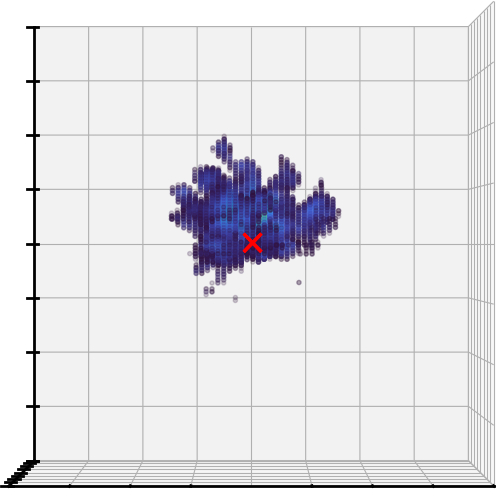}
   \put(80,85){\textbf{-90°}}  
    \end{overpic}
     \begin{overpic}[width=.28\textwidth]{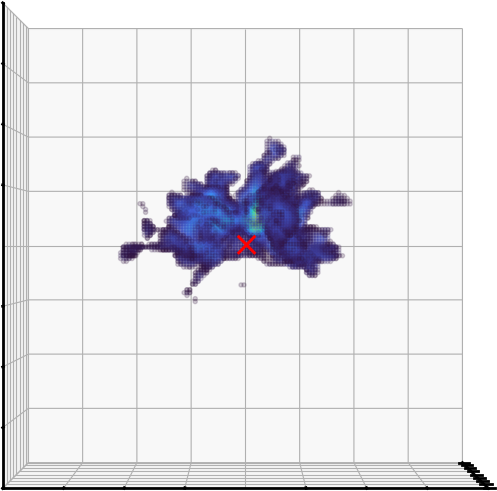}
     \put(80,85){\textbf{0°}}\end{overpic}
     \begin{overpic}[width=.28\textwidth]{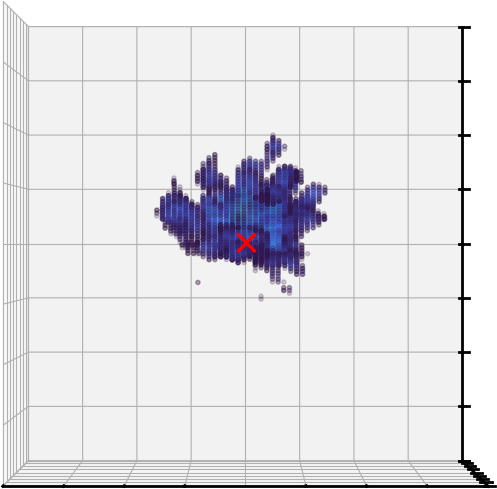}
     \put(80,85){\textbf{90°}}\end{overpic}
  \includegraphics[width=.033\textwidth]{3d_views/colorbar_turbo.png}      
    \caption{3D density maps of C+O of model SW20.8 at $t\sim1$ yr. The minimal density (dark blue) plotted is 15\% of the maximum density (yellow-red) in the grid. The axis scale of the grid is $2\times10^{15}$ cm  and the maximum grid extension is $\pm8\times10^{15}$ cm. The red cross marks the center of the explosion.}
        \label{fig:model3D_SW20.8CO}
\end{figure*}

\begin{figure*}
   \begin{overpic}[width=.28\textwidth]{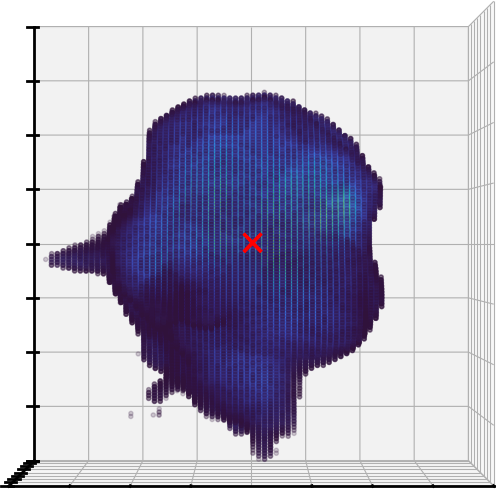}
   \put(80,85){\textbf{-90°}}  
    \end{overpic}
     \begin{overpic}[width=.28\textwidth]{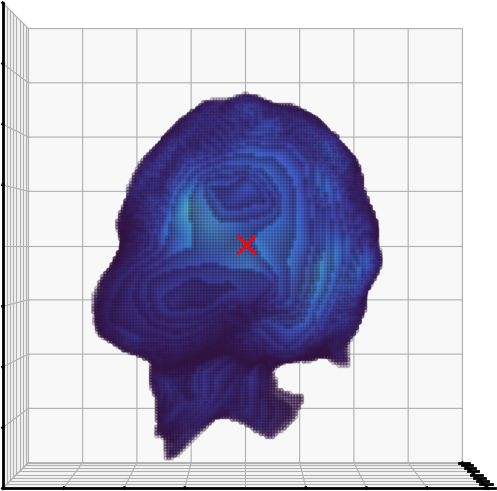}
     \put(80,85){\textbf{0°}}\end{overpic}
     \begin{overpic}[width=.28\textwidth]{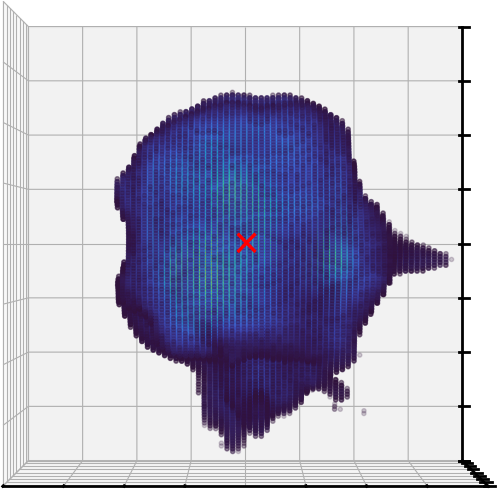}
     \put(80,85){\textbf{90°}}\end{overpic}
            \includegraphics[width=.033\textwidth]{3d_views/colorbar_turbo.png}
    \caption{3D density maps of Si+O of model M15--7b--4 at $t\sim1$ yr. The minimal density (dark blue) plotted is 15\% of the maximum density (yellow-red) in the grid. The axis scale of the grid is $2\times10^{15}$ cm  and the maximum grid extension is $\pm8\times10^{15}$ cm. The red cross marks the center of the explosion.}
      \label{fig:model3D_M157b21SiO}
\end{figure*}

\begin{figure*}
      \begin{overpic}[width=.28\textwidth]{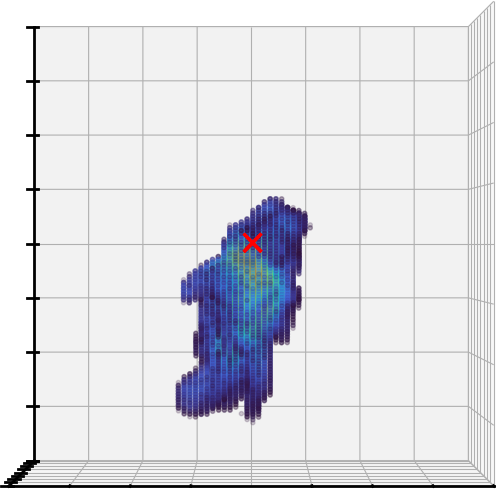}
   \put(80,85){\textbf{-90°}}  
    \end{overpic}
     \begin{overpic}[width=.28\textwidth]{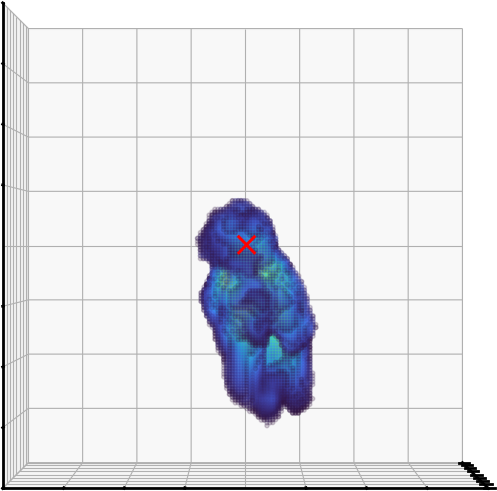}
     \put(80,85){\textbf{0°}}\end{overpic}
     \begin{overpic}[width=.28\textwidth]{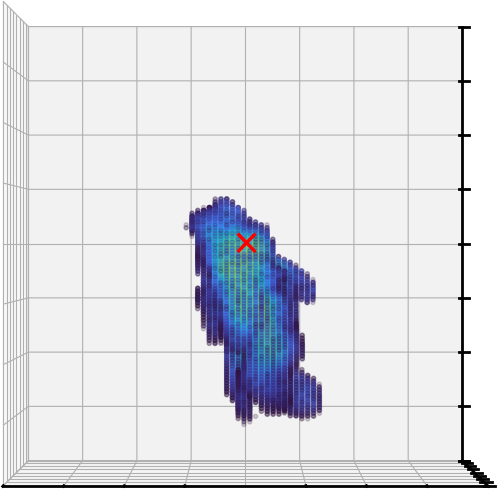}
     \put(80,85){\textbf{90°}}\end{overpic}
             \includegraphics[width=.033\textwidth]{3d_views/colorbar_turbo.png}  
    \caption{3D density maps of Si+O of model SW20.8 at $t\sim1$ yr. The minimal density (dark blue) plotted is 30\% of the maximum density (yellow-red) in the grid. The axis scale of the grid is $2\times10^{15}$ cm  and the maximum grid extension is $\pm8\times10^{15}$ cm. The red cross marks the center of the explosion.}
        \label{fig:model3D_SW20.8SiO}
\end{figure*}

\subsection{3D maps of C+O and Si+O of theoretical models}
To compare our 3D simulations visually to the observations, we searched for the best viewing angle to find 3D visualizations of the simulated mass distribution which look similar to the observed maps in Figs.\,\ref{3dmap_co21_c1}--\ref{3dmap_sio65_c2}. 
As a first step, we map the results of the 3D spherical simulations at $t\sim1$yr onto a $512^3$ Cartesian grid. These mapped data are then smoothed to cubes of size $8^3$ grid cells in order to approximate the beam size of the observations. The smoothed data are subsequently interpolated onto a $160^3$-point grid to facilitate comparison with the observations. Because there is no correspondence between the coordinates of the simulated 3D map and the sky coordinates of SN\,1987A, we attempt to find a viewing angle that best represents the features of the observational data manually.
In order to emulate the lower velocity sampling in the line of sight (see Table\,\ref{observing_log}), we only plot every second data point in the newly defined x-direction. We assume that the expansion one year after the expansion is homologous, such that the morphology does not change during the evolution until the current age of SN\,1987A. 

For the first comparison with the observations, we select model M15--7b--4, which has one of the best agreements with the CO distribution in mass fluction vs velocity plot. 
The 3D map for C+O of  M15--7b--4 is presented in Fig.\,\ref{fig:model3D_M157b21CO}. As expected, the 3D map of model M15--7b--4 resembles the observed CO morphology of SN\,1987A  (Fig.\,\ref{3dmap_co21_c1}) relatively well. 
Both observations and simulations have roughly spherical or ellipsoidal distributions with a complex, clumpy morphology superimposed. The surfaces are irregular, marked by prominent ridges, valleys, and voids as one would expect from turbulence and hydrodynamic instabilities. Protrusions, clumps or filaments extend radially from the centre and are distributed asymmetrically. These extended structures are related to Rayleigh-Taylor fingers, best visible in the panels  for $\pm90^\circ$ (top left and right in Fig.\,\ref{fig:model3D_M157b21CO}). 

After re-gridding the 3D map is manually, the qualitative comparison results in a relatively good match. Similar structures are obtained for models M15--7b--2 and M15--7b--3, while model M15--7b--1 is more spherically symmetric with fewer filaments extending from the shell-like structure.

The model that predicts the best agreement in particular with the high velocity part of the SiO mass distribution of SN\,1987A is model SW20.8 in mass fraction vs velocity plot (Fig.\ref{fig:velocity_mass_sio_c1}). However, its CO prediction is not the best (Fig.\ref{fig:velocity_mass_co21}).
As expected, the C+O morphology in 3D of this model (Fig.\,\ref{fig:model3D_SW20.8CO}) deviates significantly from the observed one. This dataset exhibits a markedly more asymmetric and irregular morphology compared to the previous two (M15-7-b-3 and -4). First, the spatial extension is much smaller. Not obvious from the mass distribution of C+O (Fig.\,\ref{fig:model3D_SW20.8CO}), the explosion is very asymmetric, and almost all C+O is in the upper hemisphere. Note that the centre of the explosion is marked with the red cross. 
In addition, the morphology is less compact and many extended fine structures stretch out from the bulk. These features stretch far beyond the central region, making the structure appear more dynamically complex and fragmented. The related clumps are larger and more isolated, and the surface is defined by sharp protrusions and voids.

In contrast to the C+O, the Si+O of model M15--7b--4 is much more spherically distributed and resides at lower velocities (Fig.\,\ref{fig:model3D_M157b21SiO}). It is basically enclosed by the C+O. This problem was already represented in the mass distributions in Figs.\,\ref{fig:velocity_mass_sio_c1} and \ref{fig:velocity_mass_sio_c2}, where the maximum velocities of Si+O are much smaller than those of C+O, in tension with observations of SN\,1987A.

From Fig.\ref{fig:velocity_mass_sio_c1}, it appears that model SW20.8 represents best the mass fraction vs velocity distribution of SiO. However, when plotting the 3D morphology in Fig.\,\ref{fig:model3D_SW20.8SiO}, we note that the distribution of Si+O does not resemble the observed 3D morphology at all. There is a single elongated clump towards the southern hemisphere. In strong contrast to the SN\,1987A data, C+O and Si+O in this model are moving in opposite directions and have qualitatively very different morphologies (very irregular with a high number of protrusions for the C+O, much more compact and elongated central structure for Si+O). Although velocity distributions enable quantitative comparisons between observed and simulated data, these 3D plots show that a similar mass distribution does not guarantee qualitatively similar 3D morphologies.

\subsection{Comparison of SiO and CO morphologies}
\label{co_sio_comparison_section}

\begin{figure*}
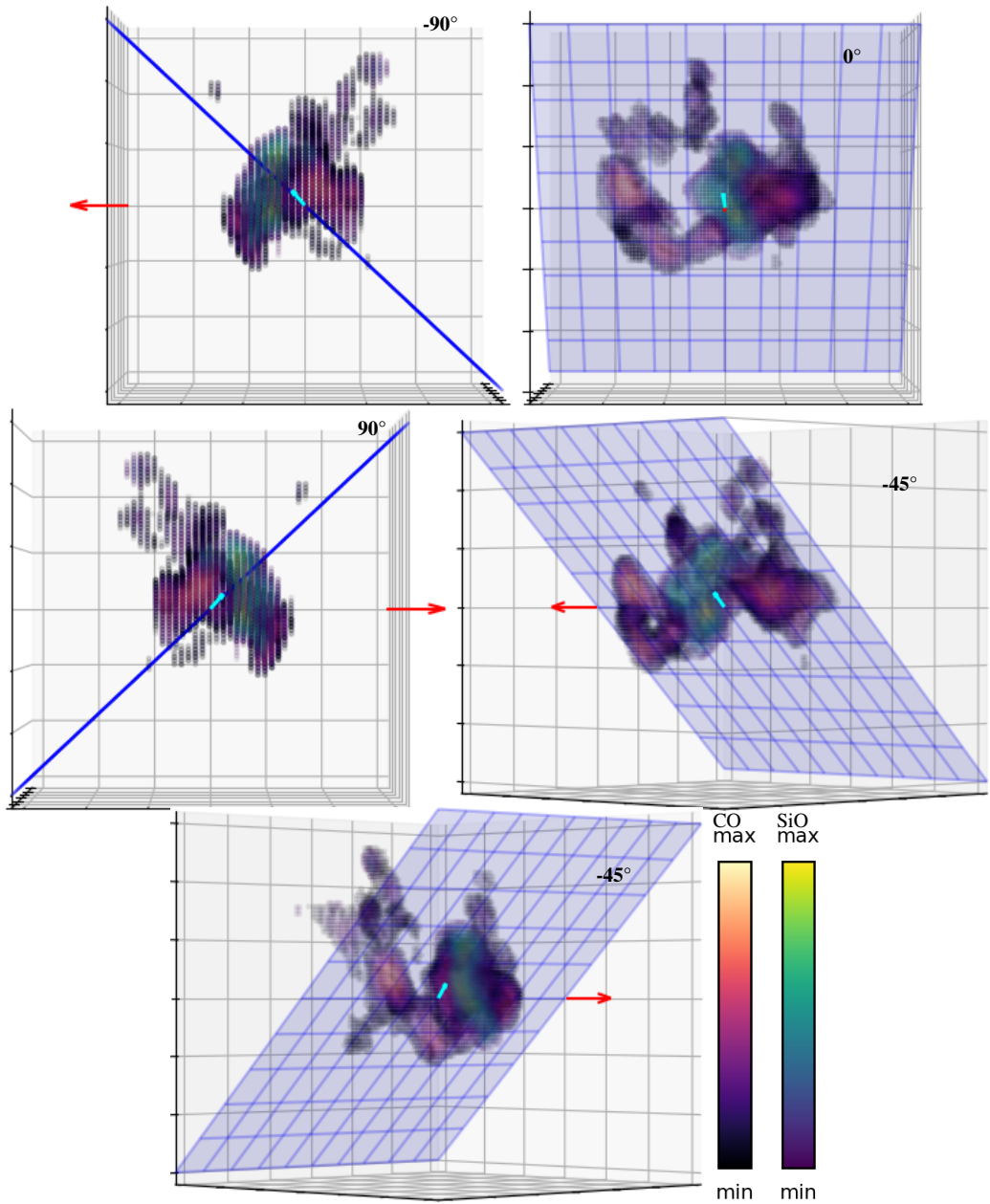

    \makepanelcombi{3d_CO-SiO_comparison}{co_sio_comparison_both_2x_alpha0.03}
    \caption{The most intense parts of the CO (dark purple-yellow) and SiO (dark purple-green) emission from the ALMA 3D maps are plotted and compared. The minimal intensity plotted is double the noise threshold used in Figs.\,\ref{3dmap_co21_c1}-\ref{3dmap_sio65_c2}. The vector from the SN explosion site to the centre of mass calculated from the CO emission is indicated in cyan. This vector points into the plane of the ring of SN\,1987A and towards the observer, with an angle of about $45^\circ$ towards the north. Box size: 30,000 AU.}
        \label{fig:co_sio_comparison}
\end{figure*}

To directly compare the relative spatial distributions of CO and SiO obtained with ALMA, we plot both molecules with different colour maps using a minimum intensity threshold set to twice the estimated noise level in Fig.\,\ref{fig:co_sio_comparison}.

We find a clumpy, shell-like structure that appears disrupted in two opposite facing directions for CO (red). This two-punctured shell was already discussed in \citet{Abellan:2017by} who termed it torus-like. From the 3D map, it appears as if the large punctures were aligned with the equatorial ring plane. Indeed, when determining the parameters of the best-fit plane to the two-punctured shell in Appendix\,\ref{ap:plane} we find an inclination angle to the equatorial ring of 84$\pm$11$^\circ$. With the new threshold, the SiO morphology (blue) is dominated by one large clump, which lies in one of the voids between major clumps forming the CO shell.

\begin{figure}
\centering
  \begin{tikzpicture}
    \node[anchor=center , inner sep=0] (bg) at (0,0)
      {\includegraphics[width=.41\textwidth]{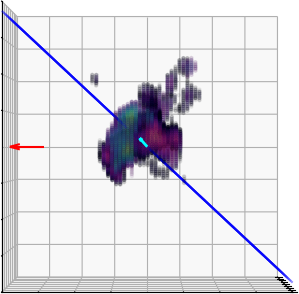}};
    \path[use as bounding box] (bg.south west) rectangle (bg.north east);
      \begin{scope}
          \clip (bg.south west) rectangle (bg.north east);
    \node[anchor=center, inner sep=0, opacity=0.5, scale=1.47, rotate=4.5, xshift=-1.mm,yshift=0.5mm] (fg) at (0,0)
      {\includegraphics[width=.41\textwidth]{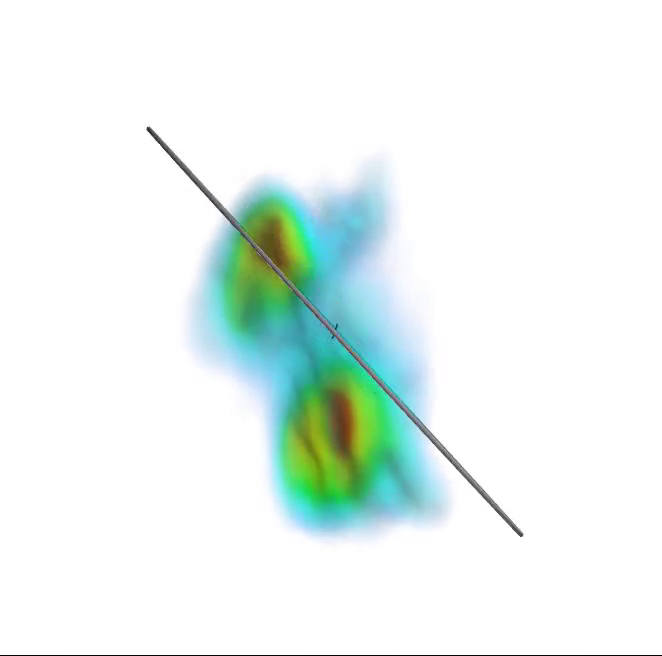}};
  \end{scope}
  \end{tikzpicture}
     \caption{Emission of CO (dark purple-orange) and SiO (dark purple-green) overlaid with the  [Fe~{\sc i}] (faint blue to yellow-red) emission of SN\,1987A obtained in \protect\cite{larsson2023}. The observer position is to the left. Box size: 40,000 AU.}
        \label{fig:co_sio_feI_comparisonI}
\end{figure}

Interestingly, \citet{2023ApJ...949L..27L} find that [Fe~{\sc i}] emission in the ejecta resembles a broken dipole, with two large clumps at velocities of approximately 2300~km\,s$^{-1}$. In Fig.\,\ref{fig:co_sio_feI_comparisonI}, we show their 3D map of [Fe~{\sc i}] emission together with our CO and SiO distributions. One  [Fe~{\sc i}]  clump lies close to the plane of the equatorial ring, while the other is located between the plane of the equatorial ring and the plane of the sky, approximately 20$^\circ$ from the ER plane. This places the clumps approximately perpendicular to the plane of the punctured CO shell. It is tempting to associate the high-density regions of CO and SiO with material pushed aside by rising clumps of heavier material moving at higher velocities. Part of the Fe-rich material synthesized during the explosion may reach sufficient speeds to penetrate the outer shells of lighter elements such as carbon and oxygen. While the distribution of neutral carbon and oxygen relative to CO and SiO remains unknown, comparisons between models and observations can be informative if this is kept in mind.

\begin{figure*}
\begin{overpic}[width=.31\textwidth]{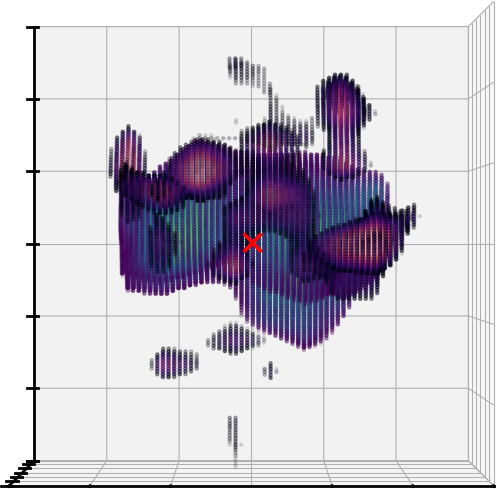}
   \put(80,85){\textbf{-90°}}  
    \end{overpic}
     \begin{overpic}[width=.31\textwidth]{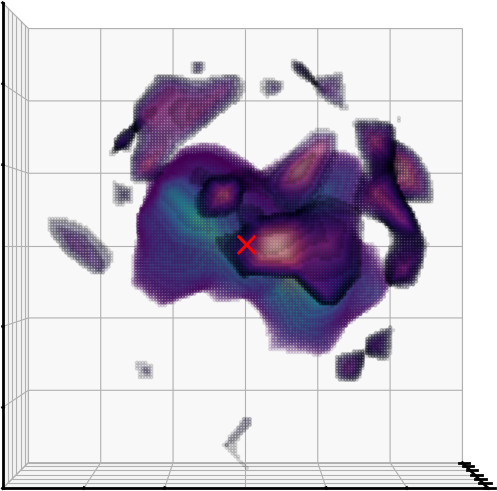}
     \put(80,85){\textbf{0°}}\end{overpic}
     \begin{overpic}[width=.31\textwidth]{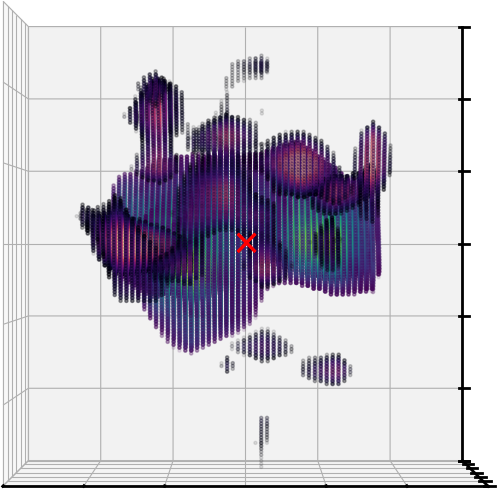}
     \put(80,85){\textbf{90°}}\end{overpic}
    \\
    \begin{overpic}[width=.41\textwidth]{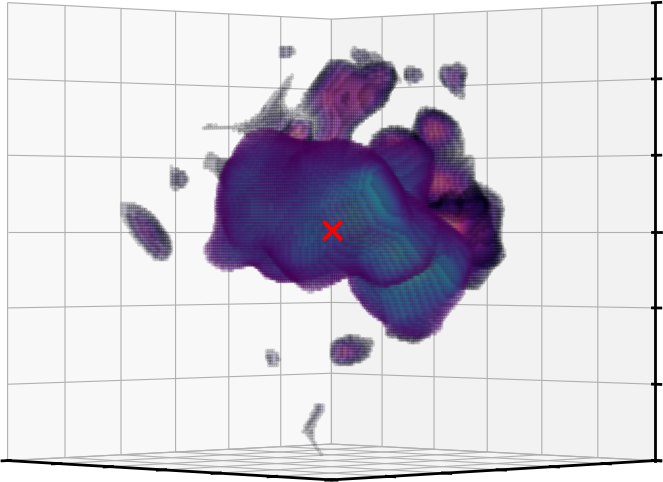} \put(80,60){\textbf{-45°}}\end{overpic}
    \begin{overpic}[width=.41\textwidth]{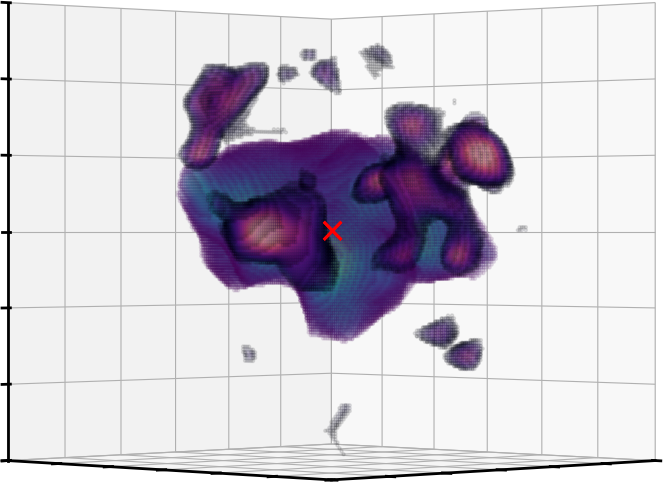}
    \put(80,60){\textbf{45°}}\end{overpic}
    \begin{overpic}[width=.032\textwidth]{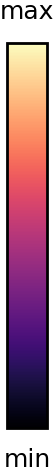}\put(0,100){{C+O}}\end{overpic}
    \hspace{1mm}
     \begin{overpic}[width=.032\textwidth]{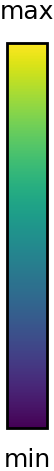} \put(0,100){{Si+O}}\end{overpic}

 \caption{3D density maps of C+O (dark-red-orange) and Si+O (dark-blue-green) of M15--7b--4 at $t\sim1$ yr. To qualitatively reproduce the observations, the minimal density plotted is 45\% of the maximum density in the grid, and the colour scale is limited to 80\% of the maximal density of the data. The axis scale of the grid is $2\times10^{15}$ cm  and the maximum grid extension is $\pm6\times10^{15}$ cm. The red cross marks the centre of the explosion.}
 \label{fig:model3DSiO_M15--7b--4}
\end{figure*}

The bBSG models reproduce best the mass fraction vs velocity distribution, in particular that of C+O. Their 3D C+O morphology is dominated by a shell-like configuration consisting of a number of small clumps. We exemplarily plot the  C+O density distribution of model M15--7b--4 (red) in Fig.\,\ref{fig:model3DSiO_M15--7b--4}. The clumps are somewhat smaller than those of CO in Fig.\,\ref{fig:co_sio_comparison}, which is related to the manually chosen resolution and smoothing of the simulation results.

Similarly to the observations (Fig.\,\ref{fig:co_sio_comparison}), the shell has punctures in the two opposite directions. Unlike C+O, the Si+O distribution (blue) does not resemble the observations to the same degree. Some parts of the Si+O reach the two-punctured shell configuration of the C+O, but the bulk of the material is further inside, expanding at lower velocities (consistent with the mass distributions in Figs.\,\ref{fig:velocity_mass_co21}-\ref{fig:velocity_mass_sio_c2}).

\begin{figure*}
\includegraphics[width=.3\textwidth]{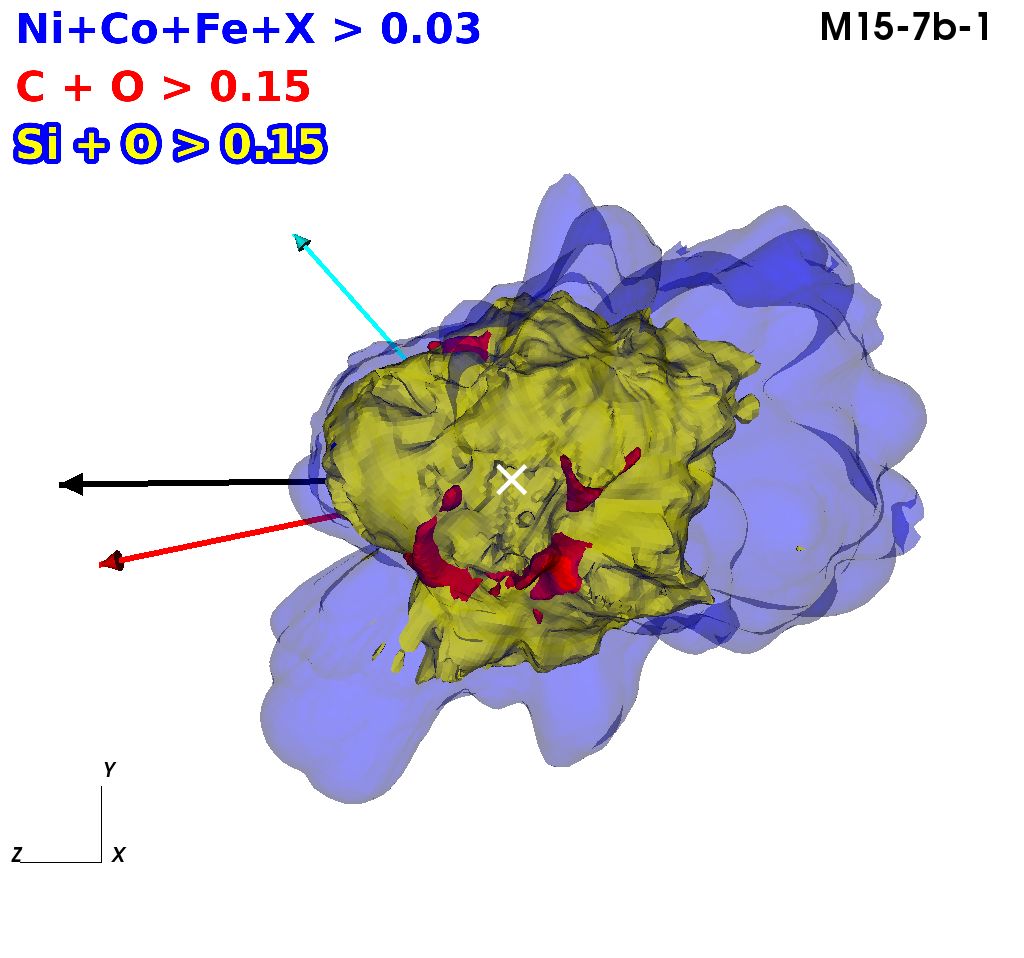}
\includegraphics[width=.3\textwidth]{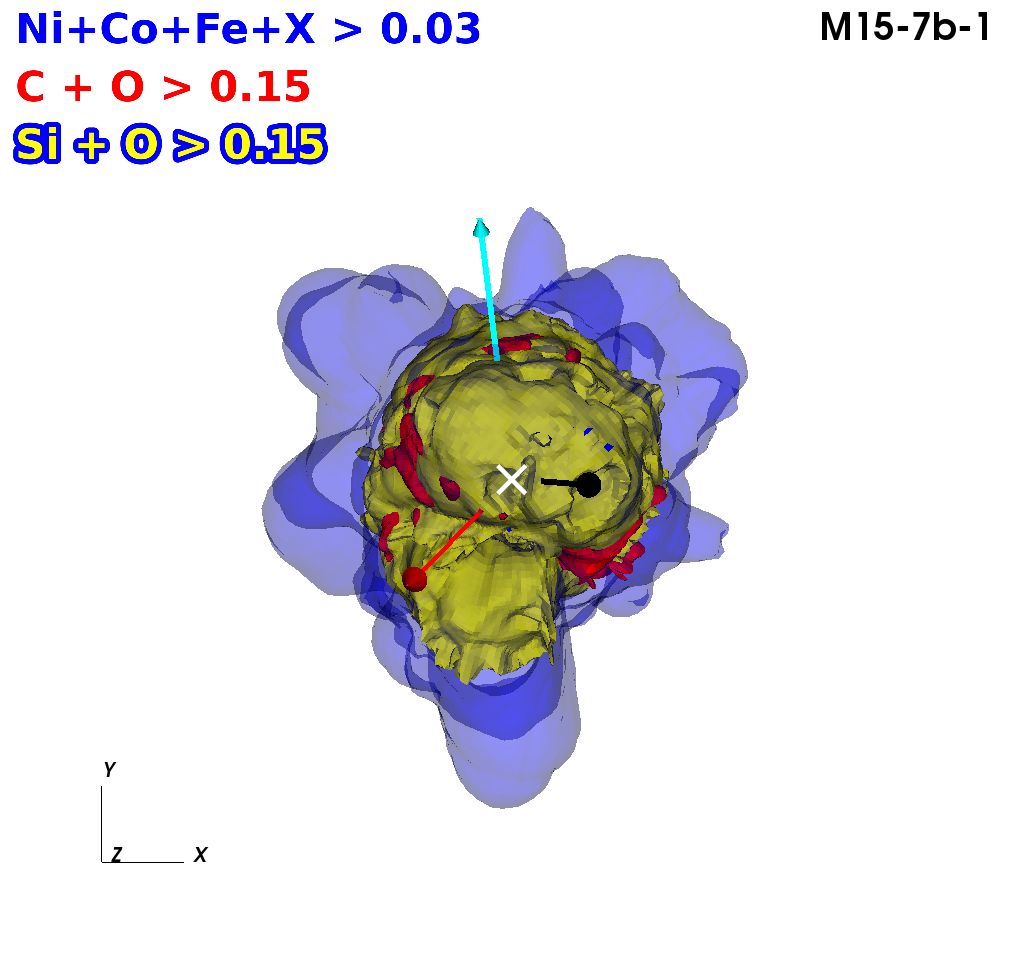}
\includegraphics[width=.3\textwidth]{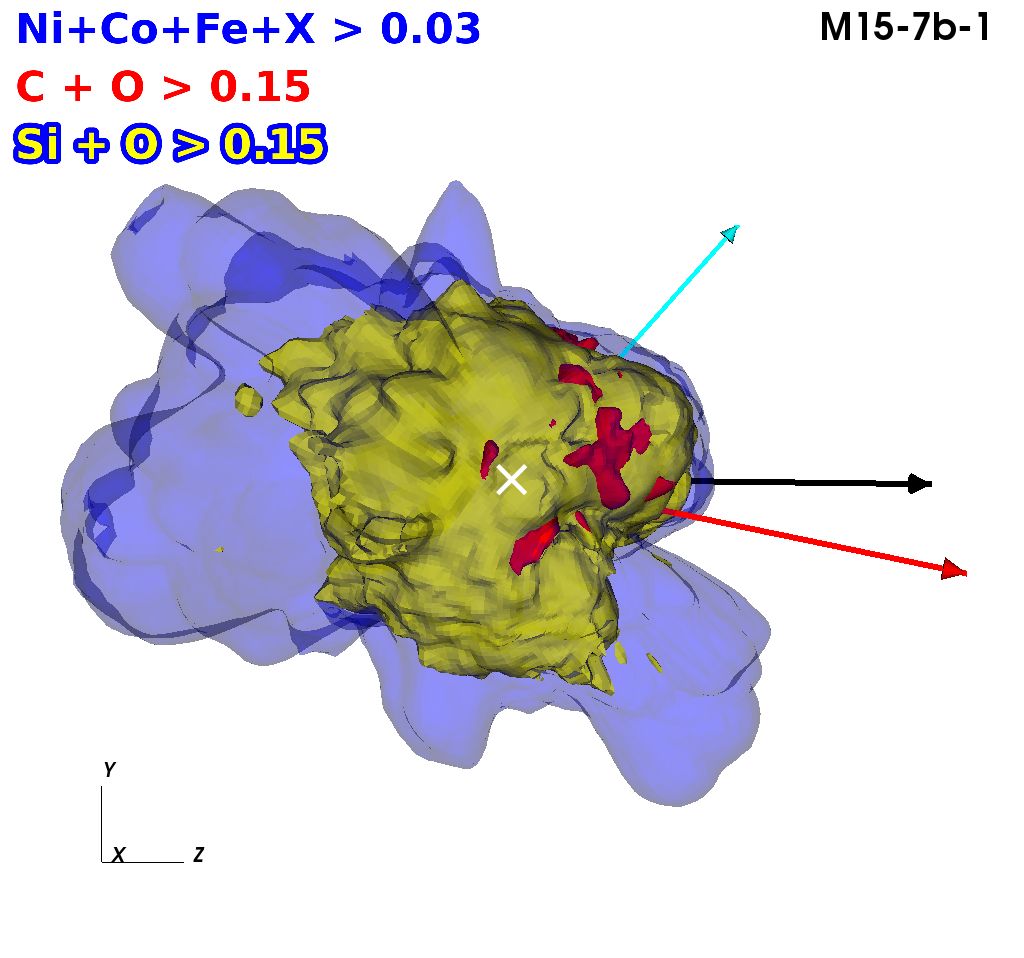}
\\
\includegraphics[width=.3\textwidth]{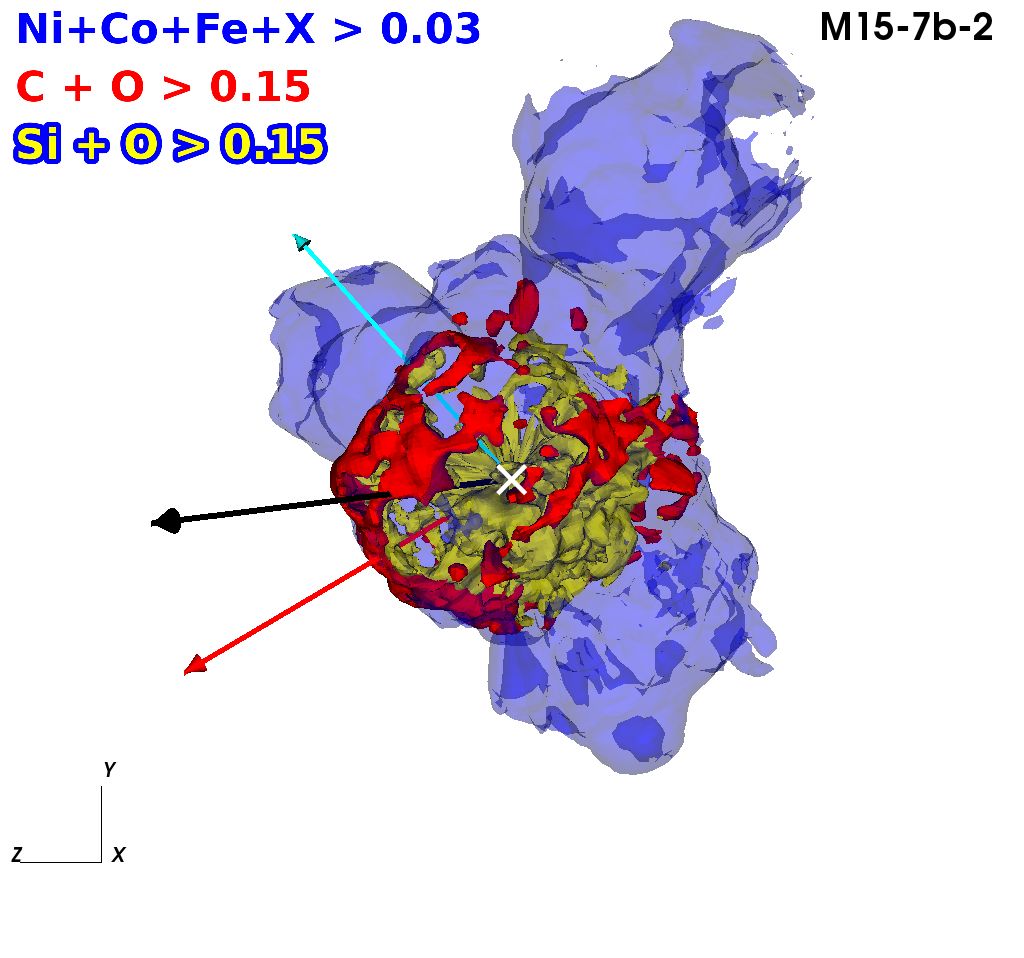}
\includegraphics[width=.3\textwidth]{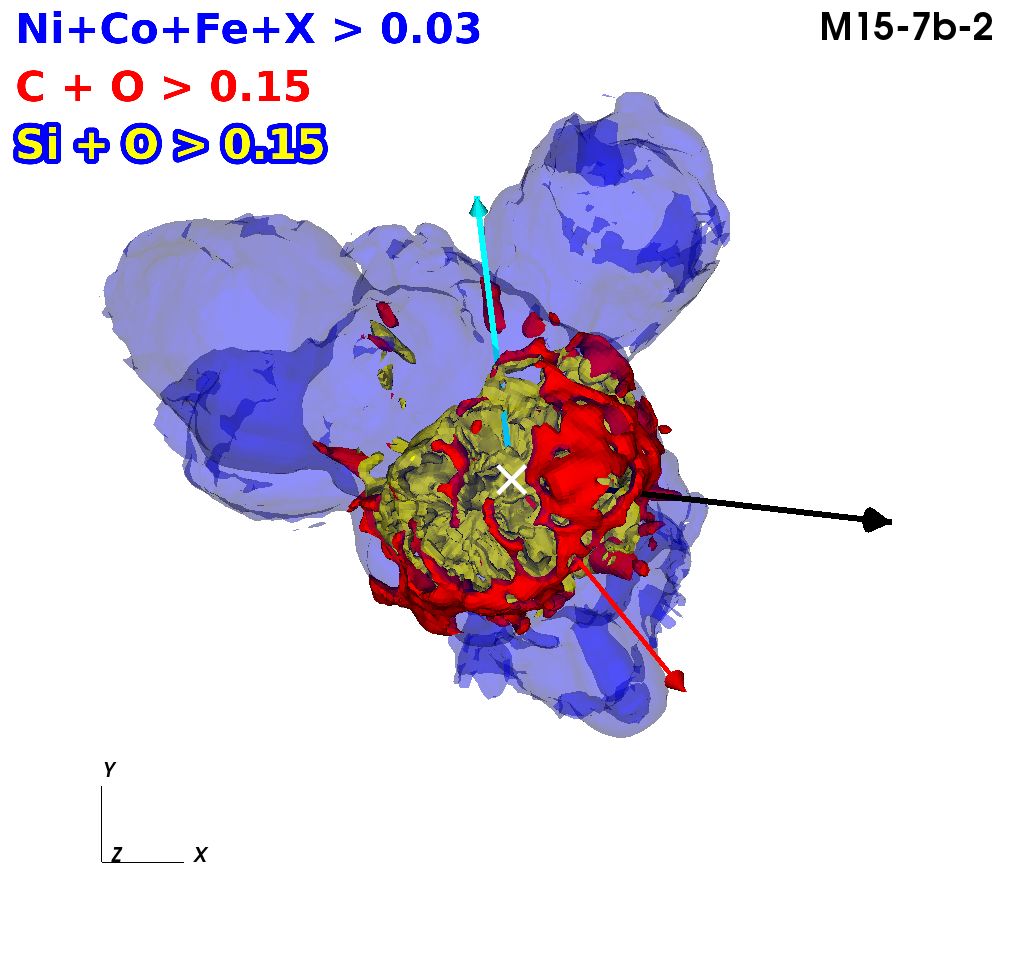}
\includegraphics[width=.3\textwidth]{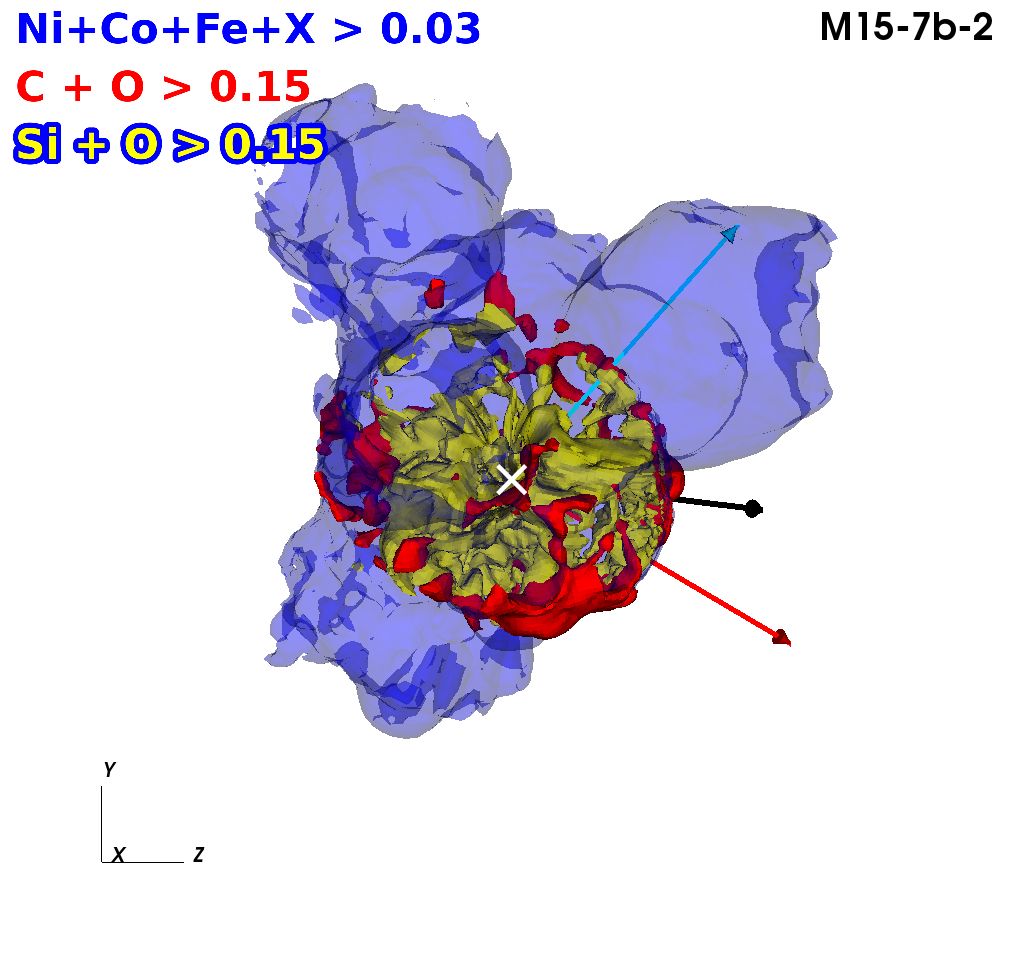}
\\
\includegraphics[width=.3\textwidth]{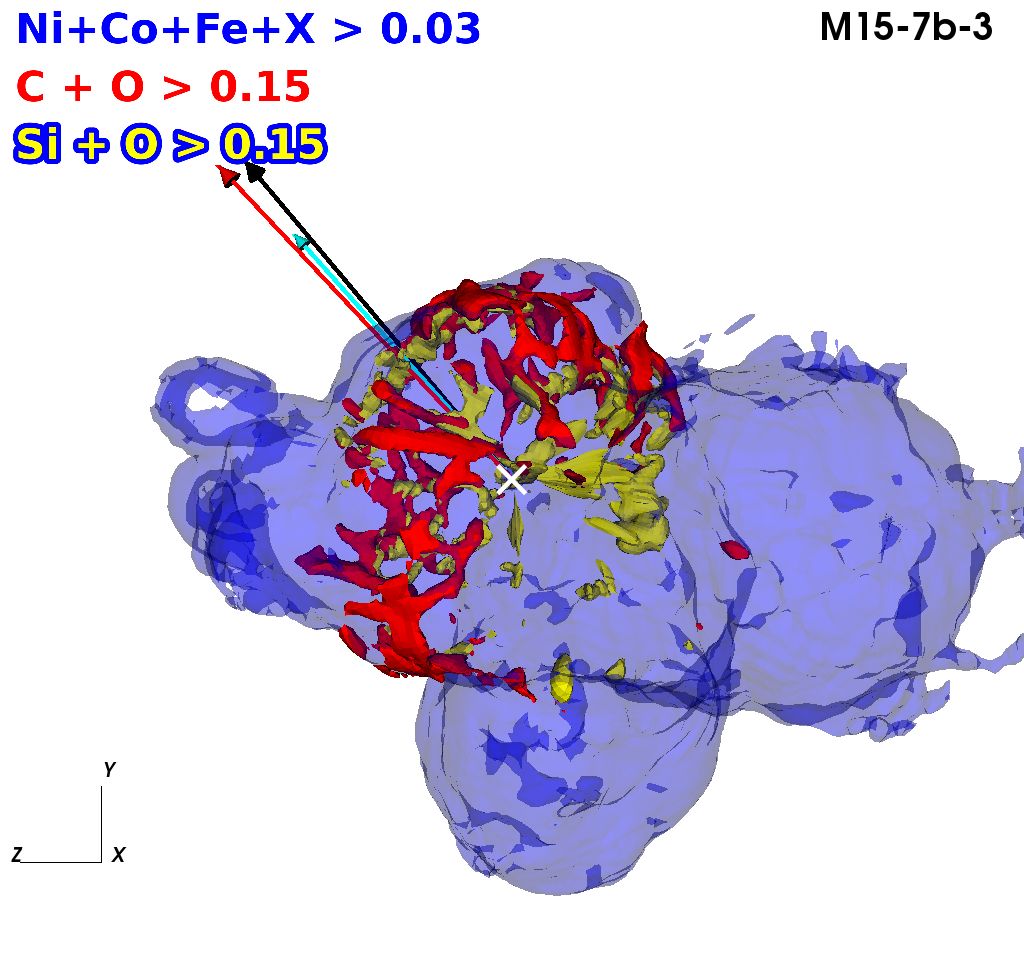}
\includegraphics[width=.3\textwidth]{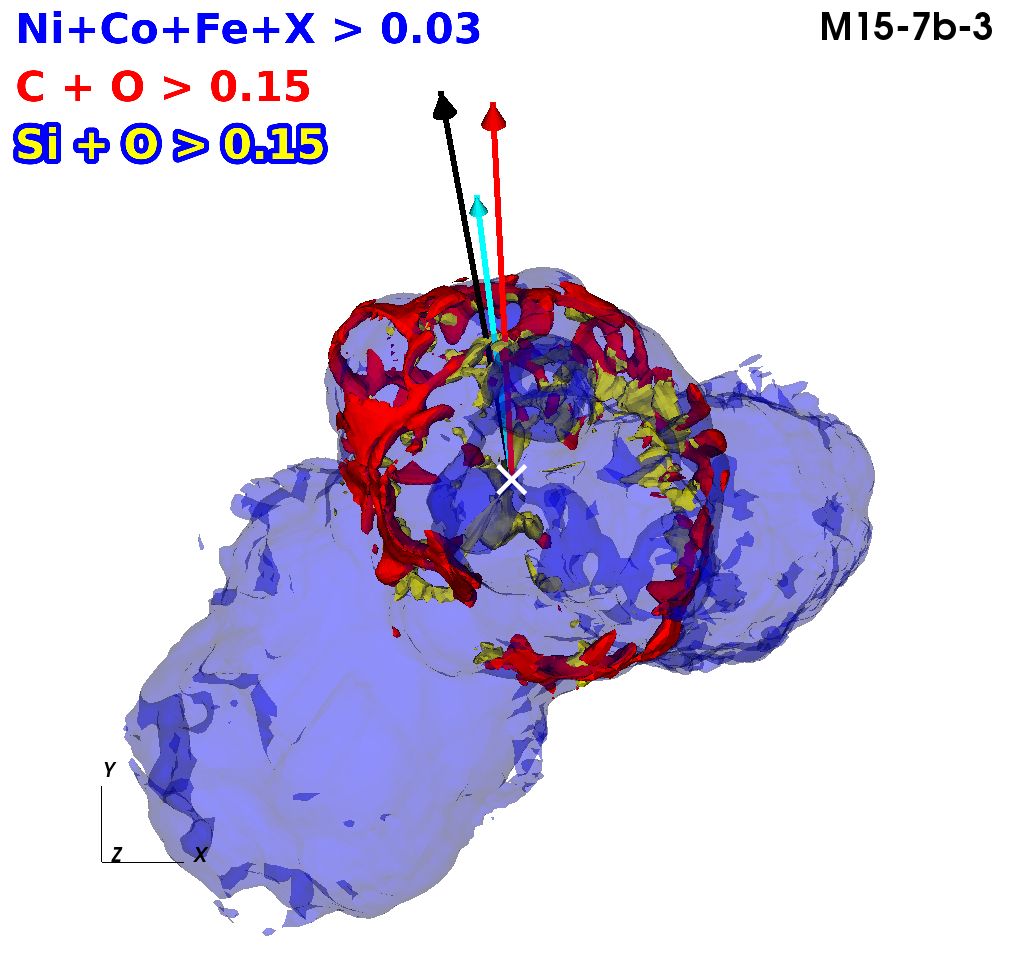}
\includegraphics[width=.3\textwidth]{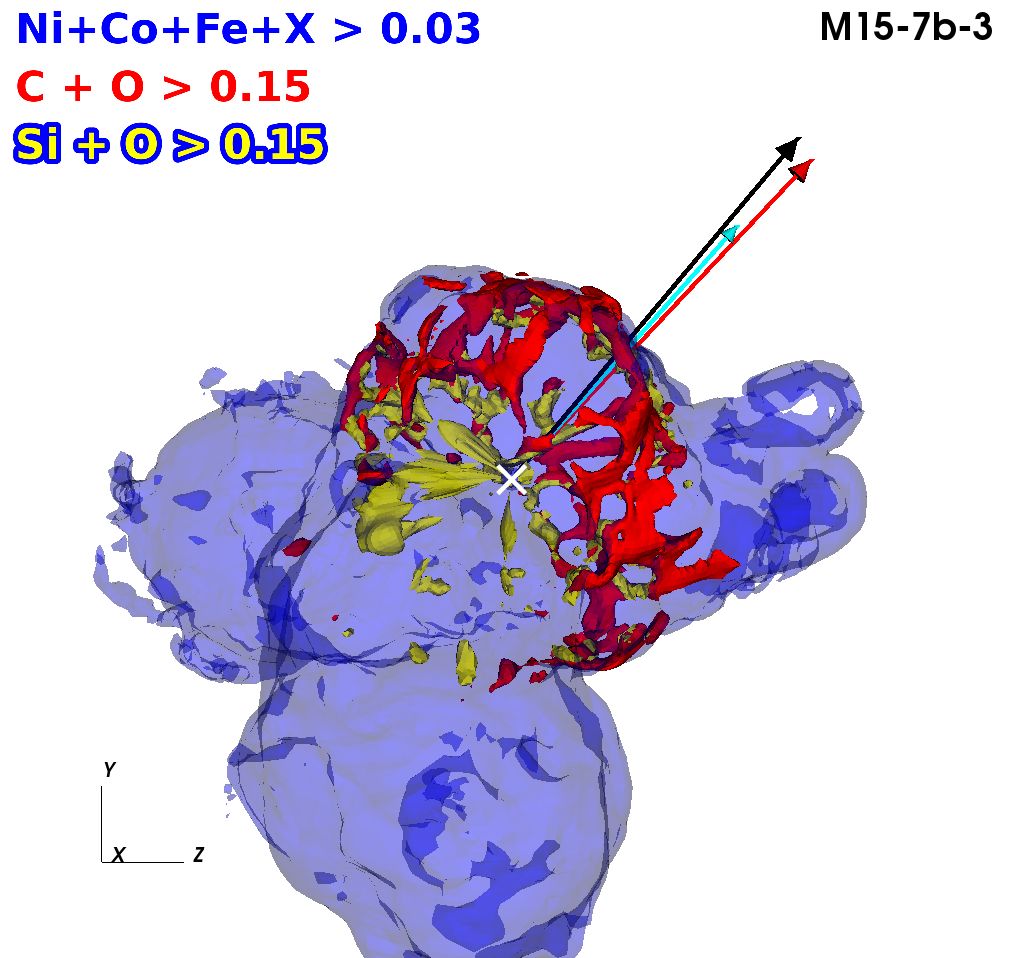}
\\
\includegraphics[width=.3\textwidth]{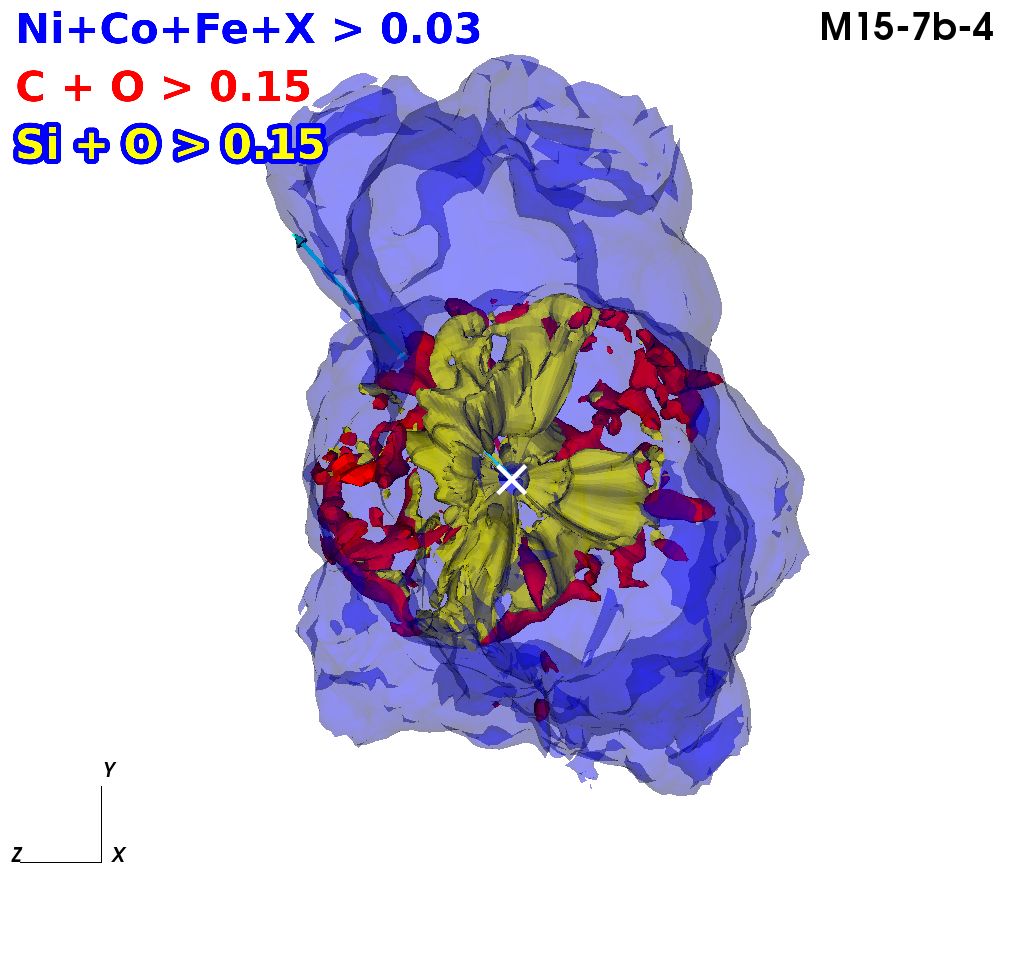}
\includegraphics[width=.3\textwidth]{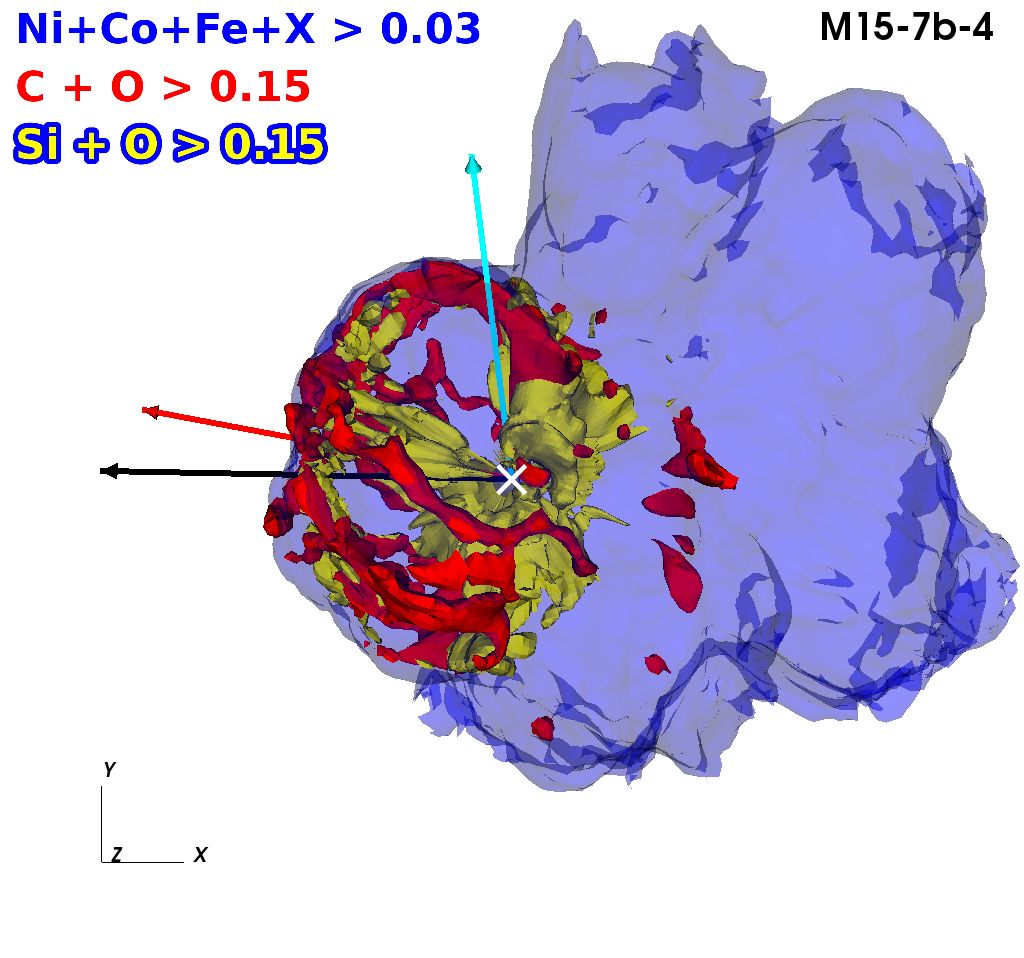}
\includegraphics[width=.3\textwidth]{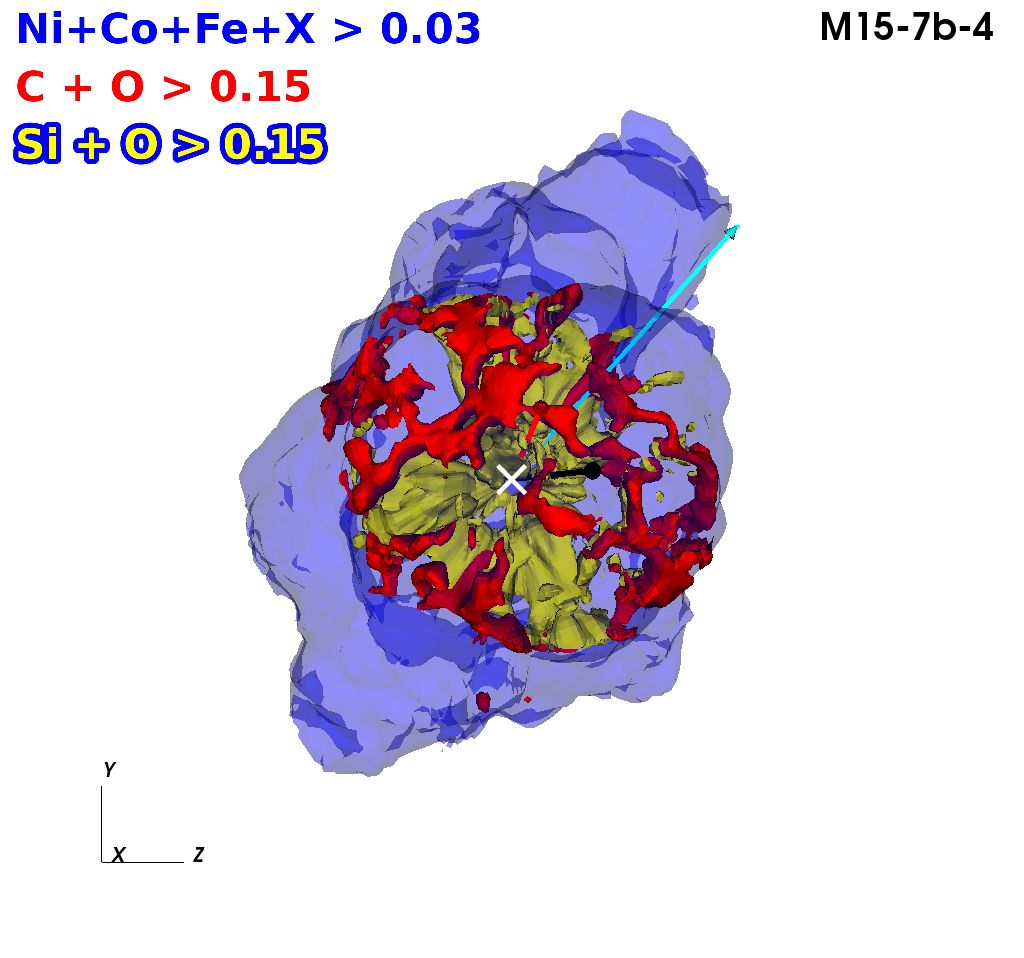}
\caption{Isocontour plots of C+O (red) and Si+O (yellow) corresponding to $15\%$ of the respective peak density. In semi-transparent blue, the sum of the heavy elements Ni, Co, Fe and the tracer X are shown above $3\%$ maximum density. The centres of the explosions are marked with white crosses. The orientation of the plots of the central column is chosen such that i) The best-fitting plane to the C+O densities has the same orientation as that of the CO observed in SN\,1987A; ii) we minimize the angle between the C+O centre of mass in the models and the centre of emission in the CO data of SN\,1987A without changing the angle of the plane. The corresponding directions are indicated as cyan (CO SN\,1987A), red (C+O centre of mass model) and black (NS kick model) arrows. In the theoretical models the direction of the centre of mass of the C+O is correlated with the kick direction of the neutron star. The central column represents the face-on view of the observer, i.e. the observer is at the positive z  direction. The panels in the left (right) column are the central ones rotated by $-90^\circ$ ($+90^\circ$). The arrows indicating kick or CO directions in some panels may be hidden behind the ejecta (e.g. bottom left panel).}\label{fig:kick1}
\end{figure*}

\begin{figure*}
\includegraphics[width=.3\textwidth]{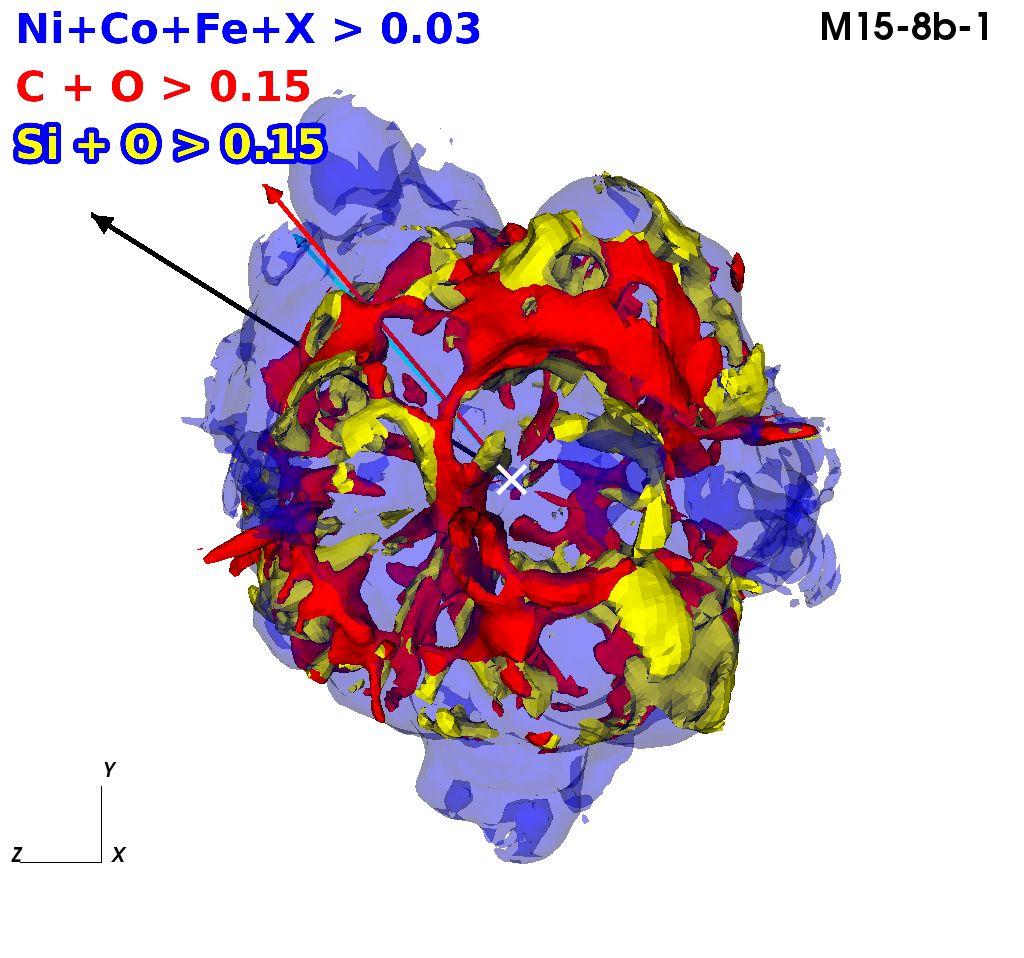}
\includegraphics[width=.3\textwidth]{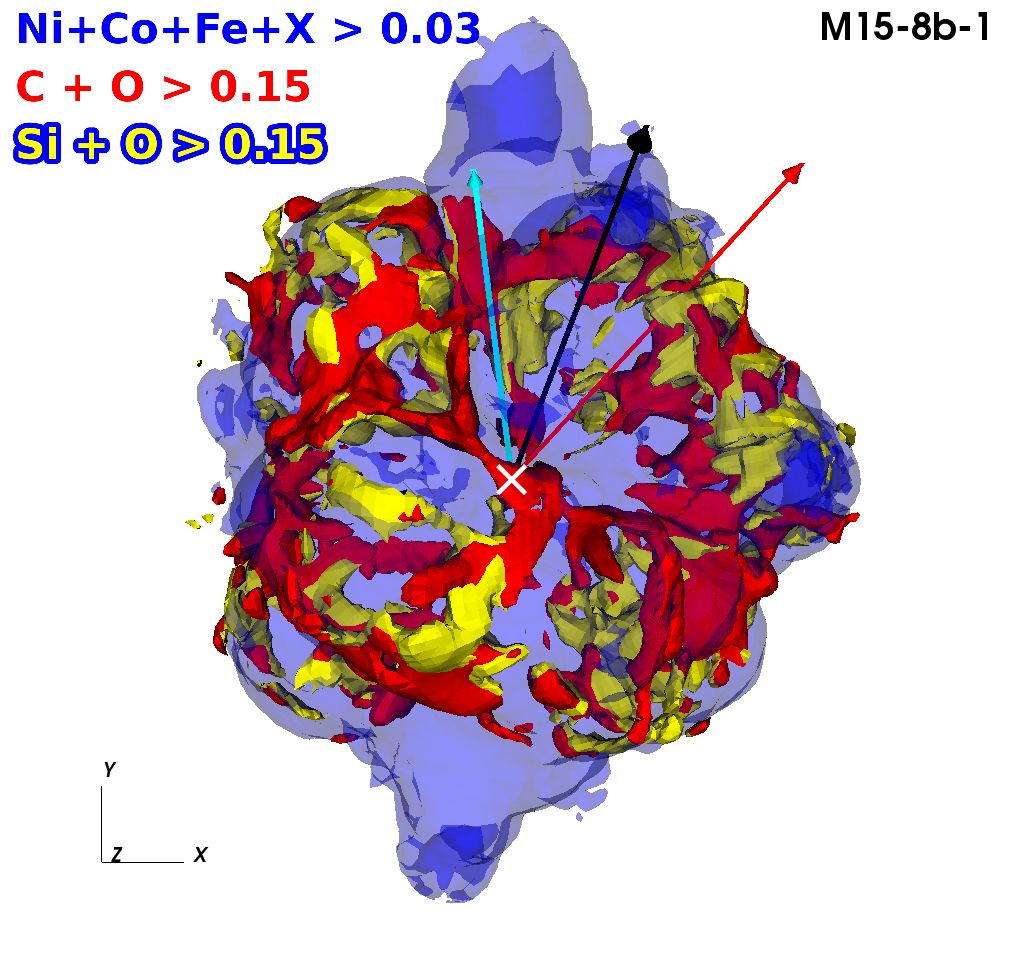}
\includegraphics[width=.3\textwidth]{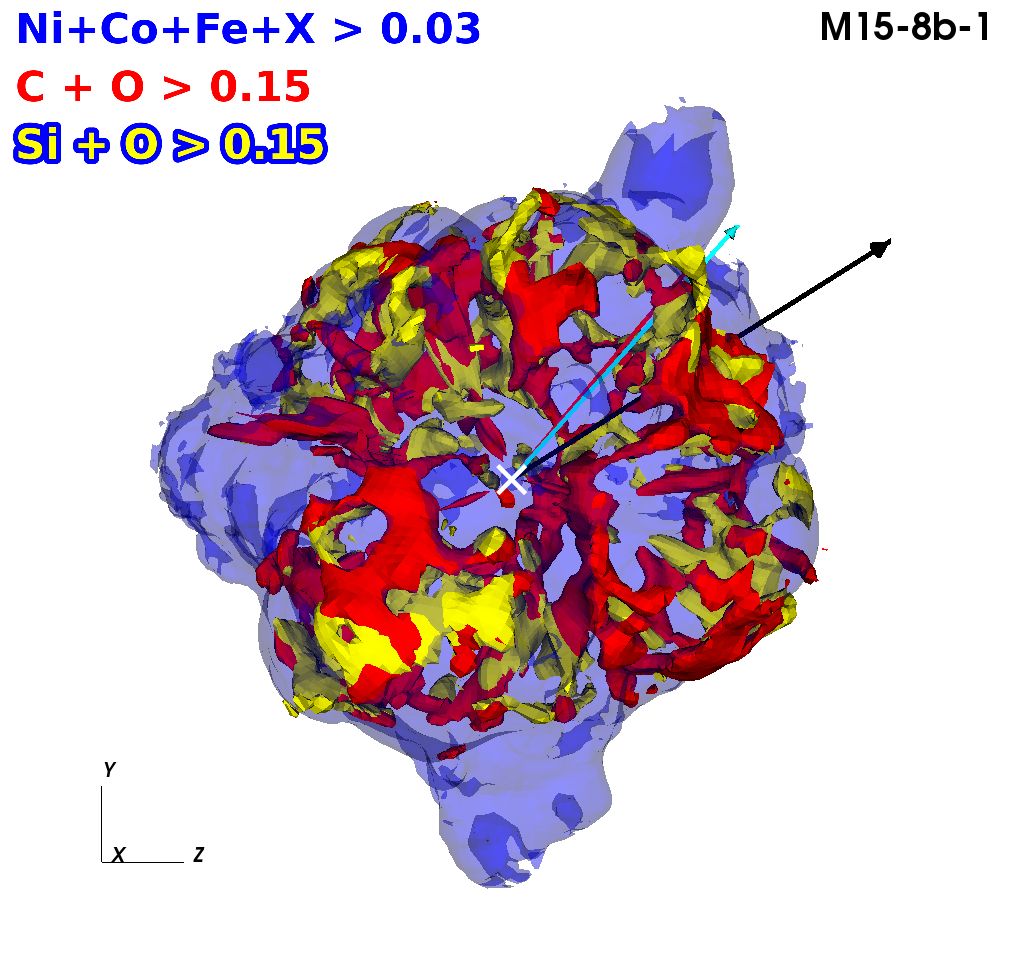}
\caption{Isocontour plots of C+O (yellow) and Si+O (red) correpsonding to $15\%$ of the peak density of model M15--8b--1. In semi-transparent blue, the sum of the heavy elements Ni, Co, Fe and the tracer X are shown. The centre of the explosion is marked with a white cross. The orientation of the plots of the central panel is chosen such that it best reproduces the observer view of SN\,1987A, as described in Fig.\,\ref{fig:kick1}. Directions of NS kick (black), centre of mass of C+O of the model (red) and CO emission of SN\,1987A (cyan) are indicated as arrows. The left (right) panel is the central one rotated by $-90^\circ$ ($+90^\circ$).
}\label{fig:kick2}
\end{figure*}

The morphologies of the other bBSG models (not shown) have very similar characteristics to M15--7b--4 for both C+O and Si+O. However, the distribution of C+O in model M15--8b--1 is more spherical, without any distinctive large-scale punctures. 
To test the hypothesis that the punctures are caused by heavy, fast moving elements, we visualize the distribution of Si+O (yellow) and C+O (red) above $15\%$ maximum density in Figs.\,\ref{fig:kick1} and \ref{fig:kick2} for all of the bBSG models. In addition, we show the sum of the densities of the tracer $^{56}$X and $^{56}$Ni and its decay products $^{56}$Co and $^{56}$Fe above $3\%$ of the respective maximum density (semi-transparent blue). For these plots, we show the results of the simulations in spherical coordinates and do not lower the resolution. Consequently, the grid cells with the highest densities of C+O (and Si+O) have higher densities than in the down-sampled case, and to represent comparable volumes, the respective percentages defining the minimum density for C+O and Si+O have to be lower. For the heavy elements, we chose $3\%$ to be able to see most of the ejecta. As anticipated, we see large Fe-rich (for simplicity we will refer to Fe-rich clumps when discussing the Ni+Co+Fe+X ejecta) clumps further out than the high density C+O ejecta. These clumps have pushed through the shells of lighter elements and pushed some material aside. This is where the high density C+O and is located. Model M15--8b--1 (Fig.\,\ref{fig:kick2}) does not present such Fe-rich clumps and is the most spherical of all models. Also the C+O and Si+O ejecta are more spherical than in the other bBSG models.

Similar to the SiO observations in SN 1987A, models M15–7b–3 and M15–8b–1 exhibit a distribution of the densest Si+O material that contains fewer clumps than the C+O component, yet retains a broadly similar overall morphology.. In contrast, the Si+O extends further inside in models M15--7b--4 and M15--7b--2, and almost completely fills the volume interior to C+O for model M15--7b--1. This different appearance may be caused by choosing a particular density threshold for Si+O that is too low in some cases (M15--7b--1, M15--7b--2 and M15--7b--4) and just sufficient in others (M15--7b--3 and M15--8b--1). We see similar effects for C+O when the corresponding density threshold is significantly decreased. More likely, this is related to insufficient fast-moving, Si+O-rich ejecta in the theoretical models (see Section\,\ref{sec:compare_distr}). The reason for this Si deficit is likely related to its low abundance in the outer shells in current progenitor models. Therefore, most of the ejected Si is produced in the centre of the explosion, and the material pushed aside by the Fe-rich protrusions contains very little Si, and hence very little Si+O.

\subsection{Neutron star kick geometry in theoretical models}\label{sec:kick}

\citet{wongwathanarat2010} showed that the NS moves in the direction opposite to that of the fastest Fe-rich ejecta. This kick is attributed to the so-called {\it tug-boat mechanism}, in which the slower-moving ejecta, remaining closer to the NS, exert a gravitational pull that accelerates the NS in their direction. In a strongly one-sided explosion, one would expect the NS to move in the opposite direction to the fastest Fe-rich ejecta. The corresponding kicks of the most interesting models of SN\,1987A progenitors (group 2) are given in Table\,\ref{tab:kicks}. As expected from the very symmetric explosion of model M15--8b--1, this model has by far the lowest kick velocity.

These kicks are plotted as the black arrows with the distribution of the ejecta in Figs.\,\ref{fig:kick1} and \ref{fig:kick2}. Confirming previous findings \citep{Wongwathanarat2013}, the kick direction never points in the direction of a large Fe-rich clump.

Instead, it seems that the kick points in the direction where a significant part of the densest C+O ejecta is located. Indeed, the direction of the centre of mass of the densest C+O, indicated by the red arrow, is correlated with the kick direction of the NS in all our models.
Here, we have considered only the cells whose C+O density reaches at least $15\%$ of the maximal density in the grid. The mass contained within this limit is on the order of $10\%$ of the total C+O mass in all of the models. 
The corresponding angles between kick and C+O centre of mass $\alpha$ are given in the last column of Table\,\ref{tab:kicks}. They are generally lower than $\alpha<33^\circ$. Only the most spherical model M15--8b--1, that has the lowest kick velocity, exceeds this value with $\alpha\sim43^\circ$. However, in this model we do not see large extended Fe-clumps. Therefore, the C+O is also distributed more symmetrically and we do not expect a strong correlation between the kick and C+O directions (Fig.\,\ref{fig:kick2}).

We have tested whether correlations between kick and C+O depend on the particular value of the minimal density considered, by changing the threshold between $5\%$ and $20\%$. In all these cases, the deviations stay below $\alpha<40^\circ$ (apart from model M15--8b--1).

\begin{table}
    \centering
    \begin{tabular}{c|c c c c c c}
 Model&M$_\mathrm{NS}$ &$v_x$ &$v_y$ &$v_z$&$|v|$ &$\alpha$\\
 & [\Msun]& \multicolumn{4}{c}{[km s$^{-1}$]}&[$^\circ$]\\
 \hline
M157b-1  & 1.577 &      266 &     -553 &      311 &       688&26\\
M157b-2  &  1.556&       336&     -367 &     -271 &       434&33\\
M157b-3  &  1.587&       373&     -415 &      377 &       673&2\\
M157b-4  &  1.492&       158&       281 &     -746&        813&11\\
M158b-1  &  1.319&      -27&      -39 &     -44&        65&43\\
    \end{tabular}
    \caption{Kicks of the neutron star of the SN explosions originating from the bBSG progenitor models. The angle $\alpha$ is the angle between the kick direction, and the centre of mass of the densest $15\%$ C+O.}
    \label{tab:kicks}
\end{table}



\begin{figure*}
    \includegraphics[width=0.47\textwidth]{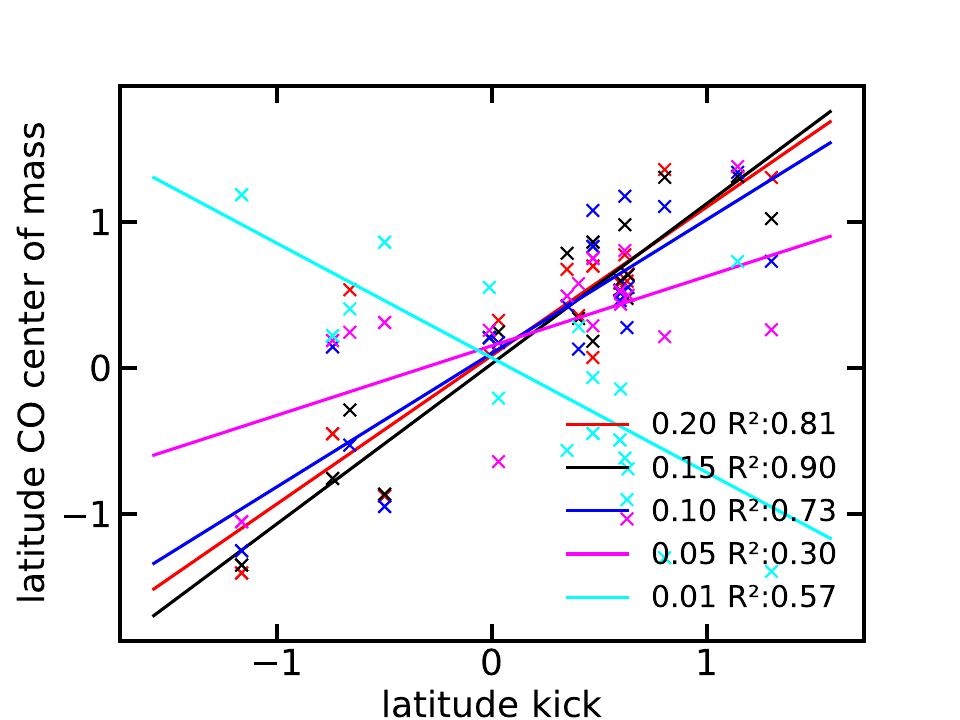}
    \includegraphics[width=0.47\textwidth]{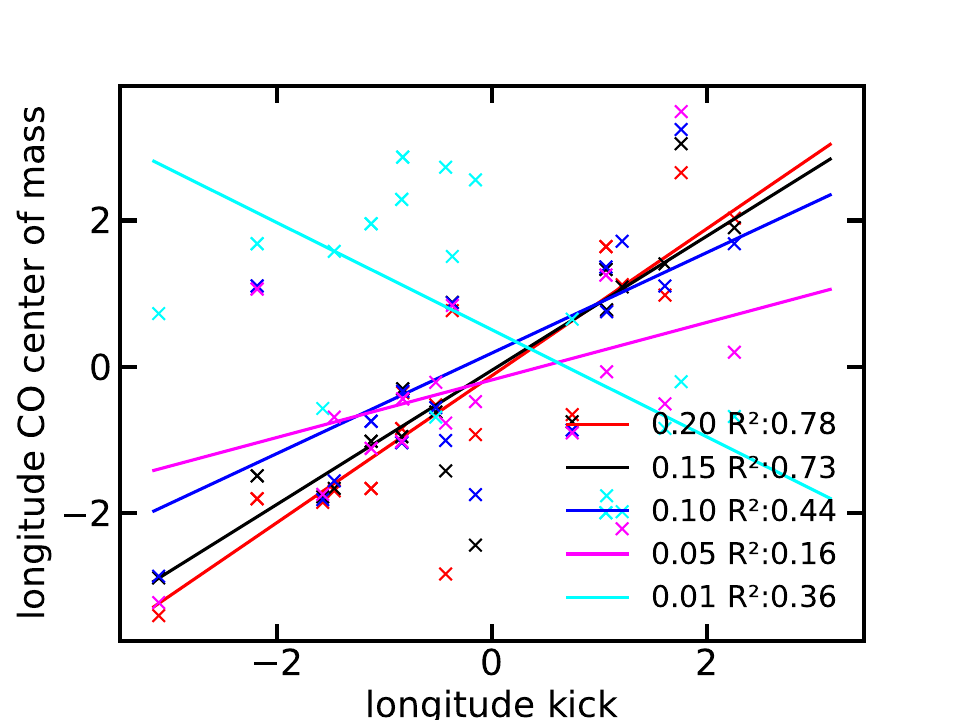}
    \caption{Angular positions of the centre of mass of the C+O material versus the angular positions of the kick direction of all models of group 2 and 13 RSG progenitors \protect\citep{Giudici2025a}. Left panel: comparison of latitudes; right panel: comparison of longitudes. Colours correspond to different thresholds of the fractional maximum density $[0.2,\ 0.15,\ 0.1,\ 0.05,\ 0.01]$ above which cells were included in the radial integration. The second number in the legend gives the goodness of the linear regression. Note that, to account for periodicity, we manually added or subtracted $2\pi$ at a few data points to improve the quality of the fit.}\label{fig:lat}
\end{figure*}

In order to further strengthen our findings, we analysed all 13 RSG models of \cite{Giudici2025a,Giudici2025b}, of which our group 3 is a subset, in the same way: determining the kick direction of the neutron star (at $t_\mathrm{pb}=2.5$s) and that of the centre of mass of the C+O with densities above a certain fraction of the maximal density in the grid (at $t\sim1\,$yr). We applied this procedure using several fractions of the threshold density, expressed as percentages of the maximum density found in the simulations: F$_{\rho\mathrm{(C+O)}}=[0.2,0.15,0.1,0.05,0.01]$. For convenience, we choose an angular grid consisting of longitude $\in[-\pi,\pi]$ and latitude $\in[-\pi/2,\pi/2]$. The corresponding angles for the kick and C+O centre of mass directions are given in Fig.\,\ref{fig:lat} for all models of groups 2 and 3. In the figure, the latitude (longitude) of the centre of the C+O mass is plotted versus the latitude (longitude) of the kick direction in the left (right) panel. For a sufficiently high density threshold, we obtain a relatively good fit with a $R^2$ coefficient of the order of $0.65-0.86$. The best coefficients are obtained for F$_{\rho\mathrm{(C+O)}}=0.15$, which justifies our previous choice for this parameter. Note that in some cases, performing a linear fit required wrapping certain angular values beyond the standard interval [$-2\pi$, $2\pi$] for longitude and [$-\pi$, $\pi$] for latitude.

\subsection*{Indications for kick direction in SN\,1987A from CO emission centre}
\begin{figure*}
\includegraphics[width=.41\textwidth]{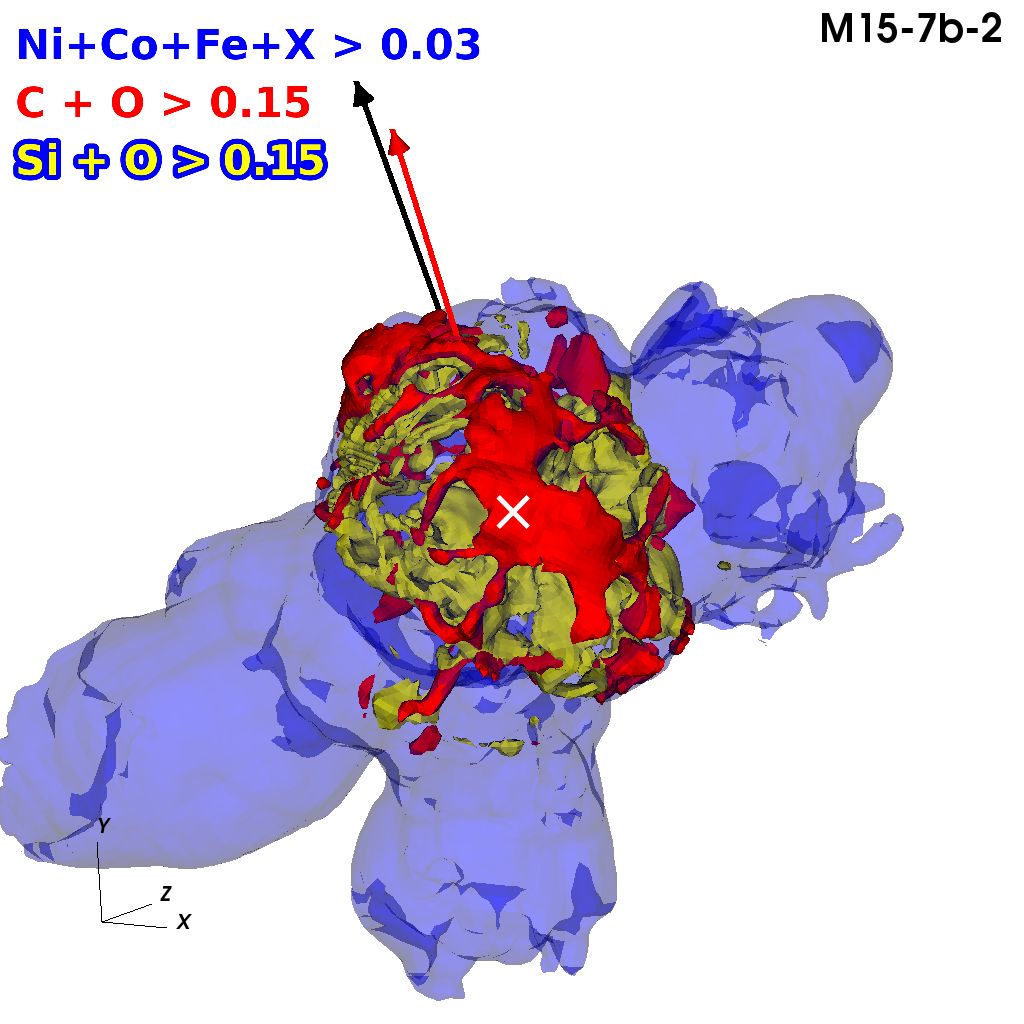}
    \begin{tikzpicture}
    \node[anchor=center, inner sep=0] (bg) at (0,0)
      {\includegraphics[width=.41\textwidth]{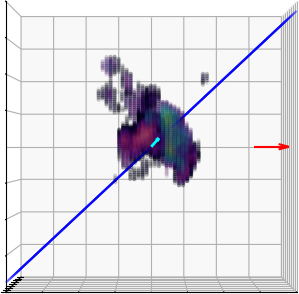}};
    \path[use as bounding box] (bg.south west) rectangle (bg.north east);
      \begin{scope}
          \clip (bg.south west) rectangle (bg.north east);
    \node[anchor=center, inner sep=0, opacity=0.5, scale=1.47, rotate=-4.5, xshift=1.mm,yshift=0.5mm] (fg) at (0,0)
      {\includegraphics[width=.41\textwidth]{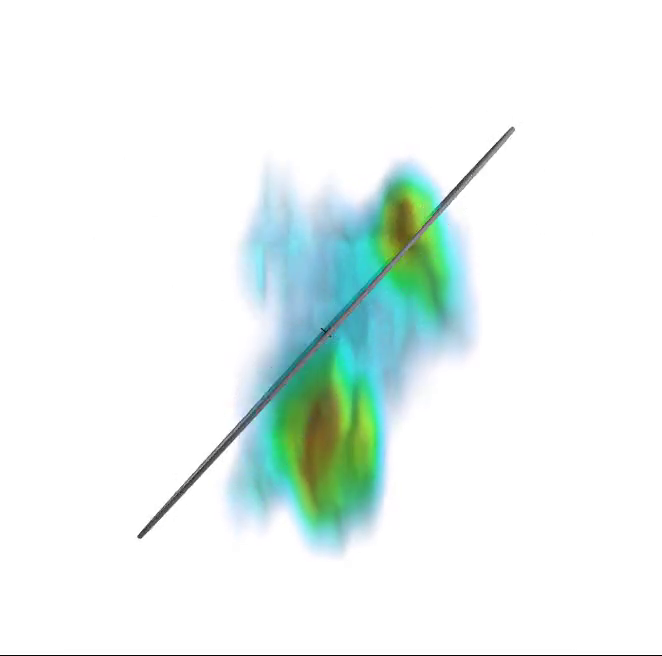}};
        \end{scope}

  \end{tikzpicture}

    \caption{Distribution of Ni+Co+Fe+X, C+O and Si+O for given minimal densities for model M15-7b-2 (left panel) and emission of CO (dark purple-orange) and SiO (dark purple-green) overlaid with the [Fe~{\sc i}] (faint blue to yellow-red) emission obtained in \protect\cite{larsson2023} (right panel, box size: 40,000 AU.). The similar morphologies of the theoretical model (left panel) and the observations of SN\,1987A (right panel) are found: large-scale Fe-rich plumes penetrating a punctured C+O shell, pushing the latter into regions into which no plume extends. The difference between the kick directions, north in the model (black arrow, left panel) and north-west in the observations (cyan arrow, right panel) is larger than the intrinsic model uncertainty ($\alpha<45^\circ$ for bBSG) alone, but well within the combined uncertainties of the model, the CO/C+O mapping, and the morphological differences. }
        \label{fig:co_sio_feI_comparisonII}
\end{figure*}

Hydrodynamic simulations predict a good correlation between the direction of the centre of C+O mass  
and the direction of the neutron star kicks. 
We now employ this correlation to estimate the neutron star kick in SN\,1987A, using ALMA measured CO distributions. The direction of the emission-weighted centre of CO lies 155~AU east, 1240~AU north and 1100~AU in front of the SN explosion site and is plotted in Fig.\,\ref{fig:co_sio_comparison} as the cyan arrow.
Its polar coordinates, where $\Theta=0$, $\phi=0$ corresponds to the line of sight, are $\Theta=8.0^\circ, \phi=48.1^\circ, r=1660$~AU.
This vector points almost towards the biggest clump of CO. Additionally, the vector to the centre of emission of SiO points towards the only large SiO clump, and is almost aligned with the direction to the CO emission centre. Interestingly, the CO vector points in a direction within the equatorial plane towards one of the extended regions containing [Fe~{\sc i}] emission lines (Figs.\,\ref{fig:co_sio_feI_comparisonI} and \ref{fig:co_sio_feI_comparisonII}) reported in \citet{larsson2023}.

To summarize our findings, the left panel of Fig.\,\ref{fig:co_sio_feI_comparisonII}
displays the viewing angle of the theoretical model that best matches the observations. In the right panel of the same figure, we show the CO, SiO and Fe observations of SN\,1987A as comparison. This is essentially the same plot as  Fig.\,\ref{fig:co_sio_feI_comparisonI} but has a different viewing angle: in Fig.\,\ref{fig:co_sio_feI_comparisonII}, the observer is to the right of the plot. The morphology of the model M15-7b-2 with the selected viewing angle resembles that of SN\,1987A. There are large, extended Fe-rich plumes extending to the north-west and south and the distribution of the densest C+O is located in regions where no Fe-rich clump is extending. The major difference to the right panel consists in the presence of one additional big Fe-plume to the south-east. Despite this additional feature, the agreement is remarkable, taking into account that it is extremely unlikely to reproduce the exact explosion dynamics of SN\,1987A in the theoretical models, because, in particular, the shock-revival phase, during which the asymmetries are seeded, is dominated by random instabilities. The striking similarity underlines that the kind of neutrino-driven explosions studied in the theoretical models presented here can explain the observed features in SN\,1987A. The difference between the indicated tentative kick directions, north in the model (black arrow, left panel) and north-west in the observations (cyan arrow, right panel) is within the uncertainties of the model.

\section{Summary and discussion}\label{sec:summary}
Our main findings regarding the distribution of molecules in the ejecta of SN,1987A are as follows:

\begin{itemize}
\item \textbf{CO}: 
The CO molecules are distributed in a clumpy asymmetric morphology. We confirm the previous indications of a central void in CO 2--1 at low velocities \citep{Abellan:2017by,Cigan:2019cl}. When increasing the intensity cut-off to double the estimated noise level, the observations show a two-punctured shell morphology of the densest material. The CO morphology is reflected in the mass distribution as a function of velocity by showing a broad peak $v^\mathrm{CO}_\mathrm{peak}\approx1840$km\,s$^{-1}$ with a slow rise and slower decaying and extended tail $v^\mathrm{CO}_\mathrm{max}\approx3980$km\,s$^{-1}$. There is no significant time evolution between the different observational epochs, consistent with a homologous expansion.

The comparison with density distributions of theoretical models reveals that models of group 1 and the RSG of group 3 generally predict a higher fractional mass component at lower velocities (<1000km\,s$^{-1}$) and lower $v^\mathrm{C+O}_\mathrm{peak}\lesssim1500$km\,s$^{-1}$ than is observed in the data. Out of group 2, only models SW19.8 and SW27.3 seem to be consistent with the observations. As expected for the correct type of progenitor star, the mass distributions of C+O for the bBSGs of group 2 agree much better with the CO distribution of SN\,1987A. In addition, also the 3D density maps of the bBSG models, in particular that of models M15--7b--2, M15--7b--3, and M15--7b--4 are in good qualitative agreement to the CO intensity map. They reproduce the general morphology, including the central void and a punctured shell morphology for the higher density cut-off. From the 3D morphologies of the different elements in Fig.\,\ref{fig:kick1}, we see that everywhere where the Fe-rich clumps or plumes extend out to large distances, they have pushed aside (and partially dragged along with itself) the C+O. As a consequence, the densest C+O structures are those not punctured by large clumps remaining in 
what was once the C+O shell of the progenitor. This is very similar to what we see in SN\,1987A, where the CO molecules emit preferentially in a plane (almost) perpendicular to the large Fe-rich plumes detected in [Fe~{\sc i}].

\item \textbf{SiO}:
The morphology of SiO is qualitatively similar to that that of CO. There are large SiO structures mostly close to locations where CO is located. These structures tend to be slightly less extended, but reach to large distances. Consequently, the mass distributions are similar to those of CO, with slightly lower $v^\mathrm{SiO}_\mathrm{peak}\approx1620$km\,s$^{-1}$ and  $v^\mathrm{SiO}_\mathrm{max}\approx3560$km\,s$^{-1}$. When increasing the cut-off intensity (Fig.\,\ref{fig:co_sio_comparison}), there remains only one dominating clump located within the two-punctured shell of CO.

Almost all models, including the bBSG models with promising results for C+O, overpredict the SiO distribution at lower velocities and arrive at too low $v^\mathrm{Si+O}_\mathrm{peak}$. Only models SW19.8 and N20 can reproduce the slow rise and both models have $v^\mathrm{SiO}_\mathrm{peak}$ closest to the observed value. However, neither model is a viable SN\,1987A progenitor. One model, SW20.8, fits the observations of SiO above $v\gtrsim2000$km\,s$^{-1}$ particularly well. However, this good match is not reflected at all in the 3D maps of Si+O, which are very different from the observational data. Furthermore, the bBSG models M15--7b--1, M15--7b--2, M15--7b--4 tend to have too much high-density Si+O inside the punctured shell. Only models M15--7b--3 and M15--8b--1 have the densest Si+O mainly located close to the locations of the C+O ejecta.

\item \textbf{Kick direction:} 
Our models (Sect.\,\ref{sec:kick} reveal a strong correlation between the centre of mass of C+O and the direction of the NS kick. Even for the least favourable cases, the angular distance between both directions is not larger than $\alpha\lesssim65^\circ$ (including the RSG) and $\alpha\lesssim45^\circ$ (only bBSG). These less aligned results are obtained for the most spherical models, which do not have a clear preferred direction of the C+O ejecta. Moreover, these models tend to have the lowest kick velocities (see Table\,\ref{tab:kicks} for model M15--8b--1). In addition to the C+O--kick correlation, we observe that the most extended Ni- or Fe-rich plumes are never close to the direction of the C+O centre of mass and mostly to the opposite direction, which is expected from the tug-boat mechanism \citep{wongwathanarat2010}.

We attribute the C+O–kick correlation to the following mechanism. The Fe-rich plumes, accelerated by Rayleigh–Taylor instabilities, expand most rapidly in the directions where the explosion is strongest. As they push outward, they interact with the overlying layers containing C and O: some C+O material in the path of the plumes is dragged along and diluted during expansion, while other material is displaced sideways. As a result, the densest C+O regions tend to lie in directions where the Fe-rich plumes do not penetrate, leading to the observed anti-correlation between the Fe-rich plumes and the densest C+O clumps. 
This is also observed in SN\,1987A (see Fig.\,\ref{fig:co_sio_feI_comparisonII}), where the extended Fe-rich plumes are located close to the plane of the ring, while the CO (and SiO) emission is mainly perpendicular to this ring. Since the kick and the densest ejecta rich in lighter elements are expected to be anti-aligned with the most extended Fe-rich structures, a correlation between the directions of the NS kick and densest C+O ejecta seems natural.

Assuming that such a correlation also holds for the ejecta of SN\,1987A, we may estimate the potential direction of its compact remnant to be in the direction of the centre of emission of CO (cyan arrow in Fig.\,\ref{fig:co_sio_comparison}). This CO direction points towards the observer with an angle of about $45^\circ$ towards the north. Such a direction is consistent with kick estimates based on observations of $^{44}$Ti by \citet{boggs2015}, a dust cloud north-east of the explosion centre moving towards the observer \citep{Cigan:2019cl}, the inferred kick direction from the integrated momentum of the innermost Fe-rich ejecta \citep{larsson2025}. However, a northward orientation would be in tension with \cite{larsson2025}, who report evidence of southward-directed emission in the [Ar~\textsc{vi}] line, possibly linked to the compact object in SN\,1987A.

If the kick was in the direction of the CO centre of mass, it would point towards the northern of the extended [Fe~{\sc i}] clumps. If the two dominant [Fe~{\sc i}] clumps \citep{larsson2023} were to represent all of the iron, a kick in the direction of one of them would be inconsistent with a tug-boat mechanism in which the kick should be in the opposite direction. However, comparison with the $^{44}$Ti observations by \citet{boggs2015} suggests that there are more iron-rich ejecta, which do not emit in [Fe~{\sc i}] (see also point $v$ below). Another striking property of the CO centre of mass direction is its alignment with the plane of the ring of SN\,1987A. It would not be surprising if the binary origin of the progenitor may have an effect that links the remnant's kick direction to the orbital plane of the binary progenitor. In this context, the near-perpendicular distribution of the densest CO with respect to the ring plane may also be of interest. However, because our model does not explicitly account for the binary evolution or the formation of CO molecules (see caveats below), the possible alignment between the NS kick and the plane remains unexplained. Clarifying its origin will require further observational or theoretical evidence.

\end{itemize}

We have seen that a major problem in the theoretical models is the too steep rise and the too low $v_\mathrm{peak}$ compared to the observed distributions. Several factors in both the observations and the theoretical modelling may help explain these discrepancies. In addition, we list below a few points of caution that deserve attention when interpreting the results discussed above:

\begin{enumerate}[label=\textbf{(\roman*)}, leftmargin=1.5em, itemsep=0.5ex]
\item \textbf{Noise at high $v$}\\
Some models have significant C+O or Si+O densities at velocities higher than the observed molecular emission. This is not worrisome, because the observations are limited by the noise in this velocity regime. Too high model predictions for the velocity could simply mean that the material is not detected, because it is either buried in the noise, or it is at such high velocities that the ejecta are excluded by the truncation of the data to avoid contamination from the ring of SN\,1987A. 

\item \textbf{Explosion energy}\\
Our approximate approach of the neutrino transport allows us to choose the explosion energy (and the energy deposition time scale) as a free parameter, and it is not calculated self-consistently. Therefore, small differences in absolute values of, for example, the location of the velocity peaks and the largest detected velocities may be cured by assuming a different explosion energy for a given model. Since the velocities scale with the square root of the explosion energy, this basically limits the possible corrections to a few tens of per cent. However, the scaling with the explosion energy is limited to the maximal value of $\sim1.5$B, representing the estimate for SN\,1987A. If the velocities are scaled so that the peaks match, the models may predict more material at high velocities than is observed. As we have seen in point (i) of this list, this is a less severe problem.

\item \textbf{$\beta$ Decay}\\
Since our model comes with uncertainties with respect to the total amount of radioactive $^{56}$Ni synthesized, the discrepancies of too low peak velocities can be cured (partially) by assuming a larger production of  $^{56}$Ni. We have discussed this exemplarily for model B15. More realistic models with more sophisticated treatment of nucleosynthesis and neutrino transport would probably lead to higher $^{56}$Ni abundances. We have studied the expected effect exemplarily for model B15$_\text{X}$ (top left panels in Figs.\,\ref{fig:velocity_mass_co21}-\ref{fig:velocity_mass_sio_c2} and Table\,\ref{summarystatistics_both}). We obtain a shift of $200$km\,s$^{-1}$ for $v^\mathrm{CO}_\mathrm{peak}$ and $v^\mathrm{SiO}_\mathrm{peak}$ when changing from the numerically-obtained amount of $^{56}$Ni in model B15 to the same amount plus all heavy tracer elements in model B15$_\mathrm{X}$. A comparison with more sophisticated models gives indications that roughly 50\% of the tracer should be added to our estimate of the synthesized $^{56}$Ni. This is our standard approach for model groups 2 and 3. From these considerations we estimate that the uncertainty of the peak velocity in the models (caused by the $\beta$-decay treatment) is about 100 km\,s$^{-1}$. 

\item \textbf{Dust}\\
The molecules CO and SiO are formed from constituents that can also form dust grains. Therefore, a lack of molecular emission in the observations may be explained by the presence of carbon- or silicon-rich dust. 
The observed dust emission in SN\,1987A is concentrated in the very central regions, with $v\lesssim1000$km\,s$^{-1}$ or $2000$km\,s$^{-1}$, depending on the estimate on the 3D dust distribution \citep{Cigan:2019cl, Matsuura2024}. This very likely causes an under-sampling of the lowest velocity components in the fractional mass distributions of the observations compared to the models. We discuss the effect in Appendix\,\ref{ap:dust}. 
However, dust is not likely to be able to explain a lack of molecular emission at higher velocities $v>2000$km\,s$^{-1}$

\item \textbf{Assumptions about ejecta}\\
To translate the observed intensities into fractional mass distributions, we assumed that the ejecta are optically thin, are in LTE and have uniform temperature. We have discussed why these should be good approximations at the
end of Section \ref{sec:mass_distributions}. However, in particular, the assumption of LTE and constant temperature may not apply perfectly in the central regions of the ejecta. If the temperature of the central regions is higher than elsewhere, the molecular mass there would be underestimated in our analysis. A more detailed radiative transfer calculation would be needed to rigorously calculate the effect of this, but given the central `hole' seen in 3D and in the 2D mass distributions, the overall impact is likely to be small.

\item \textbf{Molecule formation and destruction}\\
The simulations do not yet offer the possibility to calculate the formation of molecules self-consistently. We approximated the molecule distribution as described in Section \ref{sec:hydromodels}. This procedure may lead to an overestimation where a lot of ions or atoms are found, but where the conditions are not beneficial for molecule formation. Therefore, the strongest constraints we can claim from our comparison are related to the regions where we do not find C+O or Si+O in the models, but where we have clear emission in the observations. See also the related discussion about dust in point (iv).

\item \textbf{Binarity and magnetic fields}\\
Although our most promising models are those from binary-merger BSGs, these progenitors did not include a self-consistent treatment of all effects of an evolution in a binary system. In particular, our simulations were performed without initial rotation and without magnetic fields. Although there is no clear indication of strong magnetic fields involved in the explosion of SN\,1987A, rotation or magnetic fields would naturally provide preferred directions.

\end{enumerate}

\section{Conclusions}

We created 3D intensity maps and computed the fractional mass distributions of SiO and CO in the ejecta of SN\,1987A. Free expansion of the ejecta is confirmed by comparing two epochs of SiO observations. In both SiO and CO, clear asymmetry is observed in the 3D morphology of the 3D intensity maps, consistent with observed emission line profile asymmetries \citep{Matsuura:2017cf}. For the densest CO material, we recover a two-punctured shell-like morphology as found by \citep{Abellan:2017by} and we estimate the plane of the ejecta, which can be fit to what remains of the shell after fragmentation, to be almost perpendicular to the ring plane ($84\pm11^\circ$). 

Comparing the fractional mass distributions to the predictions of hydrodynamical models of supernova explosions, we find that the models overpredict the amount of low-velocity material. This discrepancy may be alleviated by a number of factors: increased explosion energy or higher abundances of the $\beta$-decay isotope $^{56}$Ni in the models or dust reducing the formation of molecules in SN\,1987A. The best group of models to fit the observations is formed by binary-merger BSG progenitors. Model M15--7b--4 in particular, provides a good match to the CO fractional mass distribution and overall morphology of SN\,1987A. However, there is too little Si+O at high velocities. This is not limited by observational biases or model interpretation, and hints at a problem in the progenitor model itself. If there is no Si (and O) already present in the outer layers of the progenitor, it seems hard for the explosion to accelerate enough of these elements to velocities high enough to explain the observations. 

The theoretical models show that the distribution of the densest C+O ejecta is strongly correlated with the kick direction. Keeping in mind all caveats discussed in Section\,\ref{sec:summary}, such as the details of molecule formation, observational constraints, and dust, we can estimate the expected direction of the compact remnant in SN\,1987A by associating it with the direction of the centre of emission of the brightest CO ejecta. The resulting vector points toward the observer, at an angle of 45° to the north. This direction lies in the plane of the inner ring of SN\,1987A and is perpendicular to the plane of the CO ejecta. However, since our model does not explicitly include the binary evolution or CO molecule formation, the apparent alignment between the kick and the ring plane should be regarded as tentative.

In summary, we have demonstrated the strong potential of comparing 3D observations to 3D theoretical models to gain understanding and constraining not only the explosion mechanism, but also the structure of the exploding progenitor star. Within the model uncertainties, the theoretical models based on the delayed neutrino-driven explosions of binary-merger BSGs are able to reproduce the observed features of SN\,1987A qualitatively, and to a large extent also quantitatively. Further studies including the effects of molecule and dust formation are required to confirm the findings presented here.

\section*{Acknowledgements}
This paper makes use of the following ALMA data:  ADS/JAO.ALMA\#2013.1.00280.S, ADS/JAO.ALMA\#2015.1.00631.S, ADS/JAO.ALMA\#2018.1.00717.S, ADS/JAO.ALMA\#2017.1.00789.S. ALMA is a partnership of ESO (representing its member states), NSF (USA) and NINS (Japan), together with NRC (Canada), MOST and ASIAA (Taiwan), and KASI (Republic of Korea), in cooperation with the Republic of Chile. The Joint ALMA Observatory is operated by ESO, AUI/NRAO and NAOJ.
MM acknowledges funding from RadioNet, which made it possible for her to visit ALMA ARC in Manchester.
RadioNet has received funding from the European Union's Horizon 2020 research and innovation programme under grant agreement No 730562
MM and RW acknowledge support from STFC Consolidated grant (ST/W000830/1). RW acknowledges support from the Research Ireland Pathway programme under grant number 21/PATH-S/9360.
MG and BG acknowledge the support through the Generalitat Valenciana via the grant CIDEGENT/2019/031, the grant PID2021-127495NB-I00 funded by MCIN/AEI/10.13039/501100011033 and by the European Union, 
and the Astrophysics and High Energy Physics programme of the Generalitat Valenciana ASFAE/2022/026 funded by MCIN and the European Union NextGenerationEU (PRTR-C17.I1).
TJ is grateful for support by the German Research Foundation (DFG) through the Collaborative Research Centre ``Neutrinos and Dark Matter in Astro- and Particle Physics (NDM),'' Grant No.\ SFB-1258-283604770, and under Germany's Excellence Strategy through the Cluster of Excellence ORIGINS EXC-2094-390783311.
\section*{Data availability}
The raw ALMA data are available from the ALMA archive.\footnote{\url{https://almascience.eso.org}} Reduced data products will be made available on request to the authors.
For the purpose of open access, the author has applied a CC BY public copyright licence (where permitted by UKRI, ‘Open Government Licence’ or ‘CC BY-ND public copyright licence’ may be stated instead) to any Author Accepted Manuscript version arising.
Data of simulations will be made available on the Garching SN archive (\url{https://wwwmpa.mpa-garching.mpg.de/ccsnarchive/}).



\bibliographystyle{mnras}
\bibliography{sn1987a_velocity,theory}


\appendix
\section{Noise characteristics}

\begin{figure}
    \centering
    \includegraphics[width=0.5\textwidth]{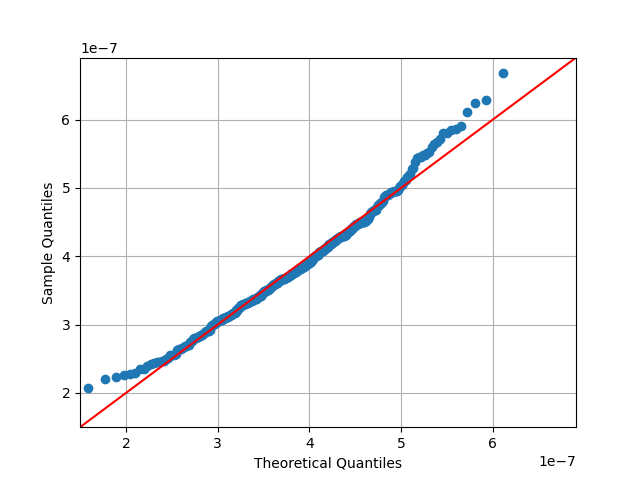}
    \caption{Q-Q plot for noise values in SiO 6--5 epoch 2 observation}
    \label{qqplot2}
\end{figure}

Fig.~\ref{qqplot2} shows one of the Q-Q plots that we produced when analysing the noise in each dataset. We analysed regions of the data outside the equatorial ring where no signal should be present. In that case, and if the noise values have a Gaussian distribution, then the points on the Q-Q plots should lie close to the $y=x$ line. We find that most points do lie close to the $y=x$ line, but points in the highest quantiles lie slightly above the line. The data shown in Fig.~\ref{qqplot2} shows the largest deviations from the $y=x$ line of the five datasets analysed. This points to a slight deviation from pure Gaussian noise, which will tend to cause uncertainties to be underestimated. However, the effect on our analysis will be minor.






\section{Uncertainty analysis}
\label{ap:noise}

We describe here the sources of uncertainty that we considered when estimating the uncertainties on our mass-velocity distributions. We show the results of our analysis for the CO 2--1 data; the results are similar for all the datasets.

\subsection{Position of the explosion site}

The most significant uncertainty arises from the uncertainty of the exact spatial location of the centre of the explosion. We estimate the associated error by varying the assumed centre of the explosion by up to $\pm5$ pixels N/S and E/W. At the distance of SN1987A, and given the different spatial sampling in the datasets analysed, this corresponds to an angular distance of up to 0.07\,arcsec, a linear distance of up to 3600 AU, and an uncertainty in velocity of up to 650~km\,s$^{-1}$. The best estimate of the position of the supernova progenitor star, Sanduleak -69$^{\circ}$~202, has an uncertainty of approximately 0.05\,arcsec (\citealt{Reynolds1995}).

\begin{figure}
\begin{overpic}[width=.5\textwidth]{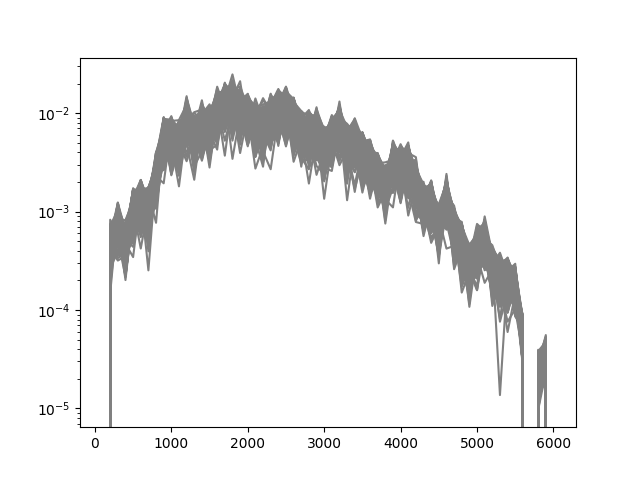}
   \put(0,30){\rotatebox{90}{$\Delta M / M_\mathrm{tot}$}}  
    \put(45,0){$v_r$ [km\,s$^{-1}$]}
    \end{overpic}
    \caption{1000 calculations of the mass-velocity curve for CO 2--1, each with the SN position taken at random from a position with $\pm$5 pixels of the nominal explosion site.}
    \label{positionuncertainty}
\end{figure}

Uncertainty in the position of the explosion site propagates into uncertainty in all velocities, and is thus the dominant source of uncertainty at low velocities. Fig.~\ref{positionuncertainty} shows 1000 analyses of the CO 2--1 data, in which the position of the SN is taken at random from a position within $\pm$5 pixels of our best estimate of the explosion site.

\subsection{Image noise}

To estimate the effect of noise in our observational data on our derived mass-velocity curves, we carried out a Monte Carlo analysis in which we repeated our analysis 1000 times. In each iteration, we added synthetic noise to our data, with a standard deviation equal to that measured from signal-free regions of our data cubes.

We create the synthetic noise by first generating pure uncorrelated normally-distributed noise values in each pixel, and then applying a Gaussian filter with a standard deviation of 3 pixels. This results in noise images which broadly resemble signal-free regions of the original data. We then add the noise images to the original data cubes and calculate the mass-velocity distributions.

\begin{figure}
\begin{overpic}[width=.5\textwidth]{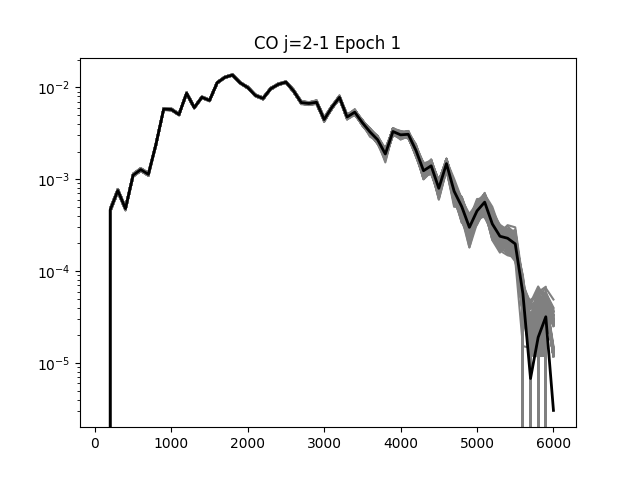}
   \put(0,30){\rotatebox{90}{$\Delta M / M_\mathrm{tot}$}}  
    \put(45,0){$v_r$ [km\,s$^{-1}$]}
    \end{overpic}
    \caption{1000 calculations of the mass-velocity curve for CO 2--1, each with synthetic noise generated as described in the text added to the original data before the analysis.}
    \label{noiseanalysis}
\end{figure}

The results of this analysis for CO J=2--1 are shown in Fig.~\ref{noiseanalysis}. The uncertainties resulting directly from noise in the images are found to be significant only at velocities greater than 3--4\,000\,km/s, and very small at velocities of $<$2000~km/s.

\subsection{Noise threshold}
\label{ap:threshold}

A final source of uncertainty that we consider is the calculation of the noise threshold used in our analysis. To estimate the magnitude of this, we vary the value used in the analysis from our fiducial value by factors between $\times$0.5 and $\times$2.

We find that, if overestimated, the noise threshold would result in high velocity material being excluded from the analysis and the high-velocity tail of the mass-velocity distributions thus being underestimated, but an underestimate of the noise threshold has little effect on the analysis. In either case, only  velocities $\gtrsim$ 3500~km\,s$^{-1}$ are affected.

\begin{figure}
\begin{overpic}[width=.5\textwidth]{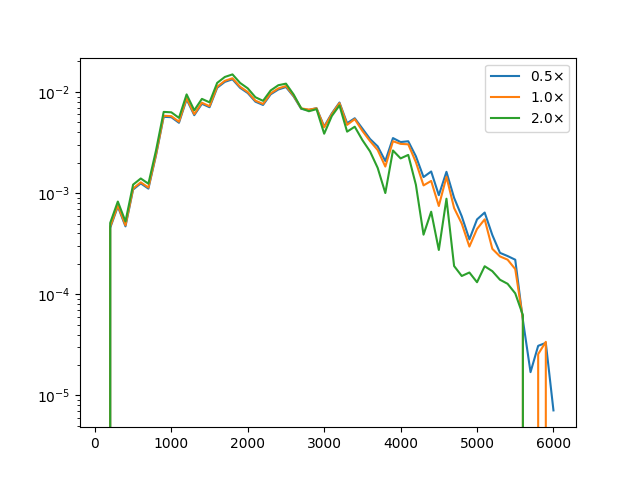}
   \put(0,30){\rotatebox{90}{$\Delta M / M_\mathrm{tot}$}}  
    \put(45,0){$v_r$ [km\,s$^{-1}$]}
    \end{overpic}
    \caption{Calculation of the mass-velocity curve for CO 2--1 using values of the noise threshold scaled by the factor shown in the plot legend.}
    \label{thresholdeffect}
\end{figure}

Fig.~\ref{thresholdeffect} shows the effect of different noise thresholds on the calculated mass-velocity curve for CO 2--1. Artificially reducing the noise threshold makes a negligible difference to the analysis, while artificially increasing it results in the exclusion of fainter, faster-moving material at velocities of $>$3000~km\,s$^{-1}$.

\section{3D distribution of dust emission}
\label{ap:dust}

The theoretical calculations of the distributions of CO and SiO assume that all the molecules are in the gas phase. In reality, some fraction of the molecules will have condensed into dust. To estimate the effect this might have when comparing the observations to theory, we calculated the distribution of dust mass with velocity using ALMA band 7 continuum data (see Table~\ref{observing_log}).

\begin{figure}
\includegraphics[width=0.5\textwidth]{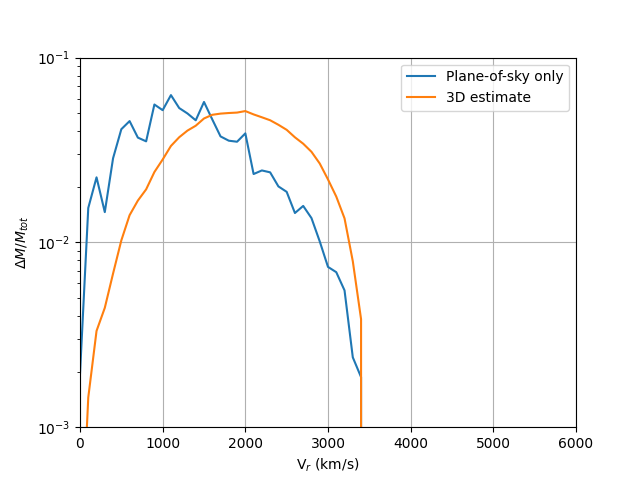}
\caption{The estimated distribution of dust with velocity. The left panel shows the relation estimated by assuming that all dust velocity is in the plane of the sky. The right panel shows the estimate made by assuming that the dust density is proportional to r$^{-2}$.}
\label{dustmassdistributions}
\end{figure}

It is not possible to obtain dust velocity information from its continuum emission. We assume that the dust is at a uniform temperature, and that its optical depth is negligible, such that the dust mass along a given line of sight is directly proportional to the observed flux. A simple limiting case then assumes that all dust velocities are in the plane of the sky. For each pixel in the dust continuum image, the velocity relative to the SN explosion location is given by simple trigonometry, and the mass is proportional to the observed flux. The estimated dust mass velocity distribution using this approximation is shown in the left panel of Fig.~\ref{dustmassdistributions}. The maximum velocity of continuum-emitting dust is found to be around 3500~km\,s$^{-1}$. Unless the dusty ejecta is significantly asymmetric and elongated along the line of sight, this will be representative of the maximum velocity overall.

A more realistic estimate needs to account for velocities along the line of sight. To proceed, we assume that the dust density declines proportionally to r$^{-2}$ with distance $r$ from the supernova explosion. At a given pixel position, if the dust shell is expanding with a velocity proportional to distance from the explosion point (ballistic expansion), then the dust velocities lie between the plane-of-sky velocity of the point and the maximum velocity of the shell, estimated at 3500~km\,s$^{-1}$. We calculate the dust mass at each velocity according to the assumed 1/r$^2$ density distribution. The dust mass-velocity relation derived using this approach is shown in the right hand panel of Fig.~\ref{dustmassdistributions}.

The left-hand panel assuming no velocity along the line of sight will systematically underestimate total velocities. The mass-velocity curve derived using this simple assumption peaks at about 1000~km\,s$^{-1}$. The right-hand panel, on the other hand, shows that a more realistic assumption of the dust velocity field along the line of sight results in the dust mass peaking at a velocity of about 2000km\,s$^{-1}$. Where many of the theoretical models have problems matching the observed molecular distributions is at velocities of $<$1000~km\,s$^{-1}$, where a number of them predict more material than is observed, and at $>$3000~km\,s$^{-}$ where they predict less material than observed. Condensation of molecules into dust therefore does not help us to resolve these discrepancies.

\section{Calculation of plane parameters}
\label{ap:plane}

As described in Section~\ref{co_sio_comparison_section}, in our 3D plots, CO and SiO have distinct spatial distributions. The SiO emission is more compact, and the more extended CO emission has a two-punctured shell-like appearance. This disrupted shell forms a plane approximately perpendicular to the plane of SN1987A's equatorial ring.

To determine the inclination of this plane, we calculated the distance of each point in the CO data from the assumed plane, and carried out a least-squares minimisation of the distance to determine the optimal parameters of the plane equation:

\begin{equation*}
    a(x-x_0) + b(y-y_0) + c(z-z_0) = 0
\end{equation*}

If we require the plane to intersect the site of the SN explosion, then (x$_0$,y$_0$,z$_0$)~=~(0,0,0), and we determine coefficients of

\begin{equation*}
\begin{aligned}
    a &= -0.137 \pm  0.04 \\
    b &= -0.454 \pm  0.12 \\
    c &= 0.707 \pm 0.19
\end{aligned}
\end{equation*}

On the other hand, if we do not require the plane to coincide with the SN explosion site, the coefficients are

\begin{equation*}
\begin{aligned}
    a &= -0.123 \pm  0.04 \\
    b &= -0.567 \pm  0.15 \\
    c &= 0.658 \pm  0.18 \\
    x_0 &= 5.004 \pm  2.20 \\
    y_0 &= 5.718 \pm  1.90 \\
    z_0 &= 4.050 \pm  1.61 \\
\end{aligned}
\end{equation*}

This places the geometric centre of the CO about 8600 AU from the site of the explosion, corresponding to a velocity of around 1500~km\,s$^{-1}$. The two planes are almost parallel, with an angle of about 8$^\circ$ between them.

Propagating the uncertainties on the plane parameters into the angle between this plane and the equatorial ring, we obtain a value of 84$\pm$11$^\circ$. The orientation of this plane perpendicular to the equatorial ring is surprising, and given the uncertainties, would have only about a 10 per cent probability of arising purely by chance.

\bsp	
\label{lastpage}
\end{document}

%% file: observing_log.tex
CO $J$=2--1  & 230.54 & 1     & 2014 Sep 02    & 10 053 & ~~8.6 & 
     \multirow{2}{*}{{\hspace{-1em}$\left.\begin{array}{l} \end{array}\right\rbrace 0.057 \times 0.042$}}  & 
     \multirow{2}{*}{27} & \multirow{2}{*}{6} & \multirow{2}{*}{100} & \multirow{2}{*}{0.04} \\
      &        & 1     & 2015 Nov 01-02 & 10 479 & ~71.7 & \\
CO $J$=6--5  & 691.47 & 1     & 2015 Sep 25    & 10 441 & ~12.1 & $0.088 \times 0.064$  & 175 & 6 & 100 & 2.82 \\
SiO $J$=5--4  & 217.10 & 1     & 2014 Sep 02    & 10 053 & ~~9.1 & 
     \multirow{2}{*}{{\hspace{-1em}$\left.\begin{array}{l} \end{array}\right\rbrace 0.058 \times 0.042$}}  & 
     \multirow{2}{*}{21} & \multirow{2}{*}{6} & \multirow{2}{*}{100} & \multirow{2}{*}{0.05} \\
             &        & 1     & 2015 Nov 01    & 10 479 & ~84.1 & \\
             &        & 2     & 2017 Nov 07    & 11 215 & 208.9 & $0.056 \times 0.034$  &  47 & 5 & 300 & 0.033 \\
SiO $J$=6--5 & 260.52 & 1     & 2014 Sep 02    & 10 053 & ~~9.1 & 
     \multirow{2}{*}{{\hspace{-1em}$\left.\begin{array}{l} \end{array}\right\rbrace 0.041 \times 0.034$}}  & 
     \multirow{2}{*}{172} & \multirow{2}{*}{6} & \multirow{2}{*}{100} & \multirow{2}{*}{0.06} \\
             &        & 1     & 2015 Nov 02    & 10 480 & ~86.2 & \\
             &        & 2     & 2019 Aug 07-08 & 11 854 & ~95.6 & $0.106 \times 0.101$  &  70 & 10 & 300 & 0.055 \\
\hline
Continuum & 315 & 1 & 2015 Jun 28 & 10 352 & \\
\hline

%% file: hemispheres_table.tex
\begin{table}
\setlength{\tabcolsep}{4pt}  
\begin{tabular}{llllll}
\hline
Transition & Epoch & N:S & E:W & Near:Far & Below:Above \\
\hline
CO 2-1 & 1 & 0.64:0.36 & 0.60:0.40 & 0.50:0.50 & 0.41:0.59 \\
CO 6-5 & 1 & 0.49:0.51 & 0.58:0.42 & 0.56:0.44 & 0.58:0.42 \\
SiO 5-4 & 1 & 0.70:0.30 & 0.63:0.37 & 0.48:0.52 & 0.37:0.63 \\
SiO 6-5 & 1 & 0.62:0.38 & 0.61:0.39 & 0.50:0.50 & 0.44:0.56 \\
SiO 5-4 & 2 & 0.54:0.46 & 0.58:0.42 & 0.56:0.44 & 0.50:0.50 \\
SiO 6-5 & 2 & 0.56:0.44 & 0.59:0.41 & 0.51:0.49 & 0.49:0.51 \\
\hline
\end{tabular}
\caption{Fraction of emission from hemispheres of the remnant of SN1987A, with the assumed supernova progenitor position lying on the dividing planes. ``Below:Above" refers to the plane of the equatorial ring, with material on the north side of the plane being considered ``above" and the rest ``below".}
\label{hemispheres}
\end{table}

%% file: distribution_parameters_observed.tex
\begin{table}
\centering
\begin{tabular}{llll}
\hline
Dataset & v$_\mathrm{peak}$ [km\,s$^{-1}$]& width [km\,s$^{-1}$]& v$_\mathrm{max}$ [km\,s$^{-1}$]\\
\hline
CO $J$=2-1      & 1830$\pm$210 & 1210$\pm$70 & 3980$\pm$40 \\
\hline
SiO $J$=5-4 ep.1 & 1730$\pm$190 & 1180$\pm$60 & 3600$\pm$30 \\
SiO $J$=5-4 ep.2 & 1600$\pm$80 & 870$\pm$60 & 3110$\pm$60 \\
SiO $J$=6-5 ep.1 & 1820$\pm$190 & 930$\pm$60 & 3220$\pm$50 \\
SiO $J$=6-5 ep.2 & 1530$\pm$100 & 1000$\pm$50 & 3510$\pm$70 \\
SiO (average)    & 1600$\pm$110 & 1080$\pm$70 & 3560$\pm$50 \\
\hline
\end{tabular}
\caption{Parameters of the velocity distributions (as defined in the text) for the observed mass distributions of CO and SiO.}
\label{summarystatistics_obs}
\end{table}

%% file: distribution_parameters_both.tex
\begin{table*}
\centering
\begin{tabular}{llll|llll}
\hline
Dataset & $v_\mathrm{peak}$ [km s$^{-1}$]& width [km s$^{-1}$]& $v_\mathrm{max}$ [km s$^{-1}$] & Dataset & $v_\mathrm{peak}$ [km s$^{-1}$]& width [km s$^{-1}$]& $v_\mathrm{max}$ [km s$^{-1}$]\\
\hline
CO $J$=2-1      & 1830$\pm$210 & 1210$\pm$70 & 3980$\pm$40 & SiO (average)    & 1600$\pm$110 & 1080$\pm$70 & 3560$\pm$50 \\
\hline
B15             &                   850 &                   600 &                  2250 & B15             &                   950 &                   700 &                  2450 \\
B15$_{\text{no}\beta}$      &                   750 &                   700 &                  2250 & B15$_{\text{no}\beta}$&                   750 &                   800 &                  2450 \\
B15X            &                  1050 &                   500 &                  2350 & B15X            &                  1150 &                   600 &                  2650 \\
L15             & \cellcolor{green}1350 & \cellcolor{green}1000 & \cellcolor{green}4050 & L15             &                  1150 &                  1500 &                  4550 \\
N20             & \cellcolor{green}1450 &                   400 &                  2050 & N20             & \cellcolor{green}1450 &                   400 &                  1950 \\
W15             &                   950 & \cellcolor{green}1400 &                  4550 & W15             &                   850 &                  1800 &                  4850 \\
\hline
M15-7b-3        & \cellcolor{green}1350 &                   800 &                  3250 & M15-7b-3        &                  1150 &                   800 &                  2550 \\
M15-7b-4        & \cellcolor{green}1550 &                   900 &                  3550 & M15-7b-4        &                  1250 &                   700 &                  2650 \\
M15-8b-1        & \cellcolor{green}1550 &                   600 &                  3250 & M15-8b-1        & \cellcolor{green}1450 &                   400 &                  1750 \\
\hline
WH12.5          & \cellcolor{green}1350 & \cellcolor{green}1000 &                  3750 & WH12.5          &                  1050 &                  1600 &                  4150 \\
SW19.8          & \cellcolor{green}1850 &                   700 & \cellcolor{green}3950 & SW19.8          & \cellcolor{green}1450 &                   400 &                  2250 \\
SW20.8          &                   950 &                   800 &                  2950 & SW20.8          &                   950 &                  1300 & \cellcolor{green}3450 \\
SW27.3          & \cellcolor{green}1850 &                   800 &                  3350 & SW27.3          &                  1050 & \cellcolor{green} 900 &                  3050 \\
\hline
\end{tabular}
\caption{Parameters of the velocity distributions (as defined in the text) for both observed and theoretical mass distributions of CO and SiO. Where the theoretical values are within 3$\sigma$ of the observed values, the table cell is highlighted in green.}
\label{summarystatistics_both}
\end{table*}